\documentclass[twocolumn,showpacs,preprintnumbers,amsmath,amssymb]{revtex4}

\usepackage{graphicx}
\usepackage{dcolumn}
\usepackage{bm}

\begin{document}


\title{The optical depth of the Universe \linebreak to ultrahigh energy cosmic
  ray scattering in the magnetized large scale structure}

\author{Kumiko Kotera}
\email{kotera@iap.fr} 
\author{Martin Lemoine}
\email{lemoine@iap.fr}
\affiliation{
Institut d'Astrophysique de Paris\\
UMR7095 - CNRS, Universit\'e Pierre \& Marie Curie,\\
98 bis boulevard Arago\\
F-75014 Paris, France}

\date{\today}

\begin{abstract}
This paper provides an analytical description of the transport of
ultrahigh energy cosmic rays in an inhomogeneously magnetized
intergalactic medium. This latter is modeled as a collection of
magnetized scattering centers such as radio cocoons, magnetized
galactic winds, clusters or magnetized filaments of large scale
structure, with negligible magnetic fields in between. Magnetic
deflection is no longer a continuous process, it is rather dominated
by scattering events. We study the interaction between high energy
cosmic rays and the scattering agents. We then compute the optical
depth of the Universe to cosmic ray scattering and discuss the
phenomological consequences for various source scenarios. For typical
parameters of the scattering centers, the optical depth is greater
than unity at $5\times10^{19}\,$eV, but the total angular deflection
is smaller than unity.  One important consequence of this scenario is
the possibility that the last scattering center encountered by a
cosmic ray be mistaken with the source of this cosmic ray. In
particular, we suggest that 
part of the correlation recently reported by the Pierre Auger
Observatory may be affected by such delusion: this experiment may
be observing in part the last scattering surface of ultrahigh
energy cosmic rays rather than their source population.  Since the
optical depth falls rapidly with increasing energy, one should probe
the arrival directions of the highest energy events beyond
$10^{20}\,$eV on an event by event basis to circumvent this effect.
\end{abstract}

\pacs{Valid PACS appear here}
                             
                             
\maketitle

\section{Introduction}

The problem of the origin of ultrahigh energy cosmic rays has
generally been expressed as the conjunction of two questions: {\it
  (i)} how can particles be accelerated to energies in excess of
$10^{20}\,$eV?  {\it (ii)} why is the source not seen in the arrival
directions of the highest energy events? Progress on the former
question has certainly been hindered by our relative lack of knowledge
on acceleration mechanisms and high energy processes in the most
powerful astrophysical objects. Regarding the latter question,
progress has been mostly limited by the scarcity of experimental data
at the highest energies, at least until very recently.

Indeed the first results of the Pierre Auger Observatory, which have
just been published, report a significant correlation of the arrival
directions of the highest energy events with a catalog of active
galactic nuclei (AGN) closer than 75~Mpc~\cite{Auger1,Auger2}. This
observation certainly marks an important step in the search for the
source of ultrahigh energy cosmic rays. However one should not
overinterpret the significance of these results. In particular, the
likelihood of the reported coincidence rests on the comparison with
isotropic arrival directions, yet the large scale structure is known
to be highly inhomogeneous at least up to 75~Mpc. Since AGN are known
to cluster with the large scale structure, one cannot exclude at
present that the observed correlation remains a coincidence if the
source itself clusters with the large scale
structure~\cite{Auger2}. More will be said on these data in
Section~\ref{sec:disc} of the present paper.

Furthermore, there exist other (and sometimes contradictory) claims in
the literature on the existence of correlations of ultrahigh energy
cosmic ray arrival directions with various source
catalogs~\cite{Tea95}, the strongest being the association with BL
Lacertae objects reported in Refs.~\cite{TT01,GTTT04,TT04} (see also
Refs.~\cite{EFS04,HiRes06}). Since the existing data is so scarce at
the highest energies, the assessment of the statistical significance
remains a difficult task. Finally, the reported evidence for
multiplets of events tends to suggest that the source lies in the
arrival direction of the events clusters. However some of these
clusters show interacting galaxies as the sole peculiar objects on the
line of sight~\cite{Uea00}, while a more recent multiplet appears
correlated with interacting clusters of
galaxies~\cite{Fea05,HiRes05}. Taken at face value, all these claims
do not allow to draw a clear and consistent picture of the source of
ultrahigh energy cosmic rays.
  
It is admitted that cosmic magnetic fields must play a key role in
this puzzle, although which role exactly is also a question that is
still seeking for an answer. And this source of uncertainty is in turn
related to our poor knowledge of the strength and the distribution of
extragalactic magnetic fields (see Ref.~\cite{K94,V97} for detailed
reviews of existing data).  There exists a rather large body of
literature on the relation between cosmic magnetic fields and
ultrahigh energy cosmic rays. Most studies have constructed models of
extragalactic magnetic fields and then resorted to Monte Carlo
simulations in order to quantify the influence of these fields on the
time, energy and angular images expected in large scale detectors.
One must however underline the analytical works of
Refs.~\cite{WM96,1999astro.ph..7060A} on cosmic ray transport in
tangled extragalactic magnetic fields of homogeneous power, those of
Refs.~\cite{2002JHEP...03..045H,2002JHEP...07..006H} which discuss the
particular effect of magnetic lensing and finally
Refs.~\cite{1979Natur.281..356W,1990A&A...232..582B,1999PhRvD..59b3001B}
which discuss diffusive transport in a magnetized supercluster.

Earlier numerical studies have addressed the phenomenology of
ultrahigh energy proton propagation in tangled magnetic fields of
homogeneous power
~\cite{1997APh.....7..213L,1997APh.....6..337M,LSOS97,1997PhRvD..56.4470S,
  1998APh.....9..221C,2000PhRvD..62i3005S,2003ApJ...586.1211Y,2004APh....21..609D}. There
has since been a trend toward more realistic magnetic field
configurations. For instance,
Refs.~\cite{1998ApJ...505L..79M,1999APh....10..141S,1999astro.ph..3124L,2001PASJ...53.1153I,2002PhRvD..65b3004I,2002PhRvD..66h3002I}
have studied the diffusive or non-diffusive propagation in a
magnetized local supercluster and Ref.~\cite{2003PhRvD..68j3004S} has
brought to light the spectral distortions induced by the interaction
of ultrahigh energy cosmic rays with a supercluster harboring large
scale regular magnetic fields. More recently, several studies have
attempted to model a realistic configuration in which the magnetic
field follows the matter density and then studied the transport of
ultrahigh energy cosmic rays in the resulting structure. In order to
construct the magnetic field,
Refs.~\cite{2003PhRvD..68d3002S,2004PhRvD..70d3007S,
  2005PhRvD..72d3009A,DGST04,DGST05,KRC07,2007PhRvD..75j3001S} have
used numerical simulations of large scale structure formation
involving a passive magnetic field whose strength was normalized to
the value measured in clusters of
galaxies. Refs.~\cite{1997astro.ph..7054M,2006ApJ...639..803T,TS07,KL07}
have rather reconstructed the extragalactic magnetic field by scaling
the field strength to the underlying density field.

In general, these studies have assumed the magnetic field to be all
pervading (albeit, with a more or less pronounced degree of
inhomogeneity) so that magnetic deflection has been modeled as a
continuous process. This assumption has been relaxed in
Ref.~\cite{MTE01} which provides numerical simulations of cosmic ray
arrival directions after scattering with fossils of radio-galaxy
lobes. Similarly, Ref.~\cite{2002astro.ph.10095B} has mentioned the
possibility of discrete cosmic ray interactions with localized regions
of enhanced magnetic fields, their discussion pointing toward clusters
of galaxies as the main scattering agents.

This picture in which ultrahigh energy cosmic ray transport occurs through
random discrete events is indeed more likely to be valid on distance
scales up to a few hundreds of Mpc as a consequence of the high degree
of clustering of matter in the Universe. For instance, even if the
magnetic field were produced in a uniform manner at high redshift (see
Ref.~\cite{W02} for a review of models of the origin of large scale
magnetic fields), then the present-day magnetic field should be highly
inhomogeneous, as a result of the amplification of the magnetic field
in the shear and compressive flows associated with the formation of
non-linear structures~\cite{1998AA...335...19R,SME04,DGST05,TS07,KRC07} (see
also~\cite{D06} for a general discussion). In these simulations, voids
in the large scale structure are essentially deprived of magnetic
field.

Furthermore, if one attributes the origin of the extragalactic
magnetic field to pollution by a sub-class of galaxies, for instance
starburst galaxies \cite{KLH99,Bea00,BVE06} or radio-galaxies
~\cite{RS68,FL01,GKW01}, the magnetic field configuration should
resemble that of a percolating process (see Ref.~\cite{BSW05} for a
clear illustration). As explained further below, if the filling factor
of the polluted regions becomes comparable to that of the filaments of
large scale structure, the filaments themselves become the scattering
agents as in Refs.~\cite{SME04,DGST05,TS07,KRC07}.

The goal of the present paper is to provide an analytical description
of ultrahigh energy cosmic ray transport in such an inhomogeneous
medium, which is modeled by scattering centers embedded in an
unmagnetized intergalactic medium.

These scattering centers comprise the filaments just as the clusters
of galaxies but also all possible regions of locally enhanced magnetic
fields, such as galactic winds, groups of galaxies, large scale
structure shocks and fossil radio-galaxy cocoons. One motivation of
the present work is thus to make progress toward a more realistic
magnetic field configuration which takes into account those localized
regions of intense magnetic activity. In order to do so, we first
sketch a census of relevant scattering centers~(Section~\ref{sec:od})
then analyse their respective influence. 

The present work is further motivated by the fact that
Refs.~\cite{SME04,DGST05,TS07,KRC07} diverge as to the conclusions
they draw on the influence of the extragalactic magnetic fields on
ultrahigh energy cosmic rays, even though they try to construct ab
initio predictions for the distribution of these large scale magnetic
fields. This difference stems from the uncertainty on the origin of
these magnetic fields, not withstanding the complexity of modeling
accurately the evolution of magnetic fields in the formation of large
scale structure. Analytical tools become useful in this context as
they allow to parametrize the influence of such magnetic fields on the
images and spectra of ultrahigh energy cosmic rays. This in turn will
help to deconvolve this effect from existing and upcoming data, and
therefore to infer useful constraints on these magnetic fields.

In the present description, magnetic deflection is no longer a
continuous process, but is instead dominated by scattering events. We
thus use the notion of the optical depth of the Universe to
ultrahigh energy cosmic ray scattering and discuss the
phenomenological consequences. In particular, we show that the
optical depth decreases very abruptly as the energy increases,
because the source distance scale decreases due to increasing energy
losses, and because the influence of cosmic magnetic fields
diminishes with increasing energy.

We argue that the energy beyond which the Universe becomes translucent
or transparent to cosmic ray scattering may be tantalizingly close to
the threshold beyond which experiments search for counterparts,
$E\simeq 4-6\times10^{19}\,$eV. This could have profound consequences
for our interpretation of existing data. For instance, if most sources
lie beyond the last scattering surface, one could mistake the
scattering centers on the last scattering surface (such as starbursts,
old radio-galaxies or giant shock waves) with the source of ultrahigh
energy cosmic rays.

The phenomenological consequences thus differ widely from the case of
continuous deflection in an all-pervading medium. We thus discuss in
some detail the expected effects and their relation to current and
future observations of cosmic ray arrival directions.

This paper is laid out as follows. In Section~\ref{sec:od}, we sketch
a census of possible scattering centers and their influence on the
optical depth of the Univers to cosmic ray scattering. We also
calculate the distance to the last scattering surface and compare it
to the expected source distance scale. In Section~\ref{sec:crt}, we
discuss the transport of cosmic rays in this strongly inhomogeneous
medium and the expected observational consequences. We notably
provide sky maps of the expected optical depth up to different
distances for our local Universe. Finally, in
Section~\ref{sec:disc}, we summarize our findings and comment on the
existing data in the framework of the present model. The physics of
the interaction of cosmic rays with scattering centers is discussed in
Appendix~\ref{sec:appA}.

\section{The optical depth of the Universe to high
  energy cosmic ray scattering}\label{sec:od}

\subsection{Scattering centers in the large scale structure}
We adopt a description in which the extragalactic magnetic field is
inhomogeneous. If this magnetic field originates from a sub-class of
galaxies, its configuration is bound to follow that of the large scale
structure since the mixing length in the Universe is small for
cosmological standards: for typical intergalactic velocities of
$\sim300\,$km/s, the length traveled in a Hubble time is only
$\,\simeq\,4\,$Mpc. Note that the mixing length is even less in
filaments, for which the typical dispersion of velocities is of order
$50\,$km/s.

Obviously, at a given energy, the total optical depth to cosmic ray
scattering is dominated by the structures with the largest $n \sigma$,
where $n$ represents the space density and $\sigma$ the cross-section
of the magnetized halo. One should thus focus on the radio halos of
radio-galaxies, the magnetized winds of star forming galaxies, and on
larger scales to clusters of galaxies and filaments as well as their
surrounding accretion shock waves.

\subsubsection{Radio halos}

Radio halos of old radio-galaxies (deemed radio ghosts) have been
already considered as possible sites of ultrahigh energy cosmic ray
scattering in Ref.~\cite{MTE01}. This study evaluates their space
density as $n_{\rm rg}\,\simeq\,10^{-2}-10^{-1}\,$Mpc$^{-3}$, the radius
of their magnetized halos as $r_{\rm rg}\,\sim\, 0.5-1\,$Mpc and their
magnetic field $B_{\rm rg}\,\sim\,1\,\mu$G. Such quasar outflows have
also been examined in detail in Ref.~\cite{FL01} as a site of magnetic
pollution of the intergalactic medium, but their results differ from
those above. These latter authors find a much lower magnetic field
strength $B_{\rm rg}\,\sim\, 10^{-9}\,$G, and a substantially larger
extent, $r_{\rm rg}\,\simeq\,1-5\,$Mpc, for a comparable space
density. With respect to the results of Ref.~\cite{FL01}, the
scattering should be dominated by the sub-population of recently
formed quasars (at redshifts $z\lesssim4$), which have $r_{\rm
  rg}\,\sim\,2-4\,$Mpc and $B_{\rm rg}\,\sim\, 3\times10^{-9}\,$G.
The main difference between these calculations results from the
different modeling of the bubble evolution. The former study assumes
that the bubble settles in pressure equilibrium in a rather dense and
hot intergalactic environment (with $\rho/\langle
\rho\rangle\,\sim\,30$ and $T\,\simeq\,10^8\,K$) while the latter
argues that the bubble expands until its velocity matches that of the
Hubble flow and takes the surrounding IGM to be much colder and less
dense ($T\,\sim\,10^4\,$K and $\rho/\langle\rho\rangle\,=\,1$). Both
fix the magnetic strength to lie at a fraction of equipartition with
thermal energy, although this fraction to equipartition
$\epsilon_B=0.5$ in Ref.~\cite{MTE01} and $\epsilon_B=0.1$ in
Ref.~\cite{FL01}; furthermore, Ref.~\cite{FL01} adopt $\epsilon_B$ as
the equipartition fraction before expansion of the bubble, assuming
that the magnetic field then decays with expansion. This study thus
neglects all possible further amplification mechanisms of $B$, hence
their estimate (at a given $\epsilon_B$) should be considered as a
lower limit. If one instead considers $\epsilon_B$ as the
equipartition fraction of the magnetic field at present, the magnetic
field strength inside the bubble can be related to the kinetic energy
of the outflow and the size of the bubble as follows:
\begin{equation}
B_{\rm rg}\,=\,5\times10^{-8}\,{\rm G}\,\left({\epsilon_B\over
  0.1}\right)^{1/2} \left({E_{\rm rg}\over 10^{59}\,{\rm
    ergs}}\right)^{1/2} \left({r_{\rm rg}\over 1\,{\rm
    Mpc}}\right)^{-3/2}\ .
\end{equation}
This latter estimate agrees with the conclusions of Ref.~\cite{GKW01}
which studies the degree of magnetization of the IGM by radio-galaxies
jets and lobes.  The outflow energy $10^{59}\,$ergs is an average energy
for a quasar population~\cite{MTE01}: it corresponds to a black hole
mass $M_{\rm BH}\,\simeq\,3\times10^7\,M_\odot$, radiating $L_{\rm
bol}\,\simeq\,3\times10^{45}\,$ergs/s over $10^7\,$yrs~\cite{FL01}. Note
however that the observational compilation of Ref.~\cite{KDLC01} leads
to slightly higher values for $E_{\rm rg}$ and $B_{\rm rg}$. These
authors have observed that the lobes of 70\% of field radio-galaxies in
their sample have a much higher energy content $\sim
10^{60}-10^{61}\,$ergs than the remaining 30\% in clusters (about
$10^{58}\,$ergs), with a typical volume $V\,\sim\, 0.03-0.3\,$Mpc$^{3}$
and inferred minimum energy magnetic field strengths in the range
$3-30\,\mu$G. If the magnetic field is to decay as $V^{2/3}$ during the
subsequent expansion of these bubbles, the final value for $B_{\rm rg}$
would be of order $0.1\,\mu$G for a typical radius $r_{\rm
rg}\,\simeq\,3\,$Mpc as above. In the following, we thus consider the
possible range of values $B_{\rm rg}\,=\,1-10\times10^{-8}\,$G and
typical radius $r_{\rm rg}\,\simeq\,1-3\,$Mpc.

Finally, Refs.~\cite{MTE01,FL01} estimate the space density of quasar
outflows from the observed density of quasars at high redshifts and
the typical duration of the quasar phase (taken as $10^7\,$yrs). Their
estimate of $\sim10^{-2}-10^{-1}\,$Mpc$^{-3}$ agrees with the recent
determinations of the black hole number density at low redshifts, in
particular $n(>10^7\,
M_\odot)\,\simeq\,2-4\times10^{-2}\,$Mpc$^{-3}$~\cite{FF05}, 
although Ref.~\cite{2007ApJ...662..808L} reports a number density
that is smaller by about an order of magnitude. In what follows, we
thus consider the range $n_{\rm rg}\,=\,3\times10^{-3}-3\times
10^{-2}\,$Mpc$^{-3}$.

\subsubsection{Magnetized galactic winds}
Galactic winds have been proposed as a source of magnetic pollution of
the intergalactic medium by various authors, see in particular
~\cite{KLH99,Bea00,BVE06}. Such outflows have been observed in
different galaxies, for instance in the starbursting nearby dwarf
galaxy M82 with wind speed $v\,\simeq\,2000\,$km/s and extension
$\sim\,10\,$kpc~\cite{SO91}, or in massive star forming Lyman break
galaxies at high redshifts with wind speed $v\,\sim\,1000\,$km/s and
extending as far as hundreds of kpc~\cite{Pea02}, maybe up to
$\,\simeq\,1\,$Mpc~\cite{Aea03} (see Ref.~\cite{H01} for a review).

Galactic winds are also a key ingredient for theoretical models which
attempt at explaining the metal enrichment of the intergalactic
medium~\cite{Aea01,Cea05,BSW05,Sea06}. At the present time, it is not
clear which galaxy type (if any) dominates the pollution. Starburst
dwarf galaxies appear more akin at producing large winds, however they
also have a smaller gaseous content and a smaller energetic
reservoir. In the following, we use the most recent simulations of
Ref.~\cite{BSW05} which detail the properties of galactic winds. This
study shows that the number of wind-blowing galaxies is relatively
insensitive to the stellar mass of the parent galaxy in the range
$10^8\,M_\odot\,\lesssim M_*\,\lesssim\,10^{10}\,M_\odot$ as a result
of the opposed influences of wind ram pressure and amount of infalling
material, and that this number falls at both ends of this mass
range. At $z\,\simeq\,0$ and in this mass range, the typical wind
radius increases slowly with galaxy mass as follows: $r_{\rm
  gw}\,\simeq\,200\,$kpc for $M_*\,=\,10^8\,M_\odot$, $r_{\rm
  gw}\,\simeq\,800\,$kpc for $M_*\,=\,10^9\,M_\odot$, and $r_{\rm
  gw}\,\simeq\,1\,$Mpc for $M_*\,=\,10^{10}\,M_\odot$. Overall, the
contribution $n_{\rm gw}r_{\rm gw}^2$ will be dominated by dwarf
galaxies of stellar mass $M_*\,\sim\,10^9\,M_\odot$. The number
density $n_{\rm gw}$ of galaxies surrounded by a wind at $z=0$ can be
derived from the filling factor $f_{\rm gw}$ of the winds;
unfortunately, this quantity appears to depend strongly on the model,
taking values between $2\times 10^{-2}$ and unity. The median value
corresponds to $f_{\rm gw}\,=\,0.1-0.2$, which gives a density $n_{\rm
  gw}\,\simeq\,f_{\rm gw}/V_{\rm
  gw}\,\simeq\,2.-5\times10^{-2}\,$Mpc$^{-3}$, with $V_{\rm
  gw}\,=\,(4\pi/3)r_{\rm gw}^3$ the wind volume. Note that this number
is comparable to the number density of galaxies of stellar mass above
$10^8-10^9\,M_\odot$. If the filling factor becomes substantially
larger, the galactic winds will overfill the filaments in which they
reside, hence the filaments themselves become the scattering centers.

Concerning the strength of $B_{\rm gw}$, Ref.~\cite{BVE06} indicates
that most winds have a magnetic field with $B_{\rm gw}\,\simeq\,
10^{-8}-10^{-7}\,$G at $z=0$, the range covering conservative and
optimistic assumptions concerning the amplification of $B_{\rm gw}$.
Such amplification may have been detected in the outflow of M82, where
a magnetic field strength as high as $10\,\mu$G~\cite{Rea92} has been
reported in the first $10\,$kpc. Ref.~\cite{Bea00} has argued that the
magnetic field could be amplified through the Kelvin-Helmholtz
instability during ejection.

\subsubsection{Clusters of galaxies}

Clusters of galaxies are rare structures in the Universe, $n_{\rm
  cg}\,\simeq\, 10^{-5}h_{70}^3\,$Mpc$^{-3}$, but they are known to
host strong magnetic fields, with $B_{\rm cg\vert c}\,\sim\,
1-10\,\mu$G in the innermost radius $r_{\rm cg\vert
  c}\,\sim\,100\,$kpc~\cite{K94,2001ApJ...547L.111C}. Measurements of
the magnetic field in the cluster outskirts are rather scarce as a
result of the smaller electron density and magnetic field
strength. The minimum energy interpretation of recent synchrotron data
nevertheless indicates that $B_{\rm cg}\,\sim\,1\,\mu$G out to $r_{\rm
  cg}\,\sim\,1$Mpc~\cite{Gea06}. Theoretical expectations tend to
differ. For instance, Ref.~\cite{DGST05} shows that $B$ varies with
cluster mass, and indicates that for a massive cluster $B_{\rm
  cg}\,\sim\,1\,\mu$G within $r_{\rm cg}\,\simeq\,0.2\,$Mpc, then
falls to $B_{\rm cg}\,\sim\,10^{-7}\,$G within $r_{\rm
  cg}\,\simeq\,1\,$Mpc, $B_{\rm cg}\,\sim\,10^{-8}\,$G within $r_{\rm
  cg}\,\simeq\,2\,$Mpc and finally $B_{\rm cg}\,\sim\,10^{-9}\,$G
within $r_{\rm cg}\,\simeq\,4-5\,$Mpc, while Fig.5 of
Ref.~\cite{Bea05} indicates more extended magnetic fields, with
$B_{\rm cg}\,\sim\,1\,\mu$G within $r_{\rm cg}\,\simeq\,1\,$Mpc, then
falls to $B_{\rm cg}\,\sim\,10^{-7}\,$G within $r_{\rm
  cg}\,\simeq\,3\,$Mpc, $B_{\rm cg}\,\sim\,10^{-8}\,$G within $r_{\rm
  cg}\,\simeq\,4\,$Mpc and finally $B_{\rm cg}\,\sim\,10^{-9}\,$G
within $r_{\rm cg}\,\simeq\,5\,$Mpc. In the following, we take these
two limits as a range for $B_{\rm cg}$ and $r_{\rm cg}$.

Note that about half of galaxies lie outside of clusters, hence one
can treat clusters of galaxies and the above field radio ghosts and
field galactic winds as distinct scattering centers.

\subsubsection{Filaments and walls of large scale structure}

Filaments or walls of large scale structure are not expected to be
sources of magnetic pollution per se. However they may be pervaded with
an average magnetic field produced in the accretion shocks surrounding
them or generated in and ejected by the galaxies they contain, provided
the filling factor of the resulting magnetic pollution in the
filament/wall volume is of order unity. In the following, we will
consider both possibilities.

If, as before, the magnetic energy density in the filament/wall is a
fraction $\epsilon_B$ of the thermal energy of the IGM, one infers a
magnetic field strength:
\begin{equation}
B_{\rm f}\,=\, 3.5\times10^{-8}\,{\rm G}\,\,
\left({\epsilon_B\over0.1}\right)^{1/2}\,
\left({\rho_{\rm f}\over10\langle\rho_{\rm b}\rangle}\right)^{1/2}\, \left({T_{\rm
f}\over 10^6\,{\rm K}}\right)^{1/2}\ ,\label{eq:Bf}
\end{equation}
$\rho_{\rm f}$ and $T_{\rm f}$ denoting the filament baryonic density
and temperature.  

The typical length scale of a filament is $l_{\rm f}\,\sim\,15\,$Mpc,
its radius $r_{\rm f}\,\sim\,1-2\,$Mpc, and the typical separation
between two filaments $d_{\rm f}\,\sim\,25\,$Mpc~\cite{Dea01}.

During the formation of non-linear structures, shock waves develop as a
consequence of the infall of material on filaments, walls and clusters
of galaxies.  Numerical simulations indicate that the typical radius of
external shock waves around filament it is of the order of $r_{\rm
sh}\,\simeq\,2-3\,$Mpc~\cite{Mea00,KRCS05}; the typical velocity of
these shock waves is of order $v_{\rm sh}\,\sim\,300-1000\,$km/s. Such
shock waves have been proposed a site of magnetic field amplification
(see for instance ~\cite{KCOR97}) and cosmic ray
acceleration~\cite{LW00,M02,Kea03}.

If the magnetic field in the shock wave vicinity corresponds to a
fraction of equipartition with the shock energy density $\rho v_{\rm
sh}^2$, one finds:
\begin{equation}
B_{\rm sh}\,\simeq\,10^{-7}\,{\rm G}\,\left({\epsilon_B\over
  0.1}\right)^{1/2}
\,\left({\rho_{\rm ext}\over \langle\rho_{\rm b}\rangle}\right)^{1/2}
\,\left({v_{\rm sh}\over 1000\,{\rm km/s}}\right)^{1/2}\ .
\end{equation}
Note that $\rho_{\rm ext}$ refers to the density of infalling material. The
estimate $\epsilon_B\sim0.1$ gives the right order of magnitude for the
inferred value of magnetic field strength $\sim100\,\mu$G in young
supernovae remnants assuming a typical interstellar medium density and
comparable shock speed~\cite{VL03,BKV03}.

If cosmic shock waves amplify the magnetic field up to the value $B_{\rm
sh}$ given above, one should then expect the filament to be endowed with
a significant fraction of $B_{\rm sh}$ out to the shock radius. In
effect, the amount of matter accreted through the shock in a Hubble time
in units of the quantity of matter contained inside the structure at the
present time can be expressed as:
\begin{eqnarray}
f_{\rm acc}&\,\simeq\,& {\rho_{\rm ext}\over \rho_{\rm in}}{v_{\rm
    sh}H_0^{-1}\over r_{\rm f}}\nonumber\\
&\,\sim\,&0.3\,\left({v_{\rm sh}\over
    1000\,{\rm km/s}}\right)\left({r_{\rm f}\over 2\,{\rm
    Mpc}}\right)^{-1}\left({\rho_{\rm f}\over
    10\langle\rho_{\rm b}\rangle}\right)^{-1}\ .
\end{eqnarray}
Note that the estimate $B_{\rm f}$ given in Eq.~(\ref{eq:Bf}) agrees
with that of $B_{\rm sh}$ to within a factor of a few (even though it
was derived through other means).

\subsection{Optical depth and last scattering surface for cosmic ray scattering}

Depending on the strength of the magnetic field in a halo and its
coherence length, the interaction of a particle may either lead to
diffusion inside the structure, at sufficiently low energy, or to a weak
deflection angle, at higher energies. The details of the interaction
between a particle and a magnetized structure is described in detail in
Appendix~\ref{sec:appA}.

\subsubsection{Homogeneously distributed scattering centers}\label{sec:homog}
Out of simplicity, we first assume that the scattering centers are
distributed homogeneously in the Universe with a typical mean free
path to interaction $d_i$, where $i$ refers to the type of scattering
center (e.g. magnetized galactic wind, radio halo, filament ...). We
will discuss in Section~\ref{sec:inhom} the influence of inhomogeneity
on the conclusions of the discussion that follows. For scattering
centers of density $n_i$ and cross-section $\sigma_i$,
$d_i=(n_i\sigma_i)^{-1}$.  The mean free path to interaction with any
scattering center is written $\overline d$:
\begin{equation}
\overline d\,=\,{1\over \sum_i n_i\sigma_i}\ .\label{eq:dlss}
\end{equation}
The optical depth to ultrahigh energy cosmic ray scattering over a path
length $l$ is then defined as:
\begin{equation}
\tau\,=\,{l\over\overline d}\,=\,l\,\sum_i\,n_{\rm i}\sigma_i\ .\label{eq:tau}
\end{equation}
To make concrete estimates, assume that one type of scattering center
dominates, with typical interaction length $d_i$:
\begin{equation}
\tau\,\simeq\,3.1\,\left({l\over 100\,{\rm Mpc}}\right)\left({d_i\over
  32\,{\rm Mpc}}\right)^{-1}\ .\label{eq:tauval}
\end{equation}
The above fiducial value $d_i=32\,$Mpc corresponds to spherical
scattering centers of density $n_i=10^{-2}\,$Mpc$^{-3}$ and radius
$r_i=1\,$Mpc; however it is also a typical value for the interaction
distance to filaments of the large scale structure.

The above optical depth characterizes the number of scatterings along
a path length $l$ but it does not provide information on the angular
spread of the cosmic ray image on the detector. Hence it is useful to
introduce an effective optical depth $\tau_{\rm eff}$, which becomes
unity when the path length $l$ is such that the particle has suffered
a deflection of order unity. If at each scattering, the squared
deflection is noted $\delta\theta^2_i$, then the number of scatterings
to achieve a deflection of order unity reads $1/\delta\theta^2_i$.
The scattering length $l_{\rm scatt}$ of cosmic rays in the medium,
which corresponds to the distance over which the deflection becomes of
order unity, can be written as:
\begin{equation}
l_{\rm scatt}\,=\,{1\over \sum_i
n_i\sigma_i\delta\theta_i^2}\ .\label{eq:lscatt}
\end{equation}
We thus define the effective optical depth $\tau_{\rm eff}$ as:
\begin{equation}
\tau_{\rm eff}\,=\,{l\over l_{\rm scatt}}\,=\,l\,\sum_i\,n_{\rm
  i}\sigma_i\delta\theta_i^2\ .\label{eq:teffdef}
\end{equation}
The angular deflection can be expressed in a simple way as a function
of the Larmor radius $r_{{\rm L}\vert i}$ of the particle in structure
$i$, of the magnetic field coherence length $\lambda_i$ of this
structure, and of the characteristic path length $\bar{r}_i$ through
the structure, which amounts to $(\pi/2)r_{\rm s}$ for a sphere of
radius $r_{\rm s}$ or $(\pi/2)^2r_{\rm f}$ for a filament of radius
$r_{\rm f}$ (see also Appendix~\ref{sec:appA}). Using the formula
provided in Appendix~\ref{sec:appA} [in particular
  Eq.~(\ref{eq:dtheta})], one can rewrite the effective optical depth
as:
\begin{equation}
\tau_{\rm eff}\,\simeq\,l\,\sum_i\,n_{\rm i}\sigma_i\left(1+ 
{2r_{{\rm L}\vert i}^2\over \bar{r}_i\lambda_i}\right)^{-1}\ .
\label{eq:taueff}
\end{equation}

Obviously, one always has $\tau_{\rm eff} < \tau$.  One should interpret
the two optical depths as follows: $\tau<1$ (which implies $\tau_{\rm
eff}<1$) means that the Universe is transparent to cosmic ray scattering
on the scale $l$, while $\tau>\tau_{\rm eff}>1$ means that the Universe
is opaque over this scale, i.e. the accumulated angular deflection is
greater than unity. The intermediate regime, $\tau > 1 >\tau_{\rm eff}$
is interesting; it corresponds to a translucent situation in which
cosmic rays suffer one to many scatterings but the accumulated angular
deflection remains smaller than unity.

The phenomenology of the cosmic ray signal on the detector then depends
on the typical source distance, which should be used for $l$, as well as
on the characteristics of the scattering agents described
above. Assuming rectilinear propagation of the particles, the source
distance scale is of order $l_{\rm max}$, the maximal distance that a
particle of energy $E$ can travel without loosing its energy. Indeed, if
the source population is continously emitting and homogeneous (the
latter being a good approximation on scales beyond a few hundred Mpc), the
flux $F(<l)$ received from sources located within a distance $l$
increases as $l$:
\begin{equation}
F(<l)\,=\, n_{\rm s}\dot N_{\rm UHECR} l\ ,\label{eq:fl}
\end{equation}
where $n_{\rm s}$ denotes the source density and $\dot N_{\rm UHECR}$
 the number of cosmic rays emitted by a source per unit time.  In the
 case of bursting sources, one finds the same scaling (see
 Ref.~\cite{WM96}):
\begin{equation}
F(<l)\,=\, \dot n_{\rm s}N_{\rm UHECR} l\ .\label{eq:flgrb}
\end{equation}
In this equation, $\dot n_{\rm s}$ should now be understood as the rate
of bursting sources per unit time and unit volume, and $N_{\rm UHECR}$
as the total number of cosmic rays emitted by a source.

Hence in both cases, most of the flux comes from sources located at
distance of order $l_{\rm max}$. In the following, we therefore
substitute $l_{\rm max}$ for $l$ in the expression of the optical
depth. We will discuss apart the particular case of rare close-by
sources. One can evaluate the distance $l_{\rm max}$ in two ways:
either as the energy loss distance $E\left\vert\,{\rm d}E/{\rm
  d}x\right\vert^{-1}$, or as the maximal distance that a particle can
travel, assuming it has been detected with energy $E$ and the maximal
energy at the source is $E_{\rm max}$. In the following, we use this
latter definition and assume $E_{\rm max}\,=\,4\times10^{20}\,$eV. The
two definitions give values that never differ by more than 40\%
however, over the energy range $10^{17}\,{\rm eV}\rightarrow
10^{20}\,$eV.

If particles diffuse rather than travel rectilinearly, the maximum
distance is instead determined by $\sqrt{2Dt_{\rm max}}$, where $D$
denotes the diffusion coefficient and $t_{\rm max}=l_{\rm max}/c$. This
will be discussed in more detail in Section~\ref{sec:opaquesign}.

\begin{figure}[th]
\includegraphics[width=0.49\textwidth]{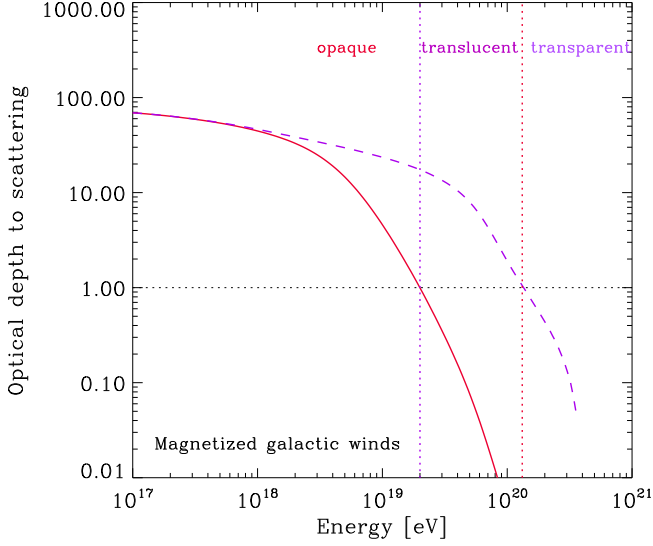} \caption{Optical depth
to cosmic ray scattering by magnetized galactic winds, with $n_{\rm
gw}=10^{-2}\,{\rm Mpc}^{-3}$, $B_{\rm gw}=3\cdot10^{-8}\,{\rm G}$,
$\lambda_{\rm gw}=50\,{\rm kpc}$, and $r_{\rm gw}=0.8\,{\rm Mpc}$. Solid
line: optical depth $\tau_{\rm eff}$ to scattering by an angle of order
unity, as defined in Eq.~(\ref{eq:taueff}); dashed line: optical depth
$\tau$ as defined in Eq.~(\ref{eq:tau}). In the energy range where
$\tau>\tau_{\rm eff}>1$, the Universe is opaque up to the energy loss
distance ; in the range where $\tau>1>\tau_{\rm eff}$, the Universe is
translucent on this distance scale, meaning that cosmic rays suffer
several to many scatterings but the total angular deflection remains
below unity; finally, at energies where $1>\tau>\tau_{\rm eff}$, the
Universe is transparent to cosmic ray scattering.}  \label{fig:1}
\end{figure}

We may now plot the optical depths to scattering $\tau$ and $\tau_{\rm
  eff}$ as functions of energy. In Fig.~\ref{fig:1}, we show an
example that ignores all scattering centers except magnetized galactic
winds, for which we assume $n_{\rm gw}\,=\,10^{-2}\,$Mpc$^{-3}$,
$r_{\rm gw}\,=\,0.8$~Mpc, $B_{\rm gw}\,=\,3\cdot10^{-8}\,$G and
$\lambda_{\rm gw}\,=\,0.05$~Mpc. The resulting optical depth $\tau$ is
shown as the dashed (blue) line, and the effective optical depth
$\tau_{\rm eff}$ as the solid (red) line. The dependence of $\tau$ on
$E$ actually reveals the dependence of $l_{\rm max}$ on $E$: $l_{\rm
  max}$ decreases sharply beyond a few $10^{19}\,$eV as a consequence
of pion production on the microwave background.  The dependence of
$\tau_{\rm eff}$ on $E$ is even more pronounced, since the number of
scatterings to achieve deflection of order unity rapidly increases
with energy, roughly as $E^2$ beyond $10^{18}\,$eV here [see
  Eq.~(\ref{eq:taueff})].  The horizontal dotted line indicates an
optical depth of order unity, while the vertical dotted lines indicate
at which energy $\tau_{\rm eff}=1$ and $\tau=1$ respectively, from
left to right. As indicated on the figure, these lines delimit the
energy ranges in which the Universe appears opaque, translucent or
transparent to cosmic ray scattering. Interestingly, for this example,
the Universe is translucent at energies close to the threshold for
pion production $E_{\rm GZK}\,\simeq\, 6\cdot
10^{19}\,$eV~\cite{G66,ZK66}.

\begin{figure}[th]
\includegraphics[width=0.49\textwidth]{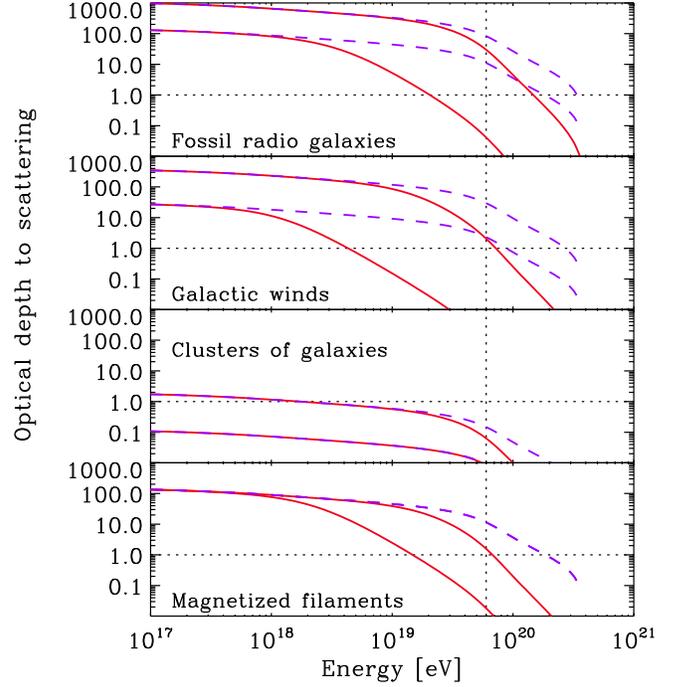} \caption{Optical
  depth to cosmic ray scattering for different types of scattering
  agents and for two different sets of parameters in each case. Solid
  lines: optical depth $\tau_{\rm eff}$ to scattering by an angle of
  order unity, as defined in Eq.~(\ref{eq:taueff}); dashed lines:
  optical depth $\tau$ as defined in Eq.~(\ref{eq:tau}). The vertical
  dotted line indicates $E=6\times10^{19}\,$eV.  Fossil radio
  galaxies: $n_{\rm rg}=3\cdot10^{-3}\,{\rm Mpc}^{-3}$, $B_{\rm
    rg}=10^{-8}\,{\rm G}$, $\lambda_{\rm rg}=100\,{\rm kpc}$, and
  $r_{\rm rg}=2\,{\rm Mpc}$ (lower curves); $n_{\rm rg}=10^{-2}\,{\rm
    Mpc}^{-3}$, $B_{\rm rg}=10^{-7}\,{\rm G}$, $\lambda_{\rm
    rg}=100\,{\rm kpc}$, and $r_{\rm rg}=3\,{\rm Mpc}$ (upper
  curves). Magnetized galactic winds: $n_{\rm gw}=10^{-2}\,{\rm
    Mpc}^{-3}$, $B_{\rm gw}=10^{-8}\,{\rm G}$, $\lambda_{\rm
    gw}=50\,{\rm kpc}$, and $r_{\rm gw}=0.5\,{\rm Mpc}$ (lower
  curves); $n_{\rm gw}=5\cdot10^{-2}\,{\rm Mpc}^{-3}$, $B_{\rm
    gw}=10^{-7}\,{\rm G}$, $\lambda_{\rm gw}=50\,{\rm kpc}$, and
  $r_{\rm gw}=0.8\,{\rm Mpc}$ (upper curves). Clusters of galaxies:
  $n_{\rm cg}=10^{-5}\,{\rm Mpc}^{-3}$, $B_{\rm cg}=10^{-6}\,{\rm G}$,
  $\lambda_{\rm cg}=100\,{\rm kpc}$, and $r_{\rm cg}=1.\,{\rm Mpc}$
  (lower curves); $n_{\rm cg}=10^{-5}\,{\rm Mpc}^{-3}$, $B_{\rm
    cg}=10^{-7}\,{\rm G}$, $\lambda_{\rm cg}=100\,{\rm kpc}$, and
  $r_{\rm cg}=4\,{\rm Mpc}$ (upper curves). Magnetized filaments of
  large scale structure: interseparation $d_{\rm f}=25\,{\rm Mpc}$,
  $B_{\rm f}=3\cdot10^{-9}\,{\rm G}$, $\lambda_{\rm f}=300\,{\rm
    kpc}$, and $r_{\rm f}=2\,{\rm Mpc}$ (lower curves); $d_{\rm
    f}=25\,{\rm Mpc}$, $B_{\rm f}=3\cdot10^{-8}\,{\rm G}$,
  $\lambda_{\rm f}=300\,{\rm kpc}$, and $r_{\rm f}=2\,{\rm Mpc}$
  (upper curves).}  \label{fig:2}
\end{figure}

In Figure~\ref{fig:2}, we show the optical depths for the various
types of scattering centers, taken in turn, and for two sets of
parameters defining their characteristics, as indicated in the
caption. In principle, one should of course sum the different optical
depths of the types of scattering centers. If, however, the pollution
of magnetized winds and radio halos permeate the filaments and nothing
else, one should of course only consider the filaments as the sole
scattering agents.  

\begin{figure}[th]
\includegraphics[width=0.49\textwidth]{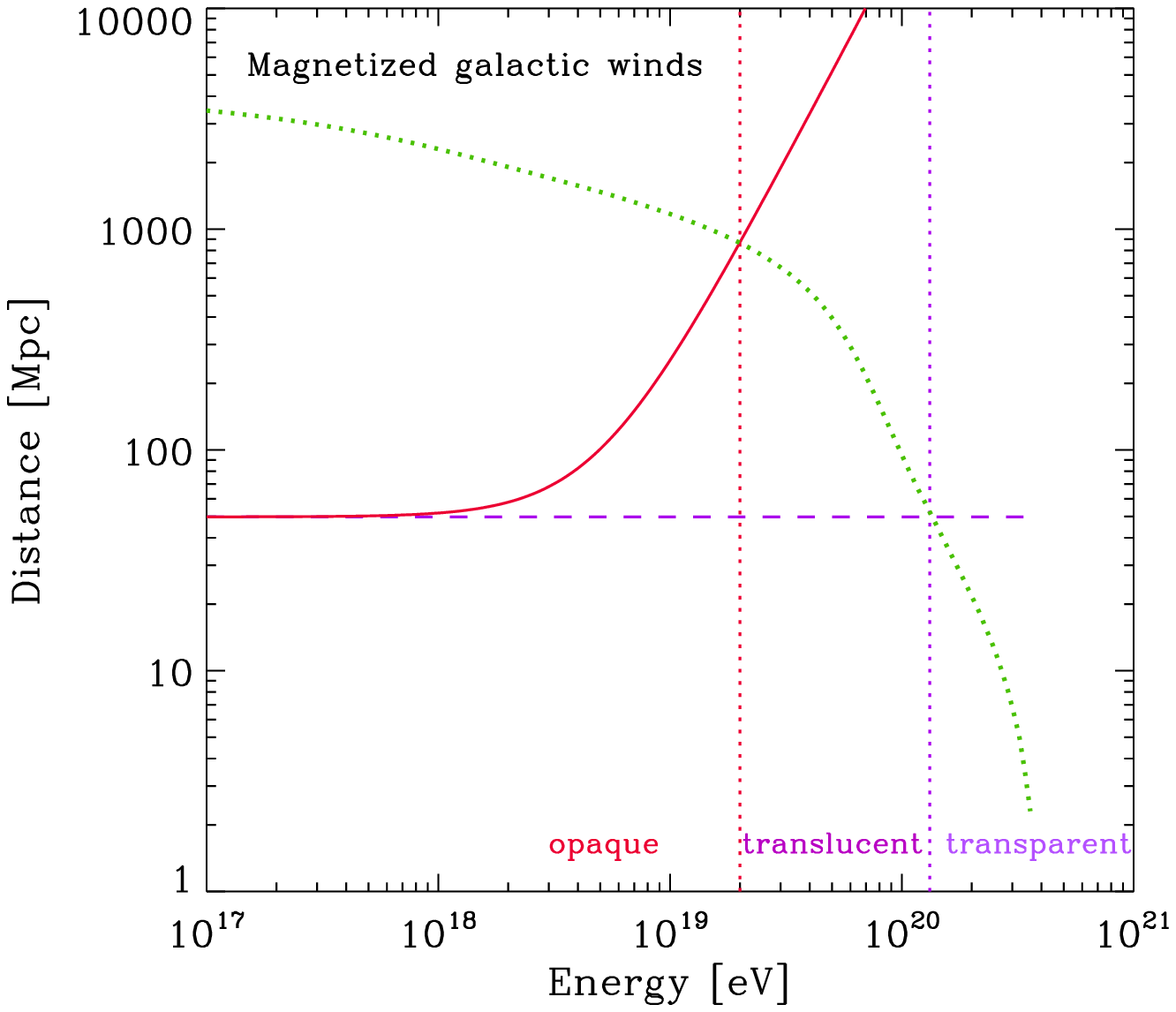} \caption{Distance to
the last scattering surface by magnetized galactic winds, with $n_{\rm
gw}=2\cdot10^{-2}\,{\rm Mpc}^{-3}$, $B_{\rm gw}=3\cdot10^{-8}\,{\rm G}$,
$\lambda_{\rm gw}=50\,{\rm kpc}$, and $r_{\rm gw}=0.8\,{\rm Mpc}$ as in
Fig.~\ref{fig:1}. Solid line: scattering length $l_{\rm scatt}$ for a
deflection of order unity, as defined in Eq.~(\ref{eq:lscatt}); dashed
line: distance $\overline d$ to the last scattering surface as defined
in Eq.~(\ref{eq:dlss}). The dotted (green) line indicates the maximal
distance to the source $l_{\rm max}$, which also gives the source
distance scale. In the energy range where $\overline d<l_{\rm
scatt}<l_{\rm max}$, the Universe is opaque ; in the range where $\overline 
d<l_{\rm max}<l_{\rm scatt}$, the Universe is translucent on the
distance scale $l_{\rm max}$, meaning that cosmic rays suffer several to
many scatterings but the total angular deflection remains below unity;
finally, at energies where $l_{\rm max}<\overline d<l_{\rm scatt}$, the
Universe is transparent to cosmic ray scattering.}  \label{fig:3}
\end{figure}

The two quantities $l_{\rm scatt}$ and $\overline d$ are shown
together with the maximal path length (or source distance scale)
$l_{\rm max}$ in Fig.~\ref{fig:3} for magnetized galactic winds as
scattering agents, with the same parameters used to construct
Fig.~\ref{fig:1}. Figure~\ref{fig:3} illustrates in a different way
the opaque, translucent or transparent nature of the Universe to
cosmic ray scattering.

\begin{figure}[th]
\includegraphics[width=0.49\textwidth]{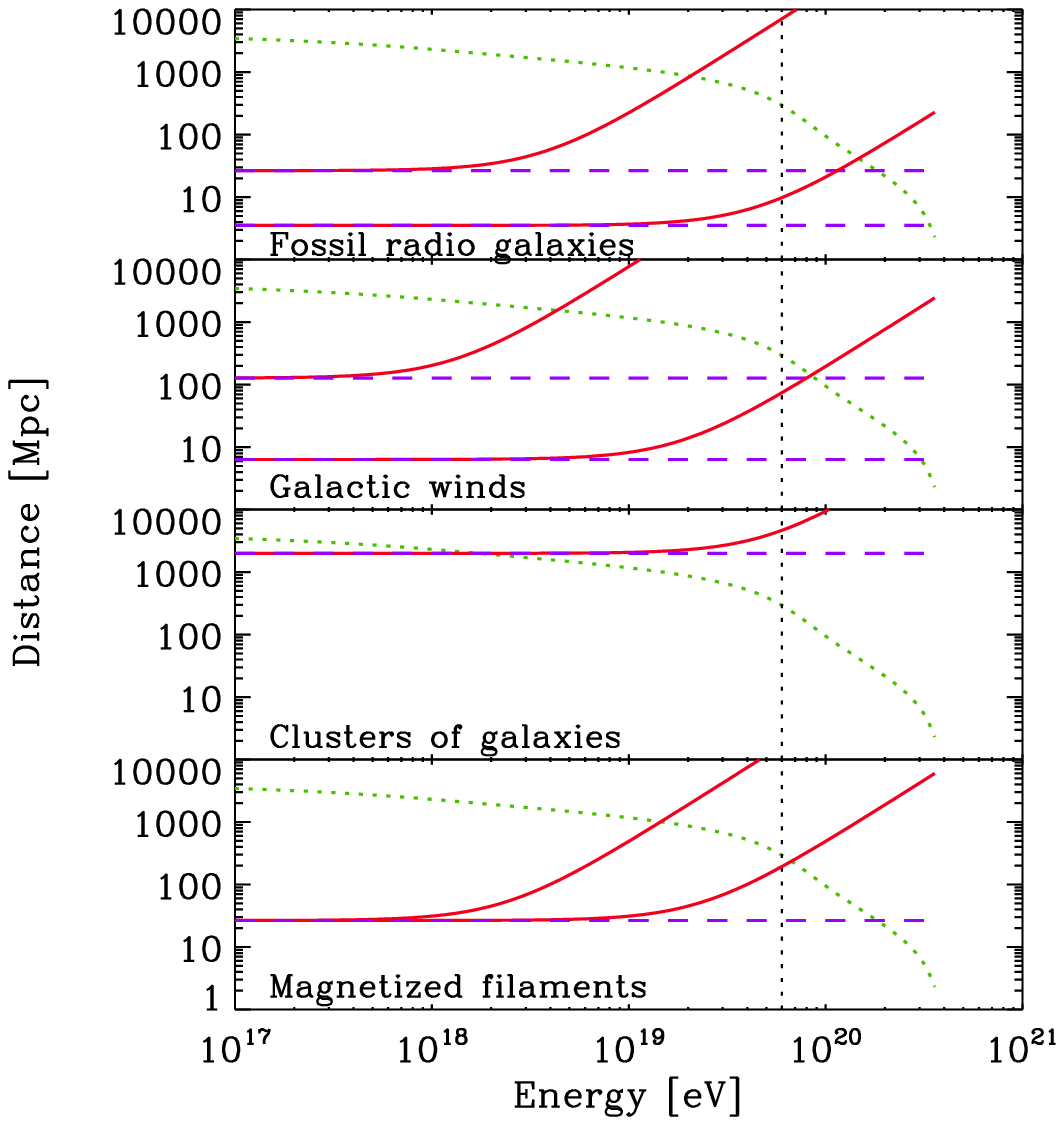} \caption{Distance to
the last scattering surface for different types of scattering agents and
for two different sets of parameters in each case. Solid lines:
scattering length $l_{\rm scatt}$ for deflection of order unity, as
defined in Eq.~(\ref{eq:lscatt}); dashed lines: distance $\overline d$ to
the last scattering surface as defined in Eq.~(\ref{eq:dlss}). The dotted
(green) line indicates the source distance scale $l_{\rm max}$. The
vertical dotted line indicates the location of $E_{\rm GZK}$.  Fossil
radio galaxies: $n_{\rm rg}=3\cdot10^{-3}\,{\rm Mpc}^{-3}$, $B_{\rm
rg}=10^{-8}\,{\rm G}$, $\lambda_{\rm rg}=100\,{\rm kpc}$, and $r_{\rm
rg}=2\,{\rm Mpc}$ (lower curves); $n_{\rm rg}=10^{-2}\,{\rm Mpc}^{-3}$,
$B_{\rm rg}=10^{-7}\,{\rm G}$, $\lambda_{\rm rg}=100\,{\rm kpc}$, and
$r_{\rm rg}=3\,{\rm Mpc}$ (upper curves). Magnetized galactic winds:
$n_{\rm gw}=10^{-2}\,{\rm Mpc}^{-3}$, $B_{\rm gw}=10^{-8}\,{\rm G}$,
$\lambda_{\rm gw}=50\,{\rm kpc}$, and $r_{\rm gw}=0.5\,{\rm Mpc}$ (lower
curves); $n_{\rm gw}=5\cdot10^{-2}\,{\rm Mpc}^{-3}$, $B_{\rm
gw}=10^{-7}\,{\rm G}$, $\lambda_{\rm gw}=50\,{\rm kpc}$, and $r_{\rm
gw}=0.8\,{\rm Mpc}$ (upper curves). Clusters of galaxies: $n_{\rm
cg}=10^{-5}\,{\rm Mpc}^{-3}$, $B_{\rm cg}=10^{-6}\,{\rm G}$,
$\lambda_{\rm cg}=100\,{\rm kpc}$, and $r_{\rm cg}=1.\,{\rm Mpc}$ (lower
curves); $n_{\rm cg}=10^{-5}\,{\rm Mpc}^{-3}$, $B_{\rm cg}=10^{-7}\,{\rm
G}$, $\lambda_{\rm cg}=100\,{\rm kpc}$, and $r_{\rm cg}=4\,{\rm Mpc}$
(upper curves). Magnetized filaments of large scale structure:
interseparation $d_{\rm f}=25\,{\rm Mpc}$, $B_{\rm
f}=3\cdot10^{-9}\,{\rm G}$, $\lambda_{\rm f}=300\,{\rm kpc}$, and
$r_{\rm f}=2\,{\rm Mpc}$ (lower curves); $d_{\rm f}=25\,{\rm Mpc}$,
$B_{\rm f}=3\cdot10^{-8}\,{\rm G}$, $\lambda_{\rm f}=300\,{\rm kpc}$,
and $r_{\rm f}=2\,{\rm Mpc}$ (upper curves).}  \label{fig:4}
\end{figure}

One may also draw the analog of Fig.~\ref{fig:2} for the distance to the
last scattering surface $\overline d$ for the different types of
scattering centers, as done in Fig.~\ref{fig:4}.

\subsection{Inhomogeneity of the large scale structure - Analytic discussion}
\label{sec:inhom}

The above results should be corrected for the presence of
inhomogeneity when the distances considered are smaller than the
inhomogeneity length $\,100\,$Mpc. In a first approach, one may assume
that all scattering centers are clustered in the filaments of large
scale structure. This affects transport in two ways: the typical
distance to an interaction becomes of order $d_{\rm f}$ rather than
$d_i$, but the typical deflection may be enhanced, as the probability
of hitting more than one scattering center during the interaction with
a filament is itself increased.

As the density of scattering centers in a filament becomes
$n_{i\vert\rm f}=n_i/f_{\rm f}$, where $f_{\rm f}\sim 5\,$\% is the
average filament filling factor in the Universe, the mean free path to
interaction inside a filament becomes $f_{\rm f}d_i$. Consequently,
the average number of interactions $N_{\rm int\vert f}$ with
scattering centers of type $i$ during the ballistic crossing of a
filament of radius $r_{\rm f}$ is:
\begin{equation}
N_{\rm int\vert f}\,=\, {\bar{r_{\rm f}}\over f_{\rm f}d_{\rm i}}\ ,
\end{equation}
where $\bar{r_{\rm f}}$ is the characteristic path length of the
particle through the filament [see the discussion that follows
  Eq.~(\ref{eq:teffdef}) and Appendix~\ref{sec:appA}]. This formula
assumes that the particle suffers a deflection angle much smaller than
unity at each interaction. The particle thus exits the filament with a
total deflection and time delay (with respect to straight line
crossing):
\begin{eqnarray}
\delta\theta_{i\vert\rm f}^2&\,=\,& N_{\rm int\vert
  f}\delta\theta_i^2\ ,\\ \delta t_{i\vert\rm f}&\,=\,& N_{\rm
  int\vert f}\delta t_i + \delta\theta_{i\vert\rm f}^2
\frac{\bar{r_{\rm f}}}{6c}\ .
\end{eqnarray}
In these equations, $\delta\theta_i$ and $\delta t_i$ denote
respectively the deflection angle and time delay consecutive to an
interaction with scattering center of type $i$, as discussed in
Appendix~\ref{sec:appA}, while $\delta\theta_{i\vert\rm f}$ and $\delta
t_{i\vert\rm f}$ give the corresponding deflection angle and time delay
after the crossing of a filament.

In the opposite diffusive regime, in which $\delta\theta_i^2\sim 1$,
the particle follows a random walk. If the interaction length $f_{\rm
  f}d_i$ in the filament is much smaller than the filament radius
$r_{\rm f}$, then the analysis of diffusive propagation in a filament
conducted in Section~\ref{sec:difffil} applies. The particle bounces
on the filament and exits on a timescale $r_{\rm f}/c$ at a distance
$\sim (f_{\rm f}d_i/r_{\rm f})^{1/2}r_{\rm f}$ away from its point of
first impact.

Note that the filling factor of the magnetized halos in the filament
is $f_i/f_{\rm f}$, with $f_i\,\simeq\, (4/3)n_i\sigma_i r_i$ the
average filling factor of scattering centers in the Universe. The
filament becomes overfilled by the halos when $f_i\gtrsim f_{\rm f}$,
or equivalently $N_{\rm int\vert f}\,\gtrsim\, (3\pi/16)^2r_{\rm
  f}/r_{\rm i}$. If this condition is satisfied, one needs not
consider the multiple interaction scenario depicted above, as it
suffices to consider the filaments themselves as the scattering
centers.

As mentioned above, the average distance to scattering is also modified
if scattering centers cluster in filaments. It becomes $d_{i,\rm f}$:
\begin{equation}
d_{i,\rm f}\,\simeq\, {d_{\rm f}\over 1-\exp(-N_{\rm int\vert f})}\ ,
\end{equation}
as the denominator in this expression represents the probability of
hitting a scattering center when the particle hits a filament.  The
quantities $d_{i,\rm f}$, $\delta\theta_{i\vert\rm f}$ and $\delta
t_{i\vert\rm f}$ suffice in principle to characterize the transport of
the particle in this structured Universe and to derive the
phenomenological consequences with respect to experimental data. To
gauge the influence of the geometry, one should compare the above
quantities to those expected for a homogeneous scattering center
distribution for typical values of the parameters. One finds:
\begin{equation}
N_{\rm int\vert f}\,\simeq\, 1.3 \left({\bar{r_{\rm f}}\over 2\,{\rm
    Mpc}}\right) \left({f_{\rm f}\over
  0.05}\right)^{-1}\left({d_i\over 32\,{\rm Mpc}}\right)^{-1}
\ .\label{eq:Nintv}
\end{equation}
For the fiducial values used in Eq.~(\ref{eq:Nintv}), $f_i/f_{\rm
  f}=0.83$, i.e. the halos barely overfill the filaments. This means
that if $r_i$ or $n_i$ is larger than the quoted values, one must
consider that the scattering centers are the filaments themselves,
with the average quantities $d_{\rm f}$, $r_{\rm f}$ and $B_{\rm f}$
discussed previously. Conversely, if $r_i$ or $n_i$ is smaller, one
must follow the above multiple interaction scheme.

Finally, one can verify that on distance scales $\gg d_{\rm f}$,
the number of interactions (hence the angular deflection and time
delay) converge toward those obtained in the homogeneous case (at
least for rectilinear propagation). In effect, the filling factor of
filaments can be written in terms of $r_{\rm f}$ and $d_{\rm f}$ as
$f_{\rm f}\,\simeq\, (\pi/2)r_{\rm f}/d_{\rm f}$, hence over a length
scale $d$, the particle suffers $N_{\rm int\vert f}d/d_{\rm
  f}\,\simeq\, d/d_{\rm i}$ interactions. Qualitatively, the number
of interactions per filament crossing compensates for the different
distance between two zones of interaction (i.e. filaments). The
effect of clustering of the scattering centers should thus be
important on distance scales $\lesssim 100-200\,$Mpc, since the
distance between two filaments is of order 30$\,$Mpc; beyond that
distance, one can use the results derived in the homogeneous limit
(Section~\ref{sec:homog}).

One cannot exclude a priori that an even more realistic description of
the hierarchical clustering of matter would produce a sophisticated
law of probability for the interaction path length, leading to
non-standard effects such as anomalous diffusion. A more realistic
description should also account for more complex distribution laws for
the scattering center parameters. Monte Carlo simulations of particle
propagation in a ``realistic'' scattering center distribution are best
suited to address such issues and to provide quantitative estimates of
the effect of inhomogeneity on the transport.

In the following section, we describe the simulations we have
performed in order to study the influence of a realistic spatial
distribution of scattering centers.  In view of the uncertainties
surrounding the origin of extragalactic magnetic fields and the
parameters describing the scattering centers, we simply describe these
latter with average values, as discussed in Section~\ref{sec:od}.

\subsection{Inhomogeneity of the large scale structure - Numerical simulations}
\label{sec:inhom-num}
We have performed our simulations using a variant of the numerical
code described in Ref.~\cite{KL07}. The simulation of the dark matter
density field has been produced by the {\tt RAMSES} code~\cite{T02},
and was kindly provided to us by S.~Colombi; its characteristics are
256$^3$ cells, with extent $280\,$Mpc, giving a grid size
$1.1\,$Mpc. For each simulation, we sample a population of scattering
centers. We adopt two physically motivated bias models: in the first
model, the scattering center density is proportional to the dark
matter density field; in the second, the same proportionality applies,
but we do not allow scattering centers to reside in regions with dark
matter density $\rho < 0.5\langle\rho\rangle$. This latter model
enhances the segregation of scattering centers in the large scale
structure.

\begin{figure}[th]
\includegraphics[width=0.49\textwidth]{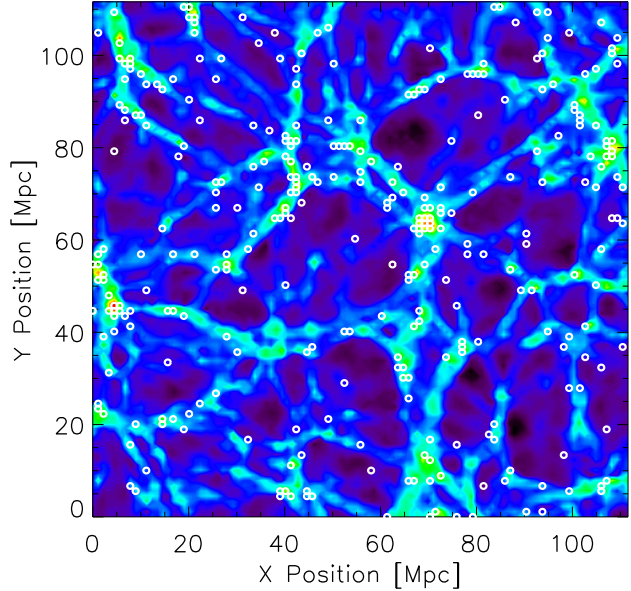} \caption{Distribution
  of scattering centers in the large scale structure (in white), in a
  model in which the density of scattering centers follows that of
  dark matter (density contrast represented in colors). The thickness
  of this slice is $1.1\,$Mpc. The average density is
  $10^{-2}\,$Mpc$^{-3}$.} \label{fig:scdist}
\end{figure}
Figure~\ref{fig:scdist} shows an example of a scattering center
distribution in a two-dimensional slice of the simulation box in the
first bias model. The segregation of scattering centers in filaments
of the large scale structure is apparent, although some tend to reside
in smaller density regions as a result of the large volume fraction
occupied by such regions. Out of simplicity, each scattering center is
modelled as a cube of the size of a cell of the simulation; each cell
in the simulation is thus occupied by zero or one scattering center.

We then follow the trajectories of cosmic rays of various energies,
using the method of Ref.~\cite{KL07}, which simulates the transport of
particles across cells of coherence of the magnetic field in both the
diffusive and non-diffusive regime. These simulations allow to compute
the various statistical properties of transport. A first effect
brought to light by these simulations is the general increase in the
length of first interaction in the inhomogeneous case, when compared
to the homogeneous scattering center distribution. This increase is of
order 40\% for the first bias model, and about 60\% for the second
bias model. It does not seem to depend strongly on the scattering
center density.

Another significant effect is related to the source environment. If
this latter is dense, as one might expect, the local scattering center
density is higher than average, and therefore the cosmic ray may
experience several interactions in the source environment in the first
megaparsecs. Accordingly, the probability distribution for the first
interaction departs from a simple exponential law: it exhibits a peak
in the first Mpc, then decreases as an exponential. These extra
interactions will not affect strongly the total deflection angle as
seen from the detector, since 1$\,$Mpc seen from 100$\,$Mpc is
subtended by an angle $0.6^\circ$. The time delay associated to this
displacement is relatively small, being of order $\simeq
r\delta^2/(2c)\,\simeq\,180\,{\rm yr}\,(r/1\,{\rm
  Mpc})(\delta/0.6^\circ)^2$ ($r$ denotes here the size of the structure
in which the source is embedded, and $\delta$ the deflection angle
associated to the displacement within this structure).

\begin{figure}[th]
\includegraphics[width=0.49\textwidth]{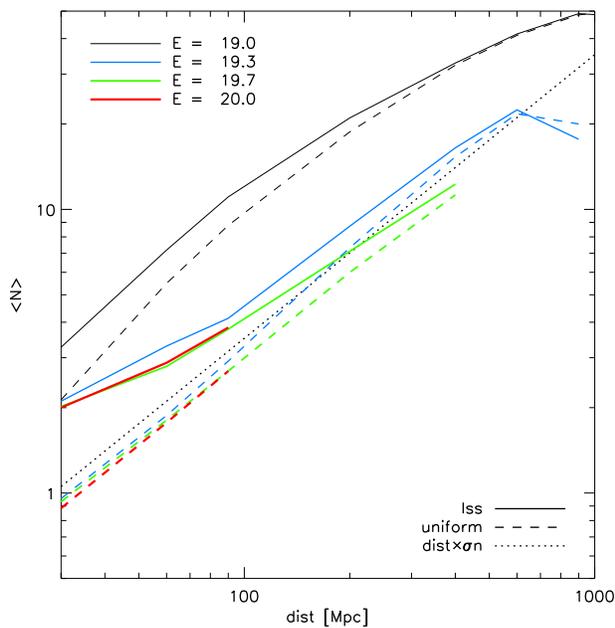} \caption{Average
  number of interactions with scattering centers as a function of distance
  traveled in a time 
  $t_{\rm max}(E)$. From top to bottom, solid lines correspond to
  different energies (in increasing order), as indicated in the
  colored version. Dashed lines indicate the numerical results for a
  homogeneous scattering center population, while solid lines
  correspond to the inhomogeneous case for which the scattering
  centers are distributed according to the dark matter density. The
  dotted line indicates the analytical homogeneous result for
  non-diffusive propagation. The scattering center density is such
  that $d_i\,=\,32\,$Mpc.  Each scattering center is endowed with a
  magnetic field $B_i=3\cdot10^{-8}\,$G and coherence length
  $\lambda_i=100\,$kpc (due to the cubic geometry of the scattering
  center, this corresponds to $B_i=2.7\times10^{-8}\,$G and
  $\lambda_i=100\,$kpc in a spherical cell of radius
  $1\,$Mpc). } \label{fig:Nint-num}
\end{figure}

This effect is apparent in Fig.~\ref{fig:Nint-num} which shows the
average number of interactions as a function of distance, for
different energies. The dashed lines indicate the corresponding trends
for a homogeneous scattering distribution, which go to zero when the
traveled distance tends to zero. On the contrary, the solid lines,
which correspond to the simulated inhomogeneous case, depart from this
scaling and indicate a fixed number of interactions, of order 2. The
exact number turns out to depend on the environment density and has a
variance of order unity. For a source in an environment of average
density, the number of such extra interactions in the source
surroundings is negligible.

Figure~\ref{fig:Nint-num} also reveals other interesting features. In
particular, one can see clearly that the average number of
interactions in the inhomogeneous case converges toward that obtained
in the homogeneous case on distance scales $\gtrsim 100-200\,$Mpc, as
expected [see the discussion that follows Eq.~(\ref{eq:Nintv})]. For
the highest energies, namely $E=10^{19.7}\,$eV and $E=10^{20}\,$eV,
there is a slight offset between the analytical prediction for $d_i$
and the homogeneous calculation; this difference is attributed to the
cubic geometry of the scattering center. Particles of lower energies,
in particular $E=10^{19}\,$eV, diffuse in the scattering center
distribution, as evidenced by the higher slope of the average number
of interactions as a function of the traveled distance $l$. One can
check in particular that $N_{\rm int}\,\sim\, (l/d_i)^2$ as
expected. At very large distances, this relation breaks down because
the trajectory is cut after a time $t_{\rm max}$; hence less and less
particles are able to travel beyond a distance $\sim (ct_{\rm
  max})^{1/2} d_i^{1/2}$. Similar features are observed in the second
bias model.

Finally, the same simulations can be used to compute the average
deflection angle as a function of energy and traveled distance. This
calculation is performed as follows. At a predetermined distance $l$,
one draws at random a certain number of ``small spheres'' positioned
on the sphere of radius $l$ around the source. These ``small spheres''
mimic the detectors located at distance $l$ from the source. There
must be a sufficient number of these ``small spheres'' to guarantee a
sufficient signal, but not so many that they would overlap, in which
case one would oversample the sphere of radius $l$.  Each time a
trajectory intersects one of these spheres, the angle between the
particle incoming direction in this sphere and the source location is
recorded. Iterating over the particles and the ``small spheres''
allows to reconstruct the probability distribution of deflection
angles. 

\begin{figure}[th]
\includegraphics[width=0.49\textwidth]{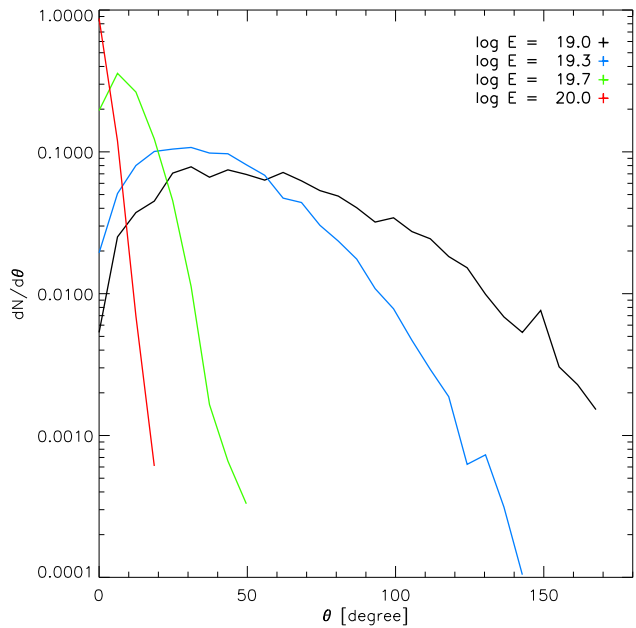} \caption{Histogram
  of deflection for different energies, as indicated. The values have
  been computed at different distances for the different energies:
  $1000\,$Mpc for $E=10^{19}\,$eV, $600\,$Mpc for $E=10^{19.3}\,$eV,
  $400\,$Mpc for $E=10^{19.7}\,$eV and $90\,$Mpc for
  $E=10^{20}\,$eV. As before, the scattering center density is such
  that $d_i\,=\,32\,$Mpc. Each scattering center is endowed with a
  magnetic field $B_i=3\cdot10^{-8}\,$G and coherence length
  $\lambda_i=100\,$kpc (due to the cubic geometry of the scattering
  center, this corresponds to $B_i=2.7\times10^{-8}\,$G and
  $\lambda_i=100\,$kpc in a spherical cell of radius
  $1\,$Mpc). } \label{fig:deflec-num}
\end{figure}

The result is shown in Fig.~\ref{fig:deflec-num} for various energies,
for the same inhomogeneous distribution of scattering centers as
above. Each cell is endowed with a magnetic field of strength
$B_i=3\times10^{-8}\,$G and of coherence length
$\lambda_i=100\,$kpc. Since the cell is cubic, of size $1.1\,$Mpc, the
deflection per interaction corresponds to that obtained for a
spherical cell of radius $1\,$Mpc and magnetic field strength
$B_i=2.7\times10^{-8}\,$G. The values shown in
Fig.~\ref{fig:deflec-num} have been computed at the following
distances: $1000\,$Mpc for $E=10^{19}\,$eV, $600\,$Mpc for
$E=10^{19.3}\,$eV, $400\,$Mpc for $E=10^{19.7}\,$eV and $90\,$Mpc for
$E=10^{20}\,$eV. These distances are representative of $l_{\rm max}$
hence of the source distance scale.  At an energy $E=10^{20}\,$eV, the
mean and median deflections are of order $3^\circ$ and $2.6^\circ$
respectively, while at $E=10^{19.7}\,$eV, they increase to $12^\circ$
and $11.5^\circ$, and become larger at smaller energies. These values
are about 30\% smaller than those expected from the analytical
calculation, given in Eq.~(\ref{eq:thetaint}) further below. This
difference can stem from the slightly different number of interactions
experienced by particles in the inhomogeneous scattering center
distribution, as compared to the homogeneous case (see
Fig.~\ref{fig:deflec-num}). The cubic geometry of scattering centers
used in our simulation can also contribute to alter the values of the
deflection angles. Obviously, these deflections could also be larger
or smaller depending on the exact values of the scattering center
characteristics, see discussion above.

To summarize this discussion on the effect of inhomogeneity, we note
the following features: when the scattering centers correlate with the
large scale structure, the probability law of first interaction and
the number of interactions departs from those obtained in the
homogeneous case, at distances $\lesssim\,100\,$Mpc. The difference
between the two cases depends on several factors: the source
environment and the bias of the scattering center distribution with
respect to the underlying dark matter distribution, in particular. It
is found however that on large scales $\gtrsim\,100\,$Mpc and in
  the weak deflection regime, one recovers the results of the
  homogeneous scattering center distribution discussed in
  Section~\ref{sec:homog}.

Extra interactions in the source environment, if sufficiently dense to
be populated by scattering centers, may increase slightly the
time delay with respect to straight line propagation but will not
modify substantially the total deflection angle.  In the
diffusive regime, scattering occurs against filaments if the
interaction length in the filament is smaller than the filament
size, or against the scattering centers, if not.

\section{Consequences for cosmic ray transport}\label{sec:crt}
In this section, we discuss the phenomenological consequences of the
above model of cosmic ray transport with respect to the signatures of
different source models, discussing in particular the absence or
existence of counterparts. We will discuss in Section~\ref{sec:disc}
the interpretation of existing data in the light of these
consequences, and in particular the recent correlation announced by
the Pierre Auger Observatory.

\subsection{Optically thin regime}\label{sec:thin}
The optically thin regime, in which $l_{\rm max}< \overline d < l_{\rm
  scatt}$, is trivial in terms of particle propagation: most particles
travel in straight line, without interacting in the intergalactic
medium, hence one should expect to see the source directly in the
arrival direction of the highest energy events.  However, in the
case of gamma-ray burst sources, the spreading of arrival times
through the interaction with cosmic magnetic fields is essential to
reconcile the gamma-ray burst rate with the rate of ultrahigh
energy cosmic ray detection~\cite{W95}. In the absence of scattering
(hence time delay), such a bursting source would be essentially
unobservable as the occurrence rate is much too low when compared to
the lifetime of the experiment. 

Independently of the source scenario, there does not exist at present
clear and unique evidence for counterpart identification, as discussed
briefly in the introduction.  Extra deflection could arise from an
all-pervading intergalactic magnetic field or the Galactic magnetic
field.  The influence of an all-pervading intergalactic magnetic field
has been discussed in previous works, see for instance
Refs.~\cite{SME04,DGST05,TS07,KRC07} for recent works. Note that our
model of magnetized filaments and non-magnetized voids may be
considered as an approximation to the more realistic magnetic field
configurations derived in these studies.

Concerning the influence of the Galactic magnetic field, existing
models suggest that the typical deflection at the highest energies,
say $\simeq10^{20}\,$eV, are probably of the order of a few
degrees~\cite{AMES02,TK07}. Hence one would need to invoke the
existence of an extended magnetized halo to provide sufficient
deflection. Alternatively, one may consider a scenario in which most
particles at the highest energies are heavy nuclei, which are more
easily deflected.

\subsection{Translucent regime}\label{sec:int}
The intermediate regime, in which $\overline d\,<\, l_{\rm max}\,<\,
l_{\rm scatt}$ is interesting, because the typical deflection is smaller
than unity, yet it could be sufficient to explain the lack of
counterpart.

\subsubsection{Transport}
Since the total deflection remains smaller than unity, one may
describe the transport as near-ballistic with a non-zero time delay as
measured relatively to straight line propagation. Furthermore, one may
use in this case the time delay and deflection formulae obtained from
random walk arguments in Ref.~\cite{WM96}, provided one accounts for
the inhomogeneity of the magnetic field. In detail, at each scattering
with scattering center $i$, the particle suffers and angular
deflection $\delta\theta_i$ and exits with a delay $\delta t_i$. 
The corresponding formulae for $\delta\theta_i$ and $\delta t_i$
are given in Eqs.~(\ref{eq:dthetanum}),(\ref{eq:dtimenum}).

The total time delay $\delta t$ acquired over a path length $l$ is
given by the sum of the time delays acquired during each scattering as
well as that resulting from the fact that the particle does not travel
in a straight line from the source to the detector. If the particle is
seen from the detector at a typical deflection angle $\delta\alpha$
away from the source direction, then the time delay associated to this
transverse displacement with respect to the line of sight is
$l\delta\alpha^2/(4c)$~\cite{1978ApJ...222..456A}. In the limit of
large optical depth $\tau\,>\,1$, this angle $\delta\alpha^2$ is
written as~\cite{1978ApJ...222..456A}:
\begin{equation}
\delta\alpha^2\,=\, {\tau\over 3}\delta\theta_i^2\ ,
\end{equation}
where $\delta\theta_i^2$ is the rms scattering angle per
scattering event.

On average, the particle interacts at every step of length $\overline
d$, with probability $\overline d/d_i$ of hitting a structure of type
$i$.  Then the total time delay and deflection acquired after traveling a
path length $l$ are:
\begin{eqnarray}
\delta\alpha^2&\,=\,& {\tau\over 3}\,\sum_i {\overline d\over
  d_i}\delta\theta_i^2\ ,\\ \delta t&\,\simeq\,&\tau\,\sum_i
            {\overline d\over d_i}\delta t_i\,+\,
            {l\delta\alpha^2\over 4c}\ .
\label{eq:tint}
\end{eqnarray}
To make simple estimates, consider the case in which one type of
scattering event dominates the scattering history. Then the typical
deflection angle reads (still assuming $\tau>1$):
\begin{eqnarray}
\delta\alpha&\,\simeq\,&1.7^\circ\, \left({\tau\over
  3}\right)^{1/2}\left({\bar{r}_i\over 2\,{\rm Mpc}}\right)^{1/2}
\times\nonumber\\ &&\,\,\,\, \left({B_i\over 10^{-8}\,{\rm
    G}}\right)\left({\lambda_i\over 0.1\,{\rm Mpc}}\right)^{1/2}
\left({E\over 10^{20}\,{\rm eV}}\right)^{-1}\ ,\label{eq:thetaint}
\end{eqnarray}
where $\bar{r}_i$ is the characteristic size of the scattering center
[see after Eq.~(\ref{eq:teffdef}) and Appendix~\ref{sec:appA}].  The
optical depth to cosmic ray scattering is related to the distance and
the geometrical characteristics of the scattering centers as in
Eq.~(\ref{eq:tauval}).  This deflection may thus be non-negligible for
typical parameters of the scattering centers discussed in the previous
section. In all cases, the arrival direction should point back to the
last scattering center encountered by the cosmic ray.  Since
scattering centers are highly magnetized regions, and as such are
probably associated with active objects such as radio-galaxies, one
may be deceived by their presence on the line of sight, and interpret
them as the source of ultrahigh energy cosmic rays. The smoking gun of
such counterfeiting is the distance scale to these objects: in this
optically thick regime, most counterparts would be located at a
distance scale $\overline d$ (which can be measured) significantly
smaller than the expected distance scale $l_{\rm max}$ (which is
known).

The associated time delay reads:
\begin{equation}
\delta t\,\simeq\,7.0\cdot10^4\,{\rm yrs}\,\left({l\over 100\,{\rm Mpc}}\right)\,
\left({\delta\alpha\over 1.7^\circ}\right)^2\ .\label{eq:tauint}
\end{equation}
The second term on the r.h.s of Eq.~(\ref{eq:tint}) indeed dominates
largely over the first.  It is easy to verify that the relation
between $\delta\alpha^2$ (the rms angle between the line of sight to
the source and the particle incoming velocity on the detector) and
$\delta\theta^2$ (the rms velocity deflection angle per scattering)
remains unchanged in the limit $\tau\,<\,1$. Obviously, however, the
solid angle of the source images cannot exceed that of the scattering
center.

Further effects related to the formation of angular images are
discussed in the following.

\subsubsection{Angular images}\label{sec:angim}
The physics of the formation of angular images has been discussed in
detail in Refs.~\cite{WM96,2002JHEP...03..045H} in the case of
ultrahigh energy cosmic rays propagating in an all-pervading irregular
magnetic field. In the model under consideration, differences may
occur when the discreteness of scattering centers cannot be
neglected. This occurs if $\tau\,\lesssim\,1$, since $\tau$ indicates
the covering factor of the scattering centers.

In the limit $\tau\,\gg\,1$, one may use the analysis of
Refs.~\cite{WM96,2002JHEP...03..045H} provided one translates the
quantities defined in these studies in terms of those relevant in the
present case, such the scattering rate per interaction determined in
Section~\ref{sec:appA}. One point of interest concerns the shape of
the angular image. As discussed in
Refs.~\cite{WM96,2002JHEP...03..045H}, the image will be centered on
the source, and broadened by an angle $\delta\alpha$, if there are
many uncorrelated paths through the scattering medium linking the
source to the detector. In the present case, this condition remains
unchanged in the limit $\tau\,\gg\,1$, i.e. it reads
$l\delta\alpha\,\gg\, \lambda_i$, with $\lambda_i$ the magnetic field
coherence length of the scattering center.  
If $l\delta\alpha\,\ll\,\lambda_i$, the image will appear displaced
from the true source position by an angle $\delta\alpha$, with a small
dispersion. As discussed in Ref.~\cite{2002JHEP...03..045H}, the
distortion of the image does not modify (on average) the flux received
from the source, in either limit considered above. This implies in
particular that the presence of scattering centers does not modify the
expected number of events, but only modifies the angular disposition
of these multiple events.

The intermediate regime $l\delta\alpha\,\simeq\,\lambda_i$ is that
were multiple images and magnetic lensing amplification effects may
become prominent (see Ref.~\cite{SLB99} for a numerical demonstration
of magnetic lensing). However, as $\lambda_i$ is unlikely to exceed a
few hundreds of kpc, this intermediate regime is to be expected only
in the limit of very small deflection:
\begin{equation}
{l\delta\alpha\over \lambda_i}\,\simeq\, 29\left({l\over 100\,{\rm
    Mpc}}\right)\left({\delta\alpha\over 1.7^\circ}\right)
\left({\lambda_i\over 100\,{\rm kpc}}\right)^{-1}\ .\label{eq:mimage}
\end{equation}
This equation indeed suggests that typical angular images should be
broadened by $\delta\alpha$ and centered on the source location.

In the limit of small optical depth ($\tau\,\sim\,1$) which becomes
all the more relevant at the highest energies $E\sim10^{20}\,$eV, some
noticeable differences can be expected. Two questions of interest are:
the shape of angular images, and the possible magnification or
demagnification of images. As we argue, these effects depend on the
hierarchy between the typical displacement $\,\simeq\, l\delta\alpha$
in the scattering center plane (oriented perpendicular to the line of
sight to the source), the size of the scattering center $r_i$, as well
as the typical distance between two scattering centers in this plane,
which is given by $(n_il)^{-1/2}$. Out of simplicity, we assume
spherical scattering centers; we will argue that the conclusions
remain unchanged for filaments.

In order to study the limit $\tau\,\lesssim\,1$, it suffices to assume
that there is only one scattering center on the line of sight to the
source. We further assume that this scattering structure is centered
on the line of sight. The shape of the angular image is here as well
determined by the ratio $l\delta\alpha_i/\lambda_i$. As we now argue,
the flux received does not deviate from that expected in the absence
of scattering, $F_0\,=\,\dot N_{\rm UHECR}/(4\pi l^2)$, provided the
scattering center is larger than the image of the source, i.e.
$\delta\beta_i\,>\,\delta\alpha_i$, denoting by
$\delta\beta_i\,\equiv\,r_i/l_1$ the typical apparent half opening
angle of the scattering center with $l_1$ the distance between the
scattering center and the detector. If the opposite inequality holds
($\delta\beta_i\,<\,\delta\alpha_i$), the flux from the source gets
demagnified through scattering. This can be seen as follows.

Each area element on the scattering structure can be assumed to dilute
an incoming unidirectional flux into a beam of solid angle
$\delta\Omega\,\simeq\,\pi \delta\theta_i^2$ (assuming small
deflection). As seen from the source, this defines a solid angle
$\delta\Omega_{\rm \vert s}$ such that, if particles are emitted
within $\delta\Omega_{\rm\vert s}$, they may be redirected toward the
detector through scattering. Then:
\begin{equation}
\delta\Omega_{\rm \vert s}\,=\,\left({l_1\over
    l}\right)^2\delta\Omega\,
\end{equation}
The ratio $l_1/l$ corresponds to the ratio between the half-opening
angle of the cone of solid angle $\delta\Omega_{\rm \vert s}$ to
$\delta\theta_i$. Effects related to the finite size of the scattering
center are considered further below.

Now, of the flux impinging on the area element, only a fraction
$\delta\Omega_{\rm d}/\delta\Omega$ is diverted away toward the
detector of solid angle $\delta\Omega_{\rm d}=A_{\rm d}/l_1^2$ and
area $A_{\rm d}$ (this solid angle is measured relative to the
scattering structure). One then finds that the flux received from the
source is:
\begin{eqnarray}
F&\,=\,& {\dot N_{\rm UHECR}\over 4\pi A_{\rm d}}{\delta\Omega_{\rm
    d}\over \delta\Omega}\,{\rm min}\left(\delta\Omega_{\rm \vert
  s},\,{\pi r_i^2\over l_2^2}\right)\,\nonumber\\ &\,=\,&F_0 \,{\rm
  min}\left[1,\,\left({l\over l_1}\right)^2{r_i^2\over
  l_2^2\delta\theta_i^2}\right]\ .\label{eq:demag}
\end{eqnarray}
In this equation, $l_2\,\equiv\,l-l_1$ represents the distance between
the source and the scattering center.  The ``min'' function has been
introduced in order to limit the angular size of the source image to
the minimum of the size produced by deflection and the size of the
scattering center (which is seen through a solid angle $\pi
r_i^2/l_2^2$ from the source).

Thus, $F=F_0$ if the solid angle $\delta\Omega_{\rm\vert s}$ is
smaller than the solid angle of the scattering structure as seen from
the source, which amounts to $\delta\alpha_i\,<\,\delta\beta_i$. This
can be traced back to the compensation between a larger source image
(which would lead to amplification) with the dilution of the signal
into a beam of solid angle $\delta\Omega$.

If, on the contrary $\delta\alpha_i\,>\,\delta\beta_i$, the source
image is demagnified by the ratio $F/F_0
\,\simeq\,\delta\beta_i^2/\delta\alpha_i^2$, i.e. by the ratio of the
solid angle of the scattering center to the solid angle that the
source image would have if the scattering center had an infinite
extent. One can generalize this result to the case of filamentary
scattering centers, by noting that the flux gets demagnified by the
ratio of the area of the scattering center to the projected area (on
the scattering plane) of the beam of solid angle
$\delta\Omega_{\rm\vert s}$. Using previous fiducial values for the
scattering centers, and assuming $l_2=l/2$, one finds:
\begin{equation}
{\delta\alpha_i \over \delta\beta_i}\,\simeq\, 0.8\,\left({\delta\alpha_i\over
  1^\circ}\right)
\left({l\over 100\,{\rm Mpc}}\right)\left({r_i\over 1\,{\rm
    Mpc}}\right)^{-1}\ .
\end{equation}
However, this result considers only the influence of one scattering
center on the line of sight. As the beam width exceeds the apparent
size of the scattering center on the line of sight, one must take into
account the possibility that a fraction of the beam interacts with
scattering centers away from the line of sight. In the limit of small
angle deflection, and still assuming
$\delta\alpha_i\,>\,\delta\beta_i$, the flux received by the detector
should be given by Eq.~(\ref{eq:demag}) above, multiplied by the
number of scattering centers of the scattering plane intercepted by
the beam of solid angle $\delta\Omega_{\rm \vert s}$. We neglect the
possible overlap of the projected areas of the scattering centers,
which corresponds to $(n_il)^{-1/2} > r_i$, or equivalently
$\tau\,<\,1$. This number of intercepted scattering structures can
then be written as:
\begin{eqnarray}
N_i&\,\simeq\,& n_ill_2^2\delta\Omega_{\rm \vert
  s}\,\nonumber\\ &\approx\,& 0.96\,\left({n_i\over 10^{-2}\,{\rm
  Mpc}^{-3}}\right) \left({l\over 100\,{\rm Mpc}}\right)^3
\left({\delta\alpha_i\over 1^\circ}\right)^2\ .
\end{eqnarray}
Hence the flux received from all intercepted scattering centers in the
limit $\delta\alpha_i\,>\,\delta\beta_i$ is:
\begin{equation}
F_{\rm tot}\,\simeq\, N_i\,F\,\simeq\,\tau\,F_0\ .\label{eq:totaldemag}
\end{equation}
This result can be understood as follows: the number of intercepted
scattering centers is the product of the surface density $n_i l$ times
the projected area (on the scattering plane) of the beam of solid
angle $\delta\Omega_{\rm\vert s}$; however, the demagnification factor
is the ratio of the scattering center area to this latter, so that the
total demagnification factor is the product of the surface density of
scattering centers times the area of one scattering center,
i.e. $\tau$.  This argument remains unchanged for filamentary
scattering centers.

Equation~(\ref{eq:totaldemag}) gives the total demagnification of the
flux from a source with one scattering structure on the line of sight,
in the limits $\delta\alpha_i\,>\,\delta\beta_i$ and $(n_i
l)^{-1/2}\,>\,r_i$ (i.e. $\tau\,<\,1$). Interestingly, the angular
image is now decomposed into $N_i$ distinct images of angular size
$\delta\beta_i$ each, of similar flux $\sim F_0\tau/N_i$, being
separated from one another by an angle of order $\delta\alpha_i$.

Note that, on average, there is neither magnification nor
demagnification of the flux, as expected. Regarding the limit
$\tau\,\gg\,1$, this effect has been discussed in
Ref.~\cite{2002JHEP...03..045H} in particular. Concerning the limit
$\tau\,<\,1$ discussed above, there are two possibilities. If
$\delta\alpha_i\,<\,\delta\beta_i$, then as shown in
Eq.~(\ref{eq:demag}) the flux is unchanged through scattering. If
$\delta\alpha_i\,>\,\delta\beta_i$, the flux of the source is
demagnified by $\tau$ through scattering, but this occurs with
probability $\simeq\,\tau$, which corresponds to the possibility of
having one scattering structure on the line of sight. There is also a
probability $1-\tau$ of seeing the source directly (without
scattering) together with echoes of flux $\tau F_0$ associated to
scattering with structures off the line of sight. Hence the total flux
is on average unchanged. Deviations from this average may occur in
certain configurations, for instance through magnetic lensing, see
Eq.~(\ref{eq:mimage}) above and Ref.~\cite{2002JHEP...03..045H}, or in
particular source scenarios, as discussed at the end of
Section~\ref{sec:direcdep} further below.

\subsubsection{Experimental signatures for continuously emitting sources}
As far as continuously emitting sources are concerned, a possibly
large angular deflection could prevent the detection of counterparts.
Indeed, values such as $d_i=30\,$Mpc and $B_i=10^{-8}\,$G suffice to
produce a deflection of order $10^\circ$ over a path length $l=l_{\rm
  max}$ at energy $4\cdot 10^{19}\,$eV, which is a generic threshold
energy used in the search for counterparts.  The strong evolution of
$\delta\theta$ with energy results from the strong evolution of
$l_{\rm max}$ with $E$ close to the threshold for pion
production. This suggests that counterparts should be found at
sufficiently high energies, which of course asks for high statistics.

Since the flux received from sources within distance $l$ scales as
$l$, one may expect to see the source in the arrival directions of a
subset $l_0/l_{\rm max}$ of all events, $l_0$ being defined as the
distance at which the typical deflection becomes comparable to the
radius within which one searches for counterparts. This number
$l_0/l_{\rm max}$ should be smaller than unity, since if it were
unity, it would mean that the total angular deflection for all sources
is very small, hence that counterparts should have been detected.

\subsubsection{Experimental signatures for bursting sources}

Regarding bursting sources, and gamma-ray bursts in particular,
Eq.~(\ref{eq:tauint}) shows that the typical time delay is sufficiently
large to explain the lack of temporal association between cosmic ray
arrival directions and gamma-ray bursts, as well as the continuous rate
of detection of high energy cosmic rays. Recall indeed that one
potential difficulty of the gamma-ray burst scenario is to explain the
near continuous detection of cosmic rays at the highest energies
$\sim10^{20}\,$eV, when the gamma-ray burst rate is only
$\sim10^{-3}\,$yr$^{-1}$ within the energy loss distance
$\sim100\,$Mpc. As noted by Waxman~\cite{W95}, this difficulty may be
overcome if the arrival time spread $\sigma_t$ of the highest energy
events is sufficiently large, i.e. $\sigma_t \gtrsim 10^{3}\,$yr at
$10^{20}\,$eV in particular. 

Following Ref.~\cite{WM96}, we note that the magnitude of
$\sigma_t/\delta t$ is influenced by the number of different
trajectories that the particle can follow from the source to the
detector. If indeed all particles follow the very same trajectory,
$\sigma_t\,\ll\,\delta t$, while if different particles may follow
different trajectories, one should expect $\sigma_t\sim\delta t$.  In
the present model, Eq.~(\ref{eq:mimage}) shows that the latter
situation is much more likely, so that $\sigma_t/\delta t\,\sim\,1$.
Furthermore, broadening of the time signal at the highest energies is
likely to be 
increased by stochastic pion production, which results in
$\sigma_t/\delta t\sim 1$~\cite{LSOS97}.  

One may also calculate the number of gamma-ray burst sources
which can contribute to the flux at a given energy $E$~\cite{WM96,W01}:
\begin{equation}
N_{\rm GRB}(E)\,\simeq\, \dot n_{\rm GRB}{2\pi\over 5}l_{\rm
max}^3\sigma_t\ .
\end{equation}
This number of apparent gamma-ray bursts in the cosmic ray sky
characterizes the amount of statistical fluctuation to expect around
the mean flux at a given energy~\cite{WM96}. Using
Eq.~(\ref{eq:tauint}), one obtains:
\begin{eqnarray}
N_{\rm GRB}(E)&\,\simeq\,&88\, \left({\tau\over
  3}\right)\left({l_{\rm max}\over 100\,{\rm Mpc}}\right)^{4}
\times\nonumber\\ &&\,\,\,\,\left({\bar{r}_i\over 2\,{\rm
    Mpc}}\right)\left({E\over 10^{20}\,{\rm
    eV}}\right)^{-2}\times\nonumber\\ &&\,\,\,\, \left({B_i\over
  10^{-8}\,{\rm G}}\right)^2\left({\lambda_i\over 0.1\,{\rm
    Mpc}}\right) \times\nonumber\\ &&\,\,\,\,\left({\dot n_{\rm
    GRB}\over 10^{-9}\,{\rm Mpc^{-3}\cdot yr^{-1}}}\right)
                  {\sigma_t\over \delta t} \ .\label{eq:ngrb}
\end{eqnarray}
The magnitude of this number of apparent sources implies that
the spectrum of ultrahigh energy cosmic rays should not reveal
statistical fluctuations until energies as large as a few
$10^{20}\,$eV, at least for these fiducial values that
  characterize the scattering centers.

\subsubsection{Direction dependent effects}\label{sec:direcdep}
Since the sources of protons with energies beyond the pion production
threshold are bound to reside within $100-200\,$Mpc, one may expect
the optical depth of scattering centers to vary with the direction of
observation, just as the density of matter. In order to discuss the
influence of such variation on existing and upcoming data, we have
constructed sky maps of the matter concentration using the PSCz
catalog of galaxies~\cite{2000MNRAS.317...55S} which presently offers
the most adequate survey for this task.

\begin{figure*}
\includegraphics[width=0.3\textwidth, angle=90]{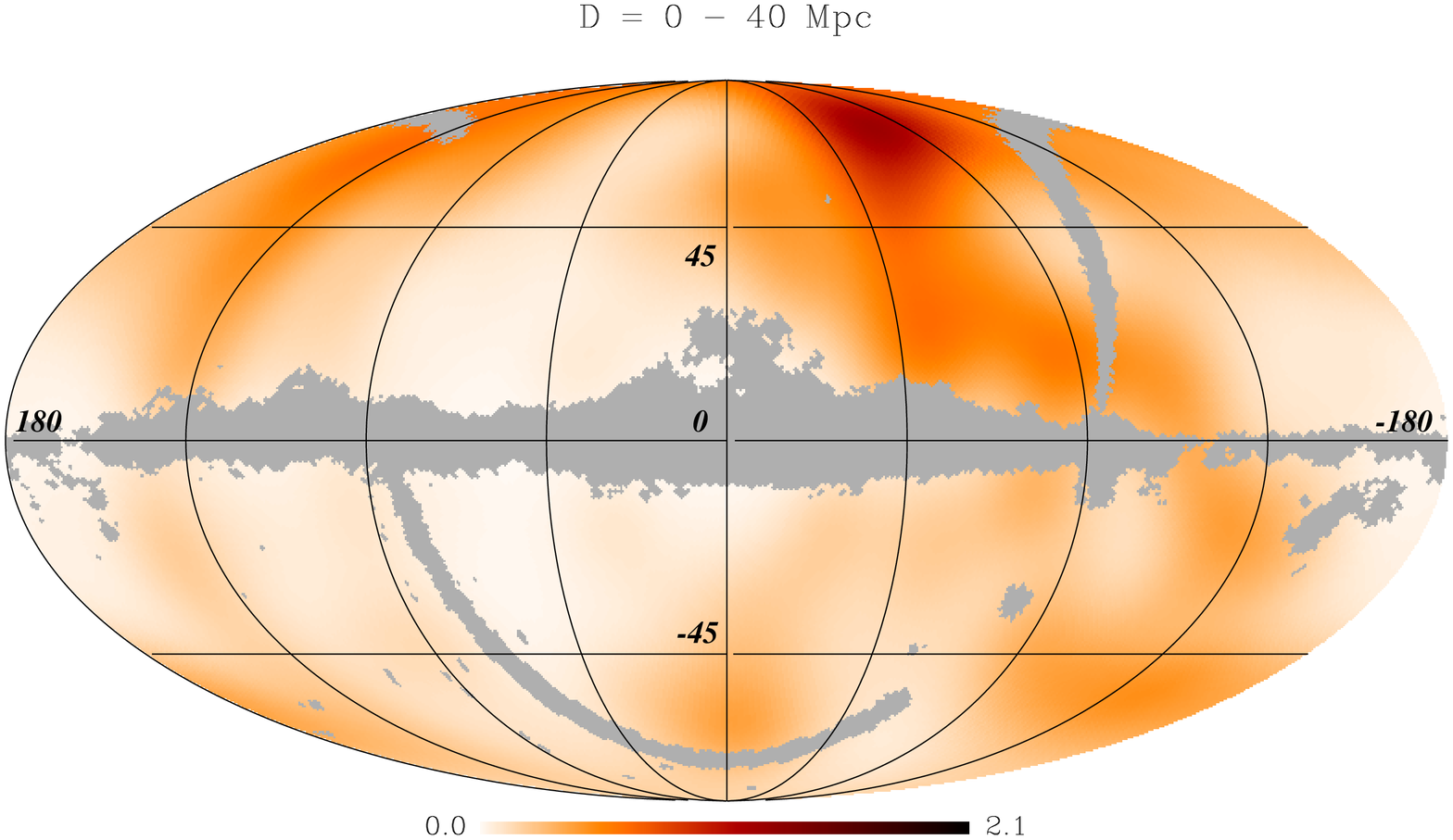} 
\includegraphics[width=0.3\textwidth, angle=90]{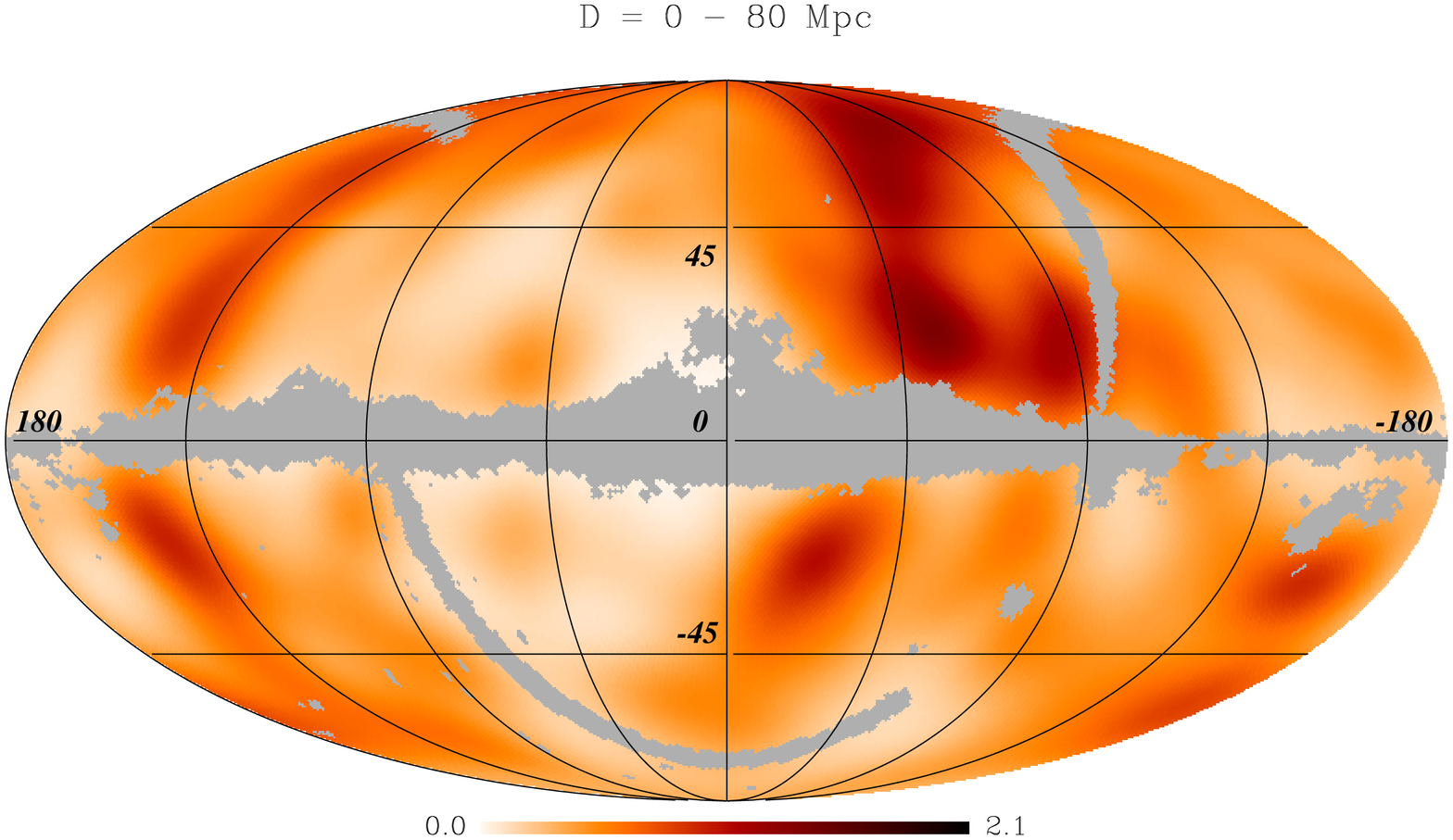} 
\includegraphics[width=0.3\textwidth, angle=90]{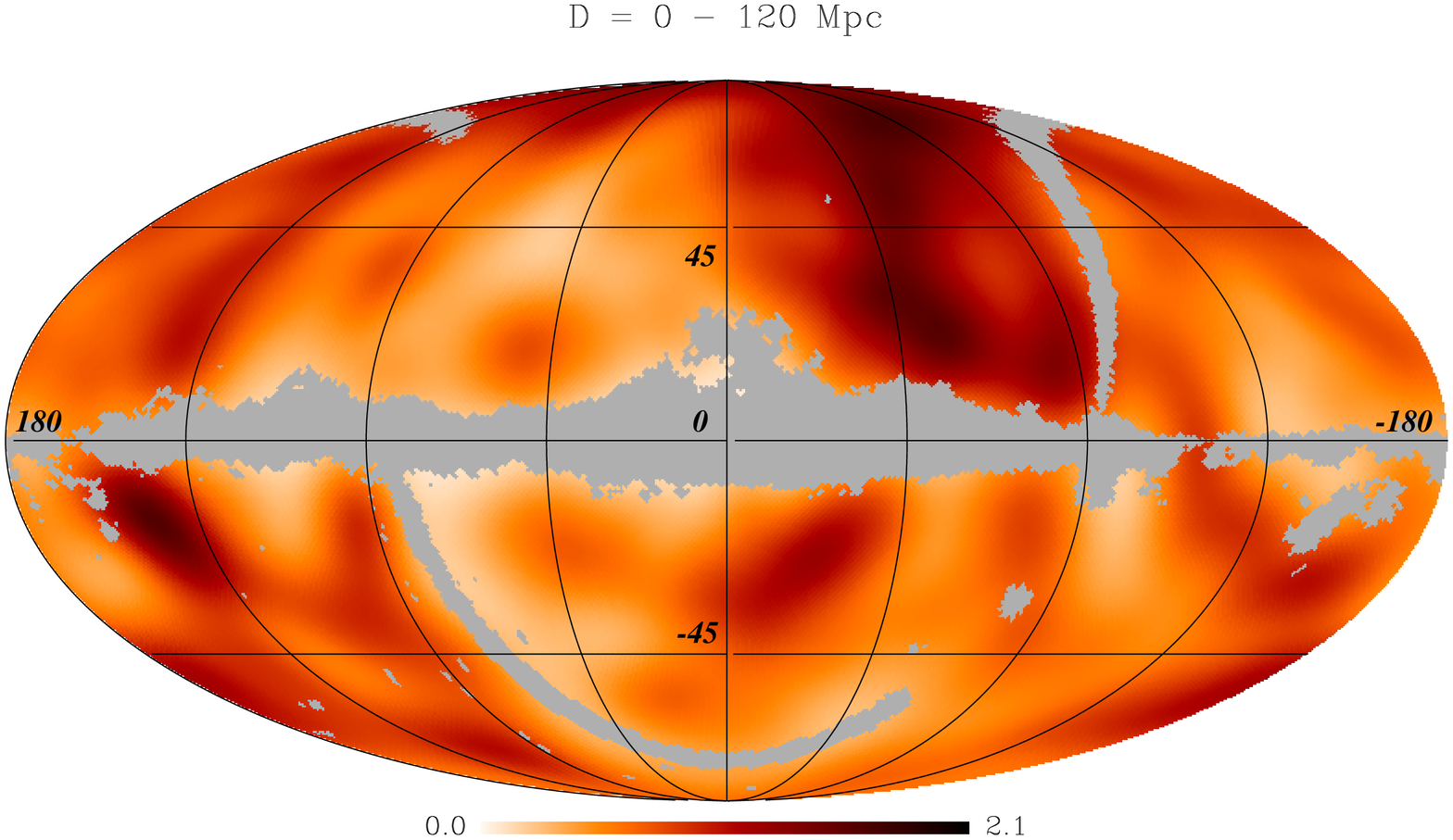} 
\includegraphics[width=0.3\textwidth, angle=90]{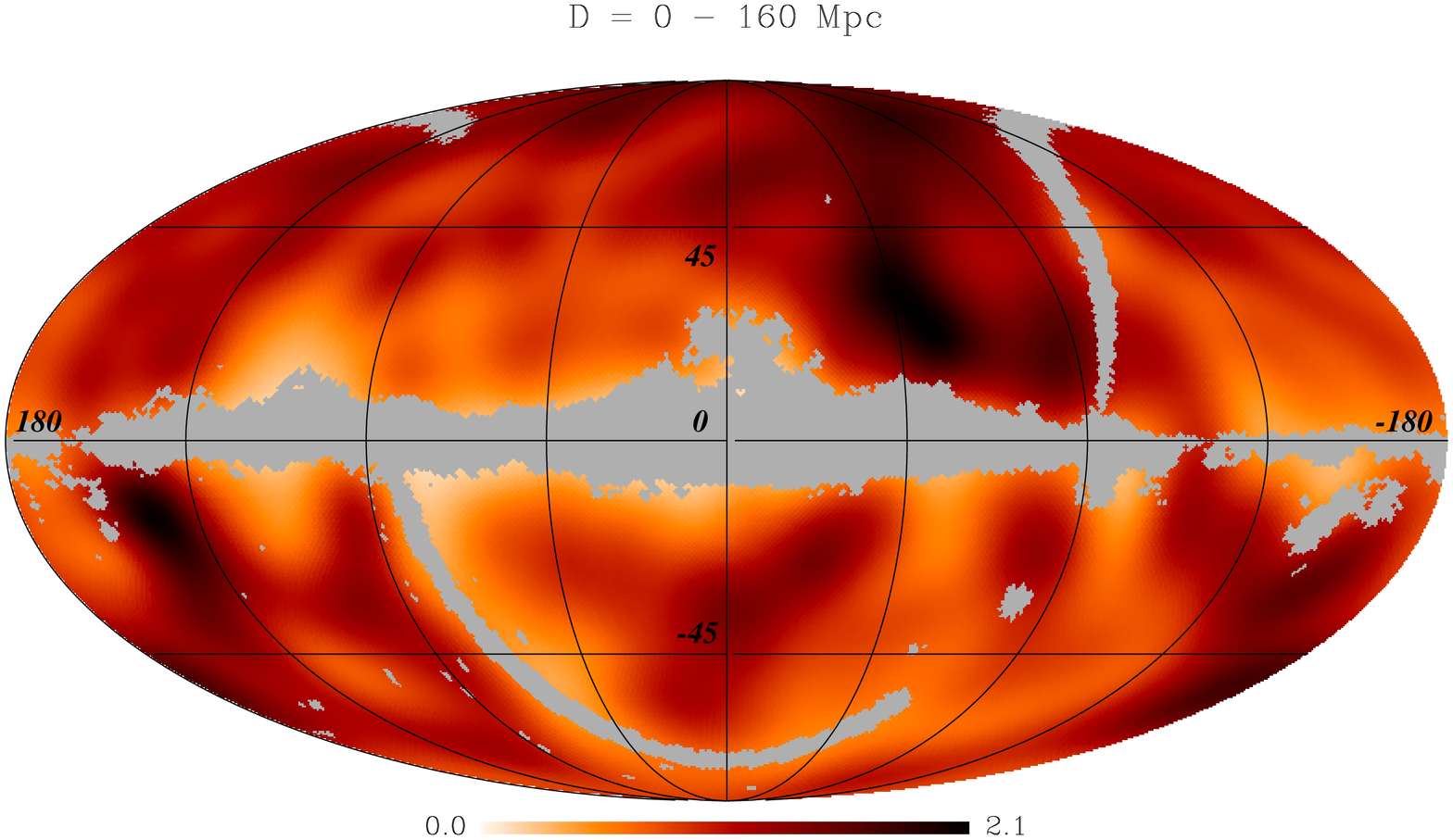} 

\caption{Integrated galaxy column density as derived from the PSCz
  catalog of galaxies up to the maximal distances $l=40\,$Mpc,
  $l=80\,$Mpc, $l=120\,$Mpc and $l=160\,$Mpc from left to right and
  top to bottom (Mollweide projection). The contours give the column
  density $N_{\rm g}$ in units of the mean column density $\langle
  N_{\rm g}\rangle\,=\,\langle n_{\rm g}\rangle \times 160$~Mpc, with
  $\langle n_{\rm g}\rangle$ the mean galaxy density. The grey mask
  indicates the regions of the sky that are not covered by the PSCz
  catalog~\cite{2000MNRAS.317...55S}.}
\label{fig:skymap}
\end{figure*}

The integrated column density of baryonic matter up to a distance $l$
is shown in Fig.~\ref{fig:skymap} for different maximal distances:
$l=40,\,80,\,120,\,160\,$Mpc (we adopt $H_0=70\,$km/s/Mpc). In order
to correct for the incompleteness of the catalog, we have followed the
prescriptions of Ref.~\cite{2000MNRAS.317...55S} and smoothed the
galaxy distribution with a variable gaussian filter, making use of the
{\tt HEALPix} library \cite{2005ApJ...622..759G}. The overall
resolution of the maps is of order $7^\circ$.

These maps provide an estimate of the optical depth to cosmic ray
scattering in the case in which the scattering centers are distributed
as the galaxies. If their distribution is biased with respect to that
of $n_{\rm g}$, for instance $n_i/\langle
n_i\rangle\,=\,b_i\left(n_{\rm g}/\langle n_{\rm g}\rangle\right)$,
then the optical depth is expressed as the following function of
$N_{\rm g}/\langle N_{\rm g}\rangle$:
\begin{equation}
\tau\,=\,\langle n_i\rangle \sigma_i l\,{\int \mathrm{d}l\,
  b_i\left(n_{\rm g}/\langle n_{\rm g}\rangle\right)\over
\int\mathrm{d}l\, n_{\rm g}/\langle n_{\rm g}\rangle}\,{N_{\rm g}\over
  \langle N_{\rm g}\rangle}\ ,\label{eq:tausky}
\end{equation}
and the prefactor $\langle n_i\rangle \sigma_i l\,=\,\langle
\tau\rangle$, see also Eq.~(\ref{eq:tauval}). The quantity $N_{\rm
  g}/\langle N_{\rm g}\rangle$ is that plotted in
Fig.~\ref{fig:skymap} for a distance $l=160\,$Mpc. This figure assumes
no bias, in which case $\tau\,=\,\langle\tau\rangle\,N_{\rm g}/\langle
N_{\rm g}\rangle$. Therefore, in order to read off $\tau$ from
Fig.~\ref{fig:skymap} and the above formula, one should use
$l=160\,$Mpc in the definition of $\langle\tau\rangle$ together with
the inferred value of $N_{\rm g}/\langle N_{\rm g}\rangle$ from
Fig.~\ref{fig:skymap}. If the bias were not trivial, meaning
$b_i(n_{\rm g}/\langle n_{\rm g}\rangle)\,\neq\,1$, its main effect
would be to increase the contrast of Fig.~\ref{fig:skymap}. Figures
for particular situations can be provided upon demand.

The above fiducial values for the scattering centers reveal an
important point: depending on the direction of observation, one may be
in a regime of small optical depth $\tau\,<\,1$ or large optical depth
$\tau\,>\,1$. This has several noteworthy consequences.

First of all, the typical deflection angle becomes itself direction
dependent. In particular, the values used in Eq.~(\ref{eq:thetaint})
correspond to $\langle\tau\rangle=3$ and $\delta\alpha\,\propto\,
\tau^{1/2}$. This simple scaling law along with Fig.~\ref{fig:skymap}
allow to estimate, as a function of the parameters characterizing the
scattering centers, the typical deflection angle in different parts of
the sky.

A sky map of deflection angles had been provided previously in
Ref.~\cite{DGST04}, using a constrained numerical simulation of the
local Universe with an all-pervading (albeit inhomogeneous) magnetic
field whose initial data was fixed at high redshift. One advantage of
the present maps shown in Fig.~\ref{fig:skymap} is to parametrize the
expected deflection in terms of the properties of the scattering
structures; in this sense, the above maps are more
general. Ref.~\cite{2006ApJ...639..803T} has also provided a similar
map, using the PSCz galaxy catalog to construct the matter density
field, and scaling the magnetic field to the matter density through
the law $B\,\propto\,\rho^{2/3}$. The exponent $2/3$ assumes isotropic
compression of the magnetic field during structure formation and it
seems that numerical simulations indicate a more sophisticated law,
with an exponent closer to $1$ (see discussion in
Refs.~\cite{D06,KL07}). Ref.~\cite{2006ApJ...639..803T} also
reconstructs the galaxy density field on small scales by repopulating
randomly the galaxy distribution using the density distribution from
the PSCz on larger scales, so that their map is influenced by this
reconstruction on scales smaller than $\sim 7^\circ$.

Following the discussion of Section~\ref{sec:angim}, the flux of a
source does not get demagnified nor magnified, up to possible magnetic
lensing effects, as it crosses a region of scattering centers,
provided the predicted apparent size of the source image does not
exceed that of the scattering structure. It will however suffer
demagnification in the opposite limit. Note that this does not
contradict the fact that an isotropic distribution of sources will
yield isotropic arrival directions on the detector provided that all
arrival directions from the detector can be backtracked to infinity.
Indeed, if a particular region of the sky is associated with a
particularly large angular deflection, the flux of any point source is
diluted by deflection through the crossing of this structure; however,
this deflection also opens a larger solid angle on the source plane,
so that a larger number of sources can contribute, and both effects
compensate each other. This fact has been discussed in particular in
Ref.~\cite{1999JHEP...08..022H} with respect to ultrahigh energy
cosmic ray propagation in the Galactic magnetic field.

As mentioned in Ref.~\cite{1999JHEP...08..022H}, one loop hole of the
above argument is the possible existence of so-called bottle orbits,
which do not connect the detector to infinity. However one does not
expect this effect to appear at the ultrahigh energies under
consideration in view of the (nearly) random and sporadic distribution
of the scattering centers throughout the Universe and in view of the
random nature of the deflection suffered at each interaction.  This
assertion could be verified using dedicated numerical simulations of
particle propagation.

Just as angular deflection, the time delay will depend on direction,
as $\delta t\,\propto \,\tau$. Although the magnitude of the time
delay (more precisely, of its variance) controls the number of
bursting sources that can be seen at a given time, it does not
influence the flux received as long as $N_{\rm GRB}\,\gg\,1$. Indeed,
a larger $\delta t$ means a larger $N_{\rm GRB}$ (at a fixed value of
$\sigma_t/\delta t$), but the flux of each gamma-ray bursts is
decreased accordingly by the larger $\sigma_t$ and both effects
compensate each other exactly. However, if at a given energy $N_{\rm
  GRB}\,\lesssim\,1$ in a certain region of the sky, one should
observe a corresponding cut-off in the energy spectrum from this
region of the sky, hence a reduced number of events.

As a clear example of the above possibility, consider a region of the
sky, of solid angle $\Delta \Omega$, in which the average optical
depth to cosmic ray scattering $\tau\,<\,1$. Then any source has a
probability $\simeq\,\tau$ of having one scattering center on the line
of sight, and therefore being seen if the time delay is sufficient. If
there is no scattering center on the line of sight (with probability
$1-\tau$), then the time delay is zero (in a first approximation), so
that the probability of observing a source within $\Delta\Omega$ and
up to a distance $l$ within the lifetime of an experiment $\Delta
t_{\rm exp}\,\sim\,10\,$years is extremely small:
\begin{eqnarray}
 P&\,=\,& {1\over 3}\Delta\Omega l^3 \dot n_{\rm GRB}\Delta\nonumber\\
  &\,\simeq\,& 3\times10^{-4}\,{\Delta\Omega\over 0.1\,{\rm str}}
{\dot n_{\rm GRB}\over 10^{-9}\,{\rm Mpc}^{-3}{\rm yr}^{-1}}
{\Delta t_{\rm exp}\over 10\,{\rm yrs}}\ .
\end{eqnarray}
Note that $0.1\,$str corresponds to a region of half-opening angle
$\,\simeq\,10^\circ$.  In practice, no source should be seen in this
particular direction unless it resides in a highly magnetized
environment [see discussion after Section~\ref{sec:inhom-num}, see
  also Eq.~(\ref{eq:dtimenum})]. As argued in Section~\ref{sec:angim},
one might see ``echoes'' of this source from scattering centers
located away from the line of sight, provided
$l\delta\alpha_i\,\gtrsim\, (n_il)^{-1/2}$. Even then, however, the
total flux of these secondary images would be demagnified by $\tau$ as
compared to that expected from the source without scattering.

In summary, the average flux expected in this solid angle
$\Delta\Omega$ is lower by a factor $\tau$ than that expected from
regions in which the optical depth is greater than unity.

Conversely, if the source is not of the bursting type, one might see
it directly in the arrival direction if this source lies in a hole of
the foreground scattering center distribution.

\subsection{Opaque regime}
The opaque regime corresponds to $\tau > \tau_{\rm eff} >1$. In this
case, cosmic rays diffuse from the source to the detector as in a
random billiard. 

The energy spectrum received from a given source is likely to be
strongly modified by the presence of strongly magnetized scattering
centers, as discussed in Ref.~\cite{2003PhRvD..68j3004S}. Roughly, one
should observe a low-energy cut-off at an energy $E_{\rm c}$ such that
$\delta\theta_i^2<1$ for $E> E_{\rm c}$ and
$\delta\theta_i^2\,\sim\,1$ at lower energies. However, when one
considers the energy spectrum received from an ensemble of sources,
whose flux interacts with an ensemble of scattering centers, one
should calculate the diffuse average flux in order to make contact
with the measured spectrum. This average spectrum should not differ
from the spectrum corresponding to rectilinear propagation if the
diffusion theorem applies~\cite{AB04} and magnetic horizon effects are
unimportant, i.e. if the distance between two sources $n_{\rm
  s}^{-1/3}$ is smaller than the energy loss distance and the
diffusion length. Otherwise, one should calculate the spectrum
following the methods of Ref.~\cite{BG06} with the diffusion
coefficient given below.

\subsubsection{Transport}
Assuming that the diffusion process obeys the normal law $\langle
r^2\rangle = 2Dt$, one may calculate the diffusion coefficient
$D$ using random walk arguments. In particular, if one neglects the
time spent in a magnetized structure in the course of an interaction,
the diffusion coefficient is related to the scattering length via the
usual law: $D\,=\, l_{\rm scatt}c$, where the scattering length
$l_{\rm scatt}$ has been defined in Eq.~(\ref{eq:lscatt}) above.

If the particle diffuses inside a structure during an interaction, then
it actually gets trapped in this structure during a certain amount of
time and exits backwards in a mirror-like fashion (see
Appendix~\ref{sec:appA}). Consider for simplicity a single scattering
agent. One may then account for the effect of time trapping by counting
the effective time taken to accomplish $N$ steps of the random walk,
which becomes $N d_i(1 + \delta t_i c/d_i)/c$. The correction decreases
$D$ by a factor $(1+\delta t_i c/d_i)$. Since the trapping time
$\delta t_i \,\simeq\, r_i/c$ is smaller than the typical distance $d_i$
between two scattering centers, this correction is not dominant.
Concerning the effect of mirroring, it suffices to note that it takes
two interactions to achieve isotropic deflection, hence this decreases
the diffusion coefficient by another factor of 2. These two corrections
thus remain of order unity.

The general scaling of this diffusion coefficient with energy is
easily grasped. At low energies (typically $E\lesssim 10^{18}\,$eV
depending on the parameters characterizing the scattering agents), it
does not depend on energy, as $l_{\rm scatt}$ simply corresponds to
the mean free path for scattering $\overline d$. In the high energy
regime, $D\propto E^2$ since the number of scatterings to achieve a
deflection of order unity scales in the same way. The above diffusion
coefficient may be used to describe the propagation of particles, as
done in Refs.~\cite{AB04,L05,AB05,BG06,KL07}. One may add that the
influence of any putative all-pervading magnetic field $B_{\rm IGM}$
may be safely neglected, even at energies of order $10^{18}\,$eV, as
long as $B_{\rm IGM}\,\lesssim\,10^{-11}\,$G, since the Larmor radius
$r_{\rm L}\,\simeq\,100\,{\rm Mpc}\,\left(E/10^{18}\,{\rm
  eV}\right)\left(B_{\rm IGM}/10^{-11}\,{\rm G}\right)^{-1}$.

In principle, a realistic distribution of magnetic field cells inside
the large scale structure might induce a scattering law with a more
complex profile than the standard exponential form adopted here, which
would furthermore depend on time in a non-trivial way so as to account
for the effect of trapping. The particle would then follow a so-called
continuous time random walk with waiting times, the properties of
which can be derived by following the methods developed in
Refs.~\cite{BG90,BHW87}. It would certainly be particularly
interesting if anomalous diffusion laws were to occur in such magnetic
field configurations.

\subsubsection{Experimental signatures for continuously 
emitting sources}\label{sec:opaquesign}
The arrival direction of high energy events will point back to the
source only if this latter is located at a distance closer than
$l_{\rm scatt}$. In the diffusive regime, the source distance scale is
no longer $l_{\rm max}$ but $\sqrt{l_{\rm scatt}l_{\rm max}}$, since
this latter gives the distance that a particle can cross before losing
its energy. Since we assume $l_{\rm max}>l_{\rm scatt}$, most of the
sources are located beyond $l_{\rm scatt}$.

  In the steady state regime, the diffusive flux received from a source
at distance $l$ scales as $1/(l_{\rm scatt}l)$, hence the flux received
from sources within $l$, with $l> l_{\rm scatt}$, scales as $l^2/l_{\rm
scatt}$. Consequently, the fraction of the flux that can be received
from sources at distances closer than $l_{\rm scatt}$ [given by
Eq.~(\ref{eq:fl})] is roughly $l_{\rm scatt}/l_{\rm max}$, just as in
the non-diffusive regime. This fraction gives the fraction of events
behind which one can hope to detect the source.

Note that the same delusive effect of finding a scattering center in
the arrival direction of cosmic rays occurs in this regime just as in the
translucent regime.

On general grounds, one expects the number of multiplets to be
significantly smaller in this case than for small deflection, since
the angular size of the image is considerably broadened.  However,
sources within the sphere of large angular scattering (for which the
Universe appears translucent) may produce images with higher
multiplicity if they exist, i.e. if $n_{\rm s}^{-1/3}\,< l_{\rm
  scatt}$. The number of events expected from a source at distance $l$
can be written as:
\begin{equation}
N_{\rm m}\,\simeq\, {N_{\rm obs} f_{\rm cov}\over n_{\rm s}4\pi l^2
l_{\rm max}}\ .
\end{equation}
In order to derive this estimate, it suffices to express the flux
received from this source, and to replace $\dot N_{\rm UHECR}$ in this
expression using Eq.~(\ref{eq:fl}). The parameter $f_{\rm cov}$
corresponds to the sensitivity of the detector in the direction
  of the source, normalized to the average sensitivity (i.e., on
  average $f_{\rm cov}=1$). One must emphasize that the above equation
  assumes that all sources have the same luminosity, which may be too
  restrictive.

Since $N_{\rm m}\propto 1/l^2$, the maximum
multiplicity $N_1$ will be associated to the closest source at
distance $\sim n_{\rm s}^{-1/3}$:
\begin{equation}
N_{\rm 1}\,\simeq\,0.1 f_{\rm cov} N_{\rm obs}{n_{\rm s}^{-1/3}\over
l_{\rm max}}\ .
\end{equation}
To provide quantitative estimates, if $n_{\rm s}\,=\,n_{\rm
  s,-5}\times10^{-5}\,{\rm Mpc}^{-3}$, the number of events expected
from the closest source at energies greater than $4\times10^{19}\,$eV
is a fraction $7\times10^{-3}\, n_{\rm s,-5}^{-1/3}$ of all observed
events. This number of events becomes a fraction $0.02 n_{\rm
  s,-5}^{-1/3}$ of $N_{\rm obs}$ above $6\times 10^{19}\,$eV.

Note that the expected multiplicity is the same in this case than that
found in the absence of magnetic fields, since we assume the source to
be within the sphere of large angular scattering. 

\subsubsection{Experimental signatures for bursting sources}
As far as bursting sources such as gamma-ray bursts are concerned,
most of the above results remains unchanged; one simply has to replace
$n_{\rm s}\dot N_{\rm UHECR}$ with $\dot n_{\rm s} N_{\rm UHECR}$.  In
the present case, the typical time spread corresponds to the diffusive
travel time, i.e. for a source at distance $l$:
\begin{equation}
\delta t \,\simeq\, {l^2\over 2 l_{\rm scatt}c}\ .
\end{equation}
Therefore the number of gamma-ray bursts sources which can contribute
to the flux at a given energy $E$, at any time, is:
\begin{equation}
N_{\rm GRB}\,\simeq\, \dot n_{\rm s}{2\pi\over 5}{l_{\rm max}^5\over 
l_{\rm scatt}c}\ .
\end{equation}

 To make concrete estimates, at $10^{20}\,$eV, $l_{\rm max}\,\simeq\,
95\,$Mpc hence $N_{\rm GRB}\,\sim\, 20$ if $l_{\rm
scatt}\,=\,20\,$Mpc, assuming $\dot n_{\rm
s}=10^{-9}\,$Mpc$^{-3}$yr$^{-1}$.  $N_{\rm GRB}$ is larger than
unity, which implies that one should not detect significant
statistical fluctuation in the energy spectrum and which explains why
one can record cosmic ray events in a near continuous manner, despite
the fact that close-by gamma-ray bursts are such rare events.  

There will of course be an energy $E_{\rm c}$ where $l_{\rm
scatt}\,=\,l_{\rm max}$, beyond which the diffusive regime will no
longer apply. In this case, one must use the formulae given in
Section~\ref{sec:int} for the translucent regime. Similarly,
regarding sources located within the sphere of large angular scattering,
i.e. at a distance $l<l_{\rm scatt}$, the phenomenological consequences
are those described in Section~\ref{sec:thin} if $l< \overline d$, or in
Section~\ref{sec:int} if $\overline d<l< l_{\rm scatt}$.

\section{Discussion}\label{sec:disc}

\subsection{Summary of present results}
The present work has provided an analytical description of ultrahigh
cosmic ray transport in highly structured extragalactic magnetic
fields. The corresponding configuration of the extragalactic magnetic
field is that of a collection of scattering centers, such as halos of
radio-galaxies or starburst galaxies, or magnetized filaments, with a
negligible magnetic field in between. Such a configuration is
generally expected in scenarios in which the magnetic field is
produced and ejected by a sub-class of galaxies, or generated at the
accretion shock waves of large scale structure. Even if the magnetic
field is rather generated at high redshift, subsequent amplification
in the shear and compressive flows of large scale structure formation
tends to produce a highly structured configuration, with strong fields
in the filaments of galaxies and weak fields in the
voids~\cite{SME04,DGST05,TS07,KRC07}.

In our description, transport of cosmic rays is modeled as a sequence
of interactions with the scattering centers, during which the particle
acquires a non-zero deflection angle and time delay (with respect to
straight line crossing of the magnetized region), see
Appendix~\ref{sec:appA}. In Section~\ref{sec:od}, we have sketched a
list of possible scattering centers and their characteristics (mean
free path to scattering, magnetic field, coherence length and
extent). We have then computed the optical depth $\tau$ of the
Universe to cosmic ray scattering as a function of energy and distance
to the source, as well as the effective optical depth $\tau_{\rm eff}$
(which is defined in such a way as to become unity when the total
angular deflection becomes unity). As discussed in
Section~\ref{sec:od}, the Universe can be translucent to cosmic
ray scattering if $\tau>1>\tau_{\rm eff}$, meaning that the total deflection is
smaller than unity but non-zero, opaque if $\tau>\tau_{\rm eff}>1$, or
even transparent if $1>\tau>\tau_{\rm eff}$. For typical values of the
scattering centers parameters, it is expected that the Universe be
translucent or opaque on the source distance scale and at
energies close to the pion production threshold. Since this energy is
that generally used by experiments as a threshold for the search for
counterparts, the above may have important phenomenological
consequences.

In particular, in the translucent or opaque regime, the closest
object lying in the cosmic ray arrival direction should be a
scattering center. Since these scattering centers are sites of intense
magnetic activity (radio-galaxies, starburst galaxies, shock waves,
...), they might be mistaken with the source. This peculiar feature
does not arise in models in which magnetic deflection is a continuous
process in an all-pervading magnetic field.  One could thus conceive
an ``ironic'' scenario, in which cosmic rays are accelerated in
gamma-ray bursts, but scatter against radio-galaxies magnetized lobes,
so that one interpret these latter as the source of cosmic rays
because they are the only active objects seen on the line of sight.
If such counterfeiting is taking place, one should observe that the
apparent distance scale to the source (actually the distance to the
last scattering surface) is smaller than the expected distance scale
to the source (as determined by the energy losses). This offers a
simple way to test for the above effect.

In the translucent regime, the source image is broadened by an
angle $\delta\alpha$ which takes values of order of a degree at energy
$10^{20}\,$eV for the fiducial values of the scattering structures
that we considered: interaction length $d_i\,\simeq\, 30\,$Mpc, extent
$r_i\,\simeq\,1\,$Mpc, magnetic field $B\,\simeq\, 10^{-8}\,$G, and
coherence length $\lambda_i\,\simeq\,0.1\,$Mpc. The average optical
depth at distance $100\,$Mpc is thus $\tau\,\simeq\,3$ for these
values. Due to the uncertainties surrounding these parameters, the
deflection could however be larger or smaller by about an order of
magnitude. In Section~\ref{sec:angim}, we have discussed effects
related to the shape of angular images when the discreteness of the
scattering centers is taken into account.

The inhomogeneous distribution of matter in the local Universe implies
that this optical depth to cosmic ray scattering should vary with the
direction of observation. In Section~\ref{sec:direcdep}, we have
provided sky maps of the integrated baryonic matter density up to
different distances, using the PSCz catalog of galaxies. These maps
allow to estimate the fluctuation of the optical depth in different
directions, hence that of the deflection angle, since
$\delta\alpha\,\propto\,\tau^{1/2}$.

In our discussion, we have taken into account the inhomogeneous
distributions of the scattering centers, see Sections~\ref{sec:inhom},
\ref{sec:inhom-num}. We have shown numerically that on path lengths
longer than $\,\sim200\,$Mpc, the effect of inhomogeneity is
negligible, as expected for a Universe that is homogeneous and
isotropic on these scales. The path length to the first interaction is
generally higher by about 40\% than in the homogeneous case if the
scattering centers distribute according to the dark matter
density. Since scattering centers tend to concentrate in filaments of
large scale structure, a particle may also experience multiple
interactions upon crossing a filament, as discussed and quantified in
Section~\ref{sec:inhom}. This explains why the number of interactions
in the inhomogeneous case converges toward that of the homogeneous
case on long path lengths.

\subsection{Recent data from the Pierre Auger Observatory}
In its first years of operation, the Pierre Auger Observatory has
already achieved the largest aperture (in ${\rm km}^2\cdot{\rm
  str}\cdot{\rm yr}$)~\cite{Auger1}, and it has recently released the
largest catalog of events above
$5.7\cdot10^{19}\,$eV~\cite{Auger2}. In this catalog, 20 out of 27
events originate from within 3 degrees of an active galactic nucleus
located within 75~Mpc.

The most straightforward interpretation is to infer that active
galactic nuclei are the sources of ultrahigh energy cosmic
rays. However, only one of the observed counterparts is of the FR-I
type (Centaurus~A), all others are more common Seyfert galaxies. From
a theoretical point of view, this is unexpected, since these common
active galactic nuclei do not seem to offer the required
characteristics for the acceleration to ultrahigh
energies~\cite{NMA95}. Even Centaurus~A, as far as its jets are
concerned, does not appear to be a likely source of ultrahigh energy
cosmic rays~\cite{CLP02}.

Furthermore, on a purely experimental level, Gorbunov and
co-authors~\cite{GTTT07} have recently pointed out an anomaly in this
observed correlation. Assuming that the AGN seen in the arrival
directions of these high energy events are the source of ultrahigh
energy cosmic rays, these authors have computed the expected flux
using the known distances to these AGN. They have observed that the
Pierre Auger Observatory has collected zero event in the direction to
the Virgo cluster, whereas at least six should be expected on the
basis of the large concentration of AGN in this direction and the
small distance scale (assuming that the cosmic rays coming from
Centaurus~A indeed originate from this object).

Ref.~\cite{GTTT07} thus argues that this observation rules out the
possibility that AGN are the sources of ultrahigh energy cosmic rays,
unless the cosmic rays seen in the direction to Centaurus~A come from
further away. However, as pointed out to us during the refereeing
process, it could also be that the absence of AGN-like source in Virgo
is a statistical fluctuation due to the small number of sources in the
local Universe, or that all AGN-like sources do not have the same
cosmic ray luminosity. One may also ponder on the possibility that the
Galactic magnetic field would exhibit a particular configuration in
the direction to Virgo (which lies toward the Galactic North Pole),
which would prevent cosmic rays from penetrating from this direction.
Hence at present, one cannot exclude formally that AGN are the source
of ultrahigh energy cosmic rays, but the data of the Pierre Auger
Observatory cannot be argued to sustend this hypothesis strongly
either.

Another interpretation suggests that sources of ultrahigh energy
cosmic rays cluster with the large scale structure, as AGN do, hence
the observed correlation with AGN is a coincidence. This hypothesis
deserves to be more carefully studied, for instance by performing
cross-correlations of the observed arrival directions with galaxy
catalogs, or by following the method introduced in
Ref.~\cite{WFP96}. However, assuming that the sources are located
close to the AGN which have been seen in the arrival directions should
not resolve the flux anomaly noted in Ref.~\cite{GTTT07}, also it
might mitigate it somewhat.

A third interpretation is to assume that at least part of the observed
correlation is accidental because the scattering centers on the last
scattering surface cluster with the large scale structure, hence with
AGN. This would alleviate this flux anomaly, since the sources would
no longer have to be associated with the AGN distribution. In
particular, the events seen to arise from the Centaurus complex might
have been deflected in its vicinity.

As mentioned previously, this scenario can be tested by comparing the
expected source distance scale with the counterpart distance
scale. Interestingly, both do not match, as the source distance scale
for particles with observed energy $6\times 10^{19}\,$eV is of the
order of 200~Mpc, significantly larger than the maximum distance of
75~Mpc for the observed counterparts.  This fact has been noted in
Ref.~\cite{Auger2}; it remained mostly unexplained, although it was
suggested in this work that both distance scales would agree if the
energy scale were raised by 30\%.

More quantitatively, one can calculate the probability that a given
event with a given observed energy originates from a certain distance,
using the fraction of the flux contributed by sources within a certain
distance at a certain energy. This probability law can be calculated
using the techniques developed in Ref.~\cite{BGG02}, then
tabulated. It is then possible to calculate the probability of seeing
20 out 27 events from a source located within 75~Mpc using the events
energies reported in Ref.~\cite{Auger2}. This probability is small,
about 3\%; the mean lies at 15 events out of 27 coming from within
75~Mpc. If one restricts the set of events to those that lie outside
the Galactic plane ($\vert b\vert > 12^\circ$), with 19 out of 21 seen
to correlate, the probability becomes marginal, of order 0.1\% (the
mean lies at 12 out 21 within 75~Mpc). Finally, if one restricts
oneself to the second set of events collected after May 27 2006, and
on those which lie outside of the Galactic plane, with 9 out of 11
seen to correlate, the probability becomes of order 10\%, with a mean
at 7 out 11 within 75~Mpc. In this latter case, the signal is less
significant, but the statistics is also smaller.  Since the above
estimates do not take into account the uncertainty on the energy, and
since they assume continuous instead of stochastic energy losses,
these numbers should be taken with caution. Nonetheless, the above
estimates agree with those of Ref.~\cite{2006JCAP...11..012H}, which
indicate that 50\% of protons with energy $E>6\times 10^{19}\,$eV
should come from distances less than $100\,$Mpc and 90\% from
distances less than $200\,$Mpc.

The above discussion suggests that, unless the energy scale is too low
or an experimental artefact is present, the inferred distance scale to
the source appears smaller than the expected source distance scale. In
light of the analysis developed in the present paper, this suggests
that part of the correlation may actually pinpoint scattering centers
correlating with AGN rather than the source of ultrahigh energy cosmic
rays. Said otherwise, the Pierre Auger Observatory may be seeing, at
least partly, the last scattering surface of ultrahigh energy cosmic
rays, rather than the source population.

In order to estimate the fraction of events that are likely to be
contaminated by such pollution, one may proceed as follows. Assume
first that the total deflection imparted to the particles with energy
$>6\times 10^{19}\,$eV is less than the $3^\circ$ radius used by the
Pierre Auger Observatory for their search. One may then calculate the
the fraction of galaxies in the PSCz catalog up to a distance
$l=200\,$Mpc, weighted appropriately, which lie within $3^\circ$ of an
AGN which is itself located closer than 75~Mpc. The distance
$l=200\,$Mpc is motivated by the fact that 90\% of events with energy
$>6\times10^{19}\,$eV originate from a distance smaller than
200~Mpc~\cite{2006JCAP...11..012H}. One should weigh each galaxy with
the selection function of the PSCz catalog at the distance $l$ of this
galaxy in order to correct for the incompleteness of the catalog; one
should also weigh each galaxy with a factor $1/l^2$ to account for
flux dilution during propagation.  In the above estimate, the PSCz
catalog is used as a tracer of the cosmic ray source population, and
one simply calculates the probability of angular coincidence with the
AGN sample. The number obtained is $0.31$, which suggests that 31\% of
events above $6\times10^{19}\,$eV could correlate with the AGN,
assuming that the PSCz galaxies provide an unbiased tracer of the
cosmic ray source population and that the magnetic deflection is much
smaller than the search radius of $3^\circ$. Note that this estimate
does not take into account the effect of the magnetic field; if one
were to restrict the angular radius to $2^\circ$ in order to account
for further possible Galactic deflection, the above fraction would
become 25\%. For reference, the probability that a random direction on
the sky falls within $3^\circ$ of an AGN (located closer than 75~Mpc)
is $0.22$ (becoming $0.11$ for a radius of $2^\circ$), which therefore
gives the covering factor on the sky of these AGN.

In order to account for magnetic deflection, one may repeat the above
procedure and calculate the probability of coincidence to within
$3^\circ$ of an AGN assuming that the event is displaced randomly by
an angle $\delta\alpha$ from the location of the galaxy drawn from the
PSCz catalog. Of course, one recovers the above result $0.31$ for
$\delta\alpha\rightarrow 0$, and the probability $0.22$ corresponding
to isotropic source distribution for $\delta\alpha \,\sim\, 1$ (in
practice, $\delta\alpha\,\gtrsim\,45^\circ$ gives a probability
$0.22$). Interestingly, the fraction of contaminated events increases
as $\delta\alpha$ becomes of order of a few degrees: it equals 39\%
for $\delta\alpha=1^\circ$, 48\% for $\delta\alpha=3^\circ$, then
decreases, being 45\% for $\delta\alpha=5^\circ$ and 43\% for
$\delta\alpha=7^\circ$, etc. If the radius of the correlation with the
AGN is restricted to $2^\circ$ to allow for further deflection in the
Galactic magnetic field, these numbers become 21\% for
$\delta\alpha=1^\circ$, 29\% for $\delta\alpha=3^\circ$ and 25\% for
$\delta\alpha=5^\circ$.

The above estimates indicate that, within the assumptions of the above
discussion, the delusion should not affect all events of the Pierre
Auger Observatory, but a significant fraction nonetheless, possibly as
high as $\simeq 50\,$\%. Moreover, it also indicates that
intergalactic magnetic deflection could be larger than $3^\circ$ and
yet produce a relatively significant false correlation with AGN. If
further data from cosmic ray experiments strengthen the observed
correlation, then the present interpretation would fail, unless some
other effects artifically enhance this false correlation.

For instance, one should point out that the above fraction of
contaminated events is likely to be enhanced if ultrahigh energy
cosmic rays originate from gamma-ray bursts. Indeed, as discussed in
Section~\ref{sec:direcdep}, one expects in this case the number of
events in regions of low foreground density to be smaller by a factor
of order $\tau$ ($\tau$ being the optical depth measured in such
directions) when compared to that coming from regions of optical depth
greater than unity. The main reason is that a given source has a
probability $\sim\,\tau$ of being located behind a scattering center
which would provide sufficient time delay for the source to become
observable wih reasonable probability. On the contrary, a nearby
gamma-ray burst with no scattering center on the line of sight has a
negligible probability of being seen during a time span of a few years
as a result of the small ocurrence rate. Although it is difficult to
give a simple estimate of the magnitude of this effect on the amount
of false correlations, one can easily see that it would tend to
increase this fraction by providing more weight to regions of high
foreground density (in which AGN are more numerous).

In Ref.~\cite{Auger2}, the Pierre Auger Observatory has discussed the
evolution of the probability of null hypothesis for an isotropic
distribution of sources with a varying search radius, maximum AGN
redshift and minimum energy (see Fig.~3 of Ref.~\cite{Auger2}). The
minimum probability (which indicates a maximal correlation with the
AGN) corresponds to a search radius $3.2^\circ$. This minimum can be
interpreted as an estimate of the amount of Galactic and intergalactic
magnetic deflection if one assumes that the source exactly correlates
with the AGN. Interestingly, our above discussion suggests that this
number may be a biased estimate and that the intergalactic deflection
could be slightly larger. The increase of the probability of null
hypothesis at larger search radii in the Pierre Auger data corresponds
to the fact that the covering factor of the search area increase
rapidly with search radius, being already 0.50 at
$6^\circ$. Concerning the redshift evolution, one would expect in the
present model that the correlation would persist to distances as large
as $200\,$Mpc if the search radius is larger than the typical
intergalactic deflection. Unfortunately, Ref.~\cite{Auger2} does not
plot this correlation beyond $100\,$Mpc. It would be interesting to
also carry out this test for different search radii. 

One should emphasize that in the present interpretation, the
correlation with AGN should not persist as the threshold energy is
decreased. Indeed, the maximum propagation distance of particles of
observed energy $4\times10^{19}\,$eV is of order $500\,$Mpc, on which
scale the Universe appears isotropic. Therefore, at these energies the
incoming flux is increasingly isotropic, and the presence of
scattering centers on the line of sight cannot induce anisotropies on
an isotropic sky distribution (see discussion in
Section~\ref{sec:angim} as well as the discussion on the application
of the Liouville theorem in Ref.~\cite{1999JHEP...08..022H}).  The
fraction of flux contributed by the isotropic background has been
estimated in Ref.~\cite{2006JCAP...01..009C} in the absence of
extragalactic magnetic field; it reaches 83\% for $E>3\times
10^{19}\,$eV, and 3.6\% for $E>5\times 10^{19}\,$eV. The strong rise
toward isotropy as the threshold energy decreases is thus clear. This
effect is present, at least qualitatively, in the data of the Pierre
Auger Observatory (see Figure 3 of Ref.~\cite{Auger2}).

Finally, it appears that comparing the apparent source distance scale
with the expected one, as we have done above, remains the most direct
and simple test of the present interpretation.  Since there is a
non-negligible degeneracy between the expected distance scale and the
energy calibration, it seems mandatory to obtain a calibration through
other methods that is as accurate as possible.

It also appears imperative to probe the arrival directions on an event
by event basis, focussing on the most energetic events. In the catalog
reported in Ref.~\cite{Auger2}, there is only one event above
$10^{20}\,$eV, whose arrival direction has a relatively small
super-Galactic latitude, $b_{\rm SG}\,\simeq\,-6.5^\circ$. In the
above scenario, one should expect to find a scattering center or the
source on the line of sight, hence it should prove useful to perform a
deep search in this direction in the radio domain, looking for traces
of synchrotron emission that would attest of the presence of a locally
enhanced intergalactic magnetic field. Many more events at higher
energies, as expected from future detectors such as Auger
North~\footnote{{\tt http://www.augernorth.org}}, would certainly help
in this regard.

A last word should be added concerning the amount of magnetic
deflection and the source models. In particular, it would be
interesting to examine whether (and to what cost) the current data
could be reconciled with ultrahigh energy cosmic rays being
accelerated in the most powerful AGN, which offer stronger ground than
Seyfert galaxies for acceleration. Such a study can only be conducted
through detailed Monte-Carlo simulations which allow for substantial
scattering angles in inhomogeneous magnetic fields.

As explained in Ref.~\cite{W01}, gamma-ray bursts are probably the
most elusive of possible ultrahigh energy cosmic ray sources, as the
strongest predictions are that no counterpart should be detected, that
the flux should show significant variations around the mean at
sufficiently high energies (a few $10^{20}\,$eV), and that multiplets
of events should be clustered in energy. Current data do not violate
any of these predictions, but it is clear that experiments with much
larger aperture at the highest energies will be needed to test such
effects.

It is certain that much physics and astrophysics of cosmic ray sources
and large scale magnetic fields remain to be unveiled by ongoing and
future detectors.

\begin{acknowledgments}
We thank S.~Colombi for providing the dark matter simulation and
H.~Atek, G.~Bou\'e, N.~Busca, Y.~Dubois, E.~Hivon, A.~Olinto, C.~Pichon
and S.~Prunet for discussions.
\end{acknowledgments}

\appendix

\section{Particle -- scattering center interaction}\label{sec:appA}
This section describes the interaction between a particle and a
scattering center in the large scale structure, then computes the
deflection angle and the trapping time in the structure before its
return to non-magnetized voids. We consider both the cylindrical
geometry, which is representative of an interaction with a filament,
and the spherical geometry, which we use to model the interaction with
a magnetized wind or cocoon. The solution of the diffusion equation in
a planar geometry can be found in Ref.~\cite{SLB99}. 

This discussion assumes that the magnetic field strength and the
diffusion coefficient are uniform in the scattering
center. Expectations in the more general non-uniform case are
discussed briefly at the end of this Appendix.

\subsection{Interaction with a sphere or a filament:\\ general results}
If the scattering length $l_{\rm sc}$ of the particle in the
scattering center is much larger than the characteristic path length
$\bar r_i$ through the structure, the particle is simply deflected by
an angle $\delta\theta_i$ and emerges after a crossing time
$t\,\simeq\,\bar r_i/c$. The characteristic size $\bar r_i$ should be
thought of as the smallest length scale of the structure,
i.e. $(\pi/2)r_{\rm s}$ for a sphere or $(\pi/2)^2r_{\rm f}$ for a
filament. The factors of $\pi/2$ account for random orientation of the
incoming direction.

The deflection angle at each interaction can be computed as
follows. Consider a spherical magnetized halo of radius $r_i$, magnetic
field $B_i$ and magnetic coherence length $\lambda_i$. The magnetic
scattering length $l_{\rm sc}$ of a particle of Larmor radius $r_{\rm
  L}$ in this structure determines the length beyond which the
particle has experienced a deflection of order unity. Hence, if
$l_{\rm sc}\,\ll\,r_i$, the particle undergoes diffusion in the
structure so that $\delta\theta_i \,\sim\, {\cal O}(1)$. In the
following Section, it is also shown that the distance traveled in the
structure is very small as compared to $r_i$, so that escape actually
takes place close to the point of entry with a mirror-like deflection
of order $\pi$ (to within $\pm\pi/2$).

If, however, $l_{\rm sc}\,\gg\,r_i$, the particle is only weakly
deflected. In order to calculate $\delta\theta^2_i$, one must specify
$l_{\rm sc}$ as a function of $r_{\rm L}$ and $\lambda_i$. The general
relationship between these quantities can be expressed as:
\begin{equation}
l_{\rm sc}\,\simeq\, \alpha \, r_{\rm L}\,\left({
r_{\rm L}\over\lambda_i}\right)^\beta\ .
\end{equation}
This equation neglects a numerical prefactor of order unity (see
Ref.~\cite{CLP02} for more details). The coefficient $\alpha$ is
directly related to the level of turbulence in the structure:
\begin{equation}
\alpha\,=\,\left({\delta B_i^2\over B_i^2}\right)^{-1}\ ,
\end{equation}
where $\delta B_i$ represents the turbulent component and $B_i$ the total
magnetic field. In the following, we assume $\alpha\,\simeq\, 1$,
meaning full turbulence, but the calculations that follow may be
generalized to $\alpha\neq1$ without difficulty. The various scenarios
of magnetic pollution discussed before do not favor the existence of
significant coherent components of the magnetic field.

Regarding the exponent $\beta$, $\beta=1$ if $r_{\rm L}\,\gg\,
\lambda_i$~\cite{GJ99,CLP02,CR04}. If, however, $r_{\rm L}\,\ll\,
\lambda_i$, then $\beta$ also depends on the shape of the turbulence
spectrum. For instance, $\beta=-2/3$ for Kolmogorov turbulence,
$\beta=0$ for scale invariant turbulence (Bohm regime).  For
simplicity, and in the absence of any knowledge of the turbulence
spectrum in the scattering centers, we assume $\beta=0$, which allows
to simplify the discussion. Again, it is possible to extend the
discussion to different values of $\beta$, albeit at the price of
slightly more complicated expressions.

Therefore, one finds the following deflection angle. If $r_{\rm
L}\,\ll\, \lambda_i$, then $l_{\rm sc}\,\simeq\,
r_{\rm L}\,\ll\,\lambda_i\,<\,r_i$, hence the particle
diffuses in the structure and exits with a deflection of order
unity. Note that the inequality $\lambda_i\,<\,r_i$ simply states that
the coherence length of the magnetic field cannot exceed the size of
the magnetic structure.

If $r_{\rm L}\,\gg\, \lambda_i$, then $l_{\rm sc}\,\simeq\,r_{\rm L}^2/
\lambda_i$. One then must consider whether $r_{\rm L}$ is larger or
smaller than $\sqrt{\lambda_ir_i}$. In the former case, $l_{\rm
  sc}\,\gg\, r_i$, hence the particle exits with a small deflection
angle $\delta\theta^2_i\,\simeq\, \bar r_i\lambda_i/(2r_{\rm
  L}^2)$~\cite{2002JHEP...03..045H}. This numerical prefactor $1/2$ is
valid for propagation in a turbulent magnetic field; it becomes $2/3$
for a randomly oriented regular magnetic field~\cite{WM96}.

In the latter case, $l_{\rm sc}\,\ll\, r_i$ hence the particle exits
with a deflection angle of order unity.

In conclusion, the deflection angle can be written in the approximate form:
\begin{equation}
\delta\theta^2_i\,\simeq\,\left(1 + {2r_{\rm L}^2\over \bar
  r_i\lambda_i}\right)^{-1}\ .
\label{eq:dtheta}
\end{equation}
Although this form is only approximate, it interpolates smoothly
between the two different regimes of interest 
$r_{\rm L}\,\ll\,\sqrt{r_i\lambda_i}$ (large deflection) and 
$r_{\rm L}\,\gg\,\sqrt{r_i\lambda_i}$ (small deflection). In the high
energy (small deflection) limit, one finds:
\begin{eqnarray}
\delta\theta_i&\,\simeq\,&1.7^\circ\,\left({\bar r_i\over 2\,{\rm
    Mpc}}\right)^{1/2} \left({B_i\over 10^{-8}\,{\rm
    G}}\right)\times\nonumber\\ &&\,\,\left({\lambda_i\over
  0.1\,{\rm Mpc}}\right)^{1/2} \left({E\over 10^{20}\,{\rm
    eV}}\right)^{-1}\ .\label{eq:dthetanum}
\end{eqnarray}

The time delay with respect to straight line crossing of the
magnetized structure can be calculated, following
Refs.~\cite{1978ApJ...222..456A,WM96,2002JHEP...03..045H}:
\begin{equation}
\delta t_i\,\simeq\,{\bar r_i\delta\theta^2_i\over 6c}\ .\label{eq:dtime}
\end{equation}
This formula is only valid for small deflection angles; the
corresponding time delay in the diffusive regime is discussed further
below. In the high energy limit $r_{\rm L}\,\gg\,\sqrt{r_i\lambda_i}$,
this gives:
\begin{eqnarray}
\delta t_i\,&\,\simeq\,& 0.93\times10^3\,{\rm yr}\,\left({\bar r_i\over 2\,{\rm
    Mpc}}\right)^2 \left({B_i\over 10^{-8}\,{\rm
    G}}\right)^2\times\nonumber\\ &&\,\,\left({\lambda_i\over
  0.1\,{\rm Mpc}}\right) \left({E\over 10^{20}\,{\rm
    eV}}\right)^{-2}\ .\label{eq:dtimenum}
\end{eqnarray}

\subsection{Diffusive interaction with a filament}\label{sec:difffil}
If $l_{\rm sc}\,\ll\,r_{\rm f}$, the particle diffuses inside the
 filament before escaping. One can assume that the particle penetrates
 a length scale $l_{\rm sc}$ inside the filament, and then enters
 the diffusive regime. The time-dependent diffusion equation can then
 be used to compute the probability of escape as a function of time,
 treating the point of first interaction (at depth $l_{\rm sc}$) as
 an impulsive source. To this effect, we describe the filament as a
 cylinder of radius $r_{\rm f}$ and infinite extension along $z$ and consider
 cylindrical coordinates $(r,\theta,z)$. For simplicity, we assume a
 spatially uniform diffusion coefficient $D_\perp$ in the plane
 perpendicular to $z$, and a spatially uniform diffusion coefficient
 $D_\parallel$ in the direction along $z$. We also neglect energy
 losses, which is justified in so far as we will show that the
 trapping time is short on the typical energy loss timescale. The
 equation for the Green's function
 $g(r,\theta,z,t;r_0,\theta_0,z_0,t_0)$ reads:
\begin{widetext}
\begin{equation}
\partial_t g\,-\,D_{\perp}{1\over r}\partial_r\left(r\partial_r
g\right) \,-\,D_\perp{1\over r^2}\partial_\theta^2g\,-\,
D_\parallel\partial_z^2 g\,=\,{1\over
r}\delta(z-z_0)\delta(r-r_0)\delta(\theta-\theta_0)\delta(t-t_0)\ .
\label{eq:diff-cyl}
\end{equation}
Here, $r_0$, $\theta_0$, $z_0$ and $t_0$ give the coordinates of the
first interaction in the filament.  One must also take into account
the appropriate boundary conditions, namely that beyond radius $r_{\rm
f}$, the volume is unmagnetized. In the theory of diffusion, such
boundary conditions can be modeled by ensuring that the solution to the
diffusion equation vanishes at a radius $r_{\rm f}$.  In order to
solve the diffusion equation in cylindrical coordinates under this
constraint, one expands the angular part of the Green's function $g$
over a basis of proper functions of the operator $\partial_\theta^2$
and the radial part over a basis of Bessel functions
$J_m(\alpha_{ms}r/r_{\rm f})$, where $\alpha_{ms}$ denotes the $s-$th
root of $J_m$. This guarantees that the boundary condition will be
satisfied. The solution of Eq.~(\ref{eq:diff-cyl}) reads:
\begin{eqnarray}
g(r,\theta,z,t;r_0,\theta_0,z_0,t_0)&\,=\,&
{1\over \pi r_{\rm f}^2}\sum_{m=-\infty}^{m=+\infty}
\sum_{s=1}^{s=+\infty}\,\,e^{im(\theta-\theta_0)}\, e^{
-\alpha_{ms}^2{\scriptstyle D_\perp\vert t-t_0\vert\over \scriptstyle
r_{\rm f}^2}} {\displaystyle e^{-{\scriptstyle \vert z-z_0\vert^2\over
\scriptstyle 4D_\parallel\vert t-t_0\vert}}\over \scriptstyle
\sqrt{4\pi D_\parallel\vert t-t_0\vert}}\,\nonumber\\ &\,\,&\,\,\,
\times {J_m\left(\alpha_{ms}{\scriptstyle r_0\over \scriptstyle r_{\rm f}}\right)
J_m\left(\alpha_{ms}{\scriptstyle r\over\scriptstyle  r_{\rm f}}\right)\over
J_{m+1}\left(\alpha_{ms}\right)^2}\ .
\end{eqnarray}
The probability of having the particle inside the filament at any time
$t>t_0$ is then given by the volume average of $g$ over the filament:
\begin{equation}
P_{\rm res}(t;t_0)\,=\,\int{\rm d}v\, g\,=\,\sum_{s=1}^{s=+\infty}\, e^{-\alpha_{0s}^2{\scriptstyle
D_\perp\vert t-t_0\vert\over\scriptstyle r_{\rm
f}^2}}{2\over
\alpha_{0s}}{J_0\left(\alpha_{0s}{r_0\over r_{\rm f}}\right)\over
J_1(\alpha_{0s})}\ .\label{eq:Pres-cyl}
\end{equation}
Through the explicit decomposition of unity over the above basis of
Bessel functions, one can verify that $P_{\rm esc}(t\rightarrow
t_0)=1$ as it should. The form of $P_{\rm res}(t;t_0)$ tends to
suggest that escape takes place on a diffusive timescale $r_{\rm
f}^2/D_\perp$; this statement is actually too naive, as shown in the
following.  The average residence time in the filament $\delta t_{\rm
f}$ is calculated as:
\begin{equation}
\delta t_{\rm f}\,=\,\int_{t_0}^{+\infty}{\rm d}t\,P_{\rm
  res}(t;t_0)\,=\,{r_{\rm f}^2\over
  2D_\perp}\sum_{s=1}^{s=+\infty}\,{4\over\alpha_{0s}^3}
  {J_0\left(\alpha_{0s}{r_0\over r_{\rm f}}\right)\over
  J_1(\alpha_{0s})}\ .\label{eq:dt-cyl}
\end{equation}
The factors in the sum on the r.h.s. of Eq.~(\ref{eq:dt-cyl}) are much
smaller than unity because the particle cannot penetrate further than
$l_{\rm sc}\,\ll\,r_{\rm f}$ into the filament before starting to
diffuse, hence $r_0\,\simeq\,r_{\rm f}\left(1-l_{\rm sc}/r_{\rm
f}\right)$. This substitution followed by the expansion of the Bessel
functions to first order in terms of $l_{\rm sc}/r_{\rm f}$ leads
to the trapping time:
\begin{equation}
\delta t_{\rm f}\,\simeq\,{r_{\rm f}l_{\rm sc}\over
  D_\perp}\,\sum_{s=1}^{s=+\infty} {2\over \alpha_{0s}^2}\ ,
\end{equation}
which is effectively smaller than the diffusive time by a factor
$l_{\rm sc}/r_{\rm f}$. Since $D_\perp\,=\,{1\over 2}l_{\rm
sc}c$, one finally obtains:
\begin{equation}
\delta t_{\rm f}\,\simeq\, {r_{\rm f}\over
  c}\,\sum_{s=1}^{s=+\infty}{4\over \alpha_{0s}^2}\,=\,{r_{\rm
    f}\over c} .
\end{equation}
Alternatively, one could calculate the residence time by averaging
Eq.~(\ref{eq:dt-cyl}) over the probability of first scattering
$P(r_0)\,\simeq\,\exp\left[-(r_{\rm f}-r_0)/l_{\rm sc}\right]/l_{\rm
sc}$, but this would lead to similar results.

We thus find that the trapping time is of order of the crossing time,
an unexpected result. Since the particle diffuses, the linear length
scale traveled in this trapping time is only $l\,\sim\, (l_{\rm
sc}/r_{\rm f})^{1/2} r_{\rm f}\,\ll\, r_{\rm f}$. Hence the
particle enters and exits the filament at about the same location,
albeit a crossing time later. In terms of angular scattering, this
interaction is thus akin to mirroring, as the particle will exit in a
direction separated by less than $\pi/2$ from the direction of entry
in the filament.

This law $\delta t_{\rm f}\,\simeq\, r_{\rm f}/c$ has been verified
numerically, using Monte Carlo simulations of the interaction of a
particle with a magnetized filament, for various coherence lengths of
the magnetic field. The numerical code used has been described in
detail in Ref.~\cite{KL07}. The results are shown in
Fig.~\ref{fig:filint} below, where it is seen that the average
residence time does not depend on the coherence length of the magnetic
field (which characterizes the diffusion coefficient, hence the
scattering length), but evolves linearly with the filament
radius. Numerically, one obtains $\delta t_{\rm f}\,\simeq\,1.3r_{\rm
  f}/c$.

\subsection{Diffusive interaction with a sphere}
The interaction with a sphere of radius $r_{\rm s}$ is quite similar
to that with a filament, although the algebra is slightly more
cumbersome. As before, we assume that the particle penetrates a length
scale $l_{\rm sc}$ before starting to diffuse in the magnetized
sphere, and adopt appropriate boundary conditions at radius $r_{\rm
s}$. The diffusion equation in spherical coordinates reads:
\begin{equation}
\partial_t g\,-\,{D\over r^2}\partial_r\left(r^2\partial_r g\right)
\,-\,{D\over
r^2\sin\theta}\partial_\theta\left(\sin\theta\partial_\theta
g\right)\,-\, {D\over r^2\sin^2\theta}\partial_\phi^2 g\,=\,{1\over
r^2\sin\theta}\delta(r-r_0)\delta(\theta-\theta_0)\delta(\phi-\phi_0)\delta(t-t_0)\
.
\label{eq:diff-sph}
\end{equation}
Its solution is written in terms of spherical harmonics and spherical
Bessel functions:
\begin{eqnarray}
g(r,\theta,\phi,t;r_0,\theta_0,\phi_0,t_0)&\,=\,&
\sum_{l=0}^{l=+\infty}\,\sum_{m=-l}^{m=+l}\,
\sum_{s=1}^{s=+\infty}
e^{-\beta_{ls}^2{\scriptstyle D\vert t -t_0\vert\over \scriptstyle
    r_{\rm s}^2}}
\,Y_{lm}(\theta,\phi)\,\overline Y_{lm}(\theta_0,\phi_0)\,\nonumber\\
& & \,\times j_l\left(\beta_{ls}{r\over r_{\rm s}}\right)
j_l\left(\beta_{ls}{r_0\over r_{\rm s}}\right){2\over
r_{\rm s}^3j_{l+1}^2(\beta_{ls})}\ .\label{eq:g-sph}
\end{eqnarray}
The notation $\beta_{ls}$ indicates the $s-$th zero of the spherical
Bessel function $j_l$. As before, the probability of residence inside
the spherical structure at time $t>t_0$ can be computed by integrating
$g$ over the volume:
\begin{equation}
P_{\rm
    res}(t;t_0)\,=\,\sum_{s=1}^{s=+\infty}e^{-\beta_{0s}^2{\scriptstyle
    D\vert t -t_0\vert\over \scriptstyle r_{\rm s}^2}}\,
    j_0\left(\beta_{0s}{r_0\over r_{\rm s}}\right)\,{2\over
    \beta_{0s}j_1(\beta_{0s})}\ .\label{eq:Pres-sph}
\end{equation}
Here as well, $P_{\rm res}(t\rightarrow t_0)=1$ as it should. Finally,
the residence time can be calculated by taking the limit
$r_0\rightarrow r_{\rm s}(1-l_{\rm sc}/r_{\rm s})$ as before and
expanding to first order in $l_{\rm sc}/r_{\rm s}$:
\begin{equation}
\delta t_{\rm s}\,=\,\int_{t_0}^{+\infty}{\rm d}t\,P_{\rm
res}(t;t_0)\,\simeq\, {r_{\rm s}\over
c}\,\sum_{s=1}^{s=+\infty}{6\over \beta_{0s}^2}\,=\,{r_{\rm s}\over c}\ .
\end{equation}
The particle bounces on the sphere, exiting at a distance $l\,\sim\,
(l_{\rm sc}/r_{\rm s})^{1/2}r_{\rm s}\,\ll\,r_{\rm s}$ away from its point of impact.

\end{widetext}

\begin{figure}[ht]
\includegraphics[width=0.49\textwidth]{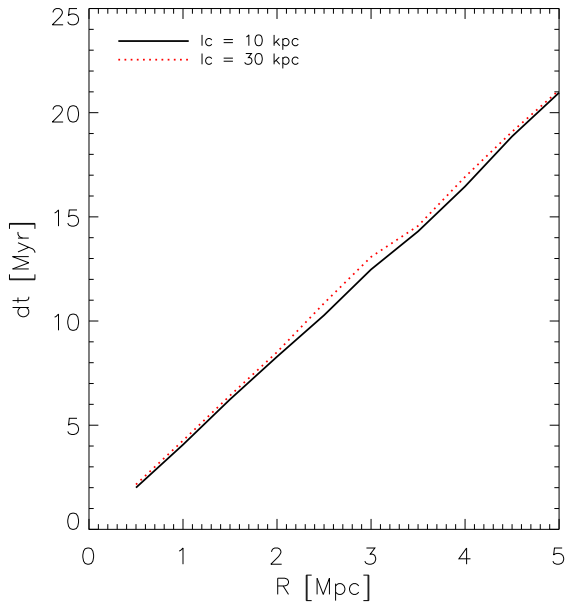} \caption{Residence
  time in a magnetized filament embedded in a non-magnetized medium
  for a particle impinging on the filament with a scattering length
  $l_{\rm sc}\,\ll\,r_{\rm f}$, as a function of the radius of the
  filament. The scattering length is a function of the coherence
  length of the magnetic field $\lambda$, that corresponds to the
  modelling of Kolmogorov turbulence inside the filament, i.e. $l_{\rm
    sc}\,\propto\, \lambda^{2/3}$, see
  Ref.~\cite{KL07}.}  \label{fig:filint}
\end{figure}

\subsection{Non uniform magnetic field}
The above discussion has assumed that the magnetic field and the
diffusion coefficient are uniform in the scattering center. If the
length scale of variation of the magnetic field, $l_B=\vert \nabla
B^2/B^2\vert^{-1}$ is ``small enough'', the discussion becomes more
intricate as the scattering length of the cosmic ray becomes itself
space dependent, and meaningless if it is larger than
$l_B$. Nevertheless, one may expect the following to occur.

If the scattering length as measured everywhere in the scattering
center is larger than its size $r$, then the total deflection will
remain much smaller than unity. Its value will be given by an average
of order $\langle r l_B/ r_{\rm L}^2\rangle$, where the average is
to be taken on $r_{\rm L}$ (through its spatial dependence via $B$) on
the trajectory. This estimate assumes that the particle is deflected
by $\delta\theta^2 \sim (l_B/r_{\rm L})^2$ every $l_B$. Its
corresponds to the estimate of the above discussion if $\lambda$ is
replaced by $l_B$ and if $B$ is understood as the average magnetic
field. The crossing time will remain unchanged, of order $r/c$.

If the scattering length is everywhere smaller than $r$, then the
above results should not be modified, i.e. the particle will bounce on
the scattering center with a trapping time of order $r/c$.

Consider now the intermediate case, for instance that where the
scattering center has a core with a magnetic field such that $l_{\rm
  sc}$ becomes smaller than the size of the core $r_{\rm c}$,
surrounded by an envelope with $B$ such that $l_{\rm sc}\gtrsim
r$. With probability $\sim (r_{\rm c}/r)^2$, the particle may cross
the envelope and bounce on the core; in this case the deflection angle
is of order unity and the total crossing time of order $r/c$. With
probability $\sim 1- (r_{\rm c}/r)^2$, the particle may also cross the
envelope without interacting with the core and suffer a deflection
smaller than unity as calculated above; the crossing time remains the
same. The typical deflection angle over many interactions of for many
particles is of course given by the average of these two
possibilities.

\bibliography{KL08}

\begin{thebibliography}{109}
\expandafter\ifx\csname natexlab\endcsname\relax\def\natexlab#1{#1}\fi
\expandafter\ifx\csname bibnamefont\endcsname\relax
  \def\bibnamefont#1{#1}\fi
\expandafter\ifx\csname bibfnamefont\endcsname\relax
  \def\bibfnamefont#1{#1}\fi
\expandafter\ifx\csname citenamefont\endcsname\relax
  \def\citenamefont#1{#1}\fi
\expandafter\ifx\csname url\endcsname\relax
  \def\url#1{\texttt{#1}}\fi
\expandafter\ifx\csname urlprefix\endcsname\relax\def\urlprefix{URL }\fi
\providecommand{\bibinfo}[2]{#2}
\providecommand{\eprint}[2][]{\url{#2}}

\bibitem[{\citenamefont{Abraham et~al.}(2007)}]{Auger1}
\bibinfo{author}{\bibfnamefont{J.}~\bibnamefont{Abraham}} \bibnamefont{et~al.}
  (\bibinfo{collaboration}{Pierre Auger}), \bibinfo{journal}{Science}
  \textbf{\bibinfo{volume}{318}}, \bibinfo{pages}{938} (\bibinfo{year}{2007}),
  \eprint{arXiv:0711.2256}.

\bibitem[{\citenamefont{Abraham et~al.}(2008)}]{Auger2}
\bibinfo{author}{\bibfnamefont{J.}~\bibnamefont{Abraham}} \bibnamefont{et~al.}
  (\bibinfo{collaboration}{Pierre Auger}), \bibinfo{journal}{Astropart. Phys.}
  \textbf{\bibinfo{volume}{29}}, \bibinfo{pages}{188} (\bibinfo{year}{2008}),
  \eprint{arXiv:0712.2843}.

\bibitem[{\citenamefont{{Stanev} et~al.}(1995)\citenamefont{{Stanev},
  {Biermann}, {Lloyd-Evans}, {Rachen}, and {Watson}}}]{Tea95}
\bibinfo{author}{\bibfnamefont{T.}~\bibnamefont{{Stanev}}},
  \bibinfo{author}{\bibfnamefont{P.~L.} \bibnamefont{{Biermann}}},
  \bibinfo{author}{\bibfnamefont{J.}~\bibnamefont{{Lloyd-Evans}}},
  \bibinfo{author}{\bibfnamefont{J.~P.} \bibnamefont{{Rachen}}},
  \bibnamefont{and} \bibinfo{author}{\bibfnamefont{A.~A.}
  \bibnamefont{{Watson}}}, \bibinfo{journal}{Physical Review Letters}
  \textbf{\bibinfo{volume}{75}}, \bibinfo{pages}{3056} (\bibinfo{year}{1995}),
  \eprint{arXiv:astro-ph/9505093}.

\bibitem[{\citenamefont{{Tinyakov} and {Tkachev}}(2001)}]{TT01}
\bibinfo{author}{\bibfnamefont{P.~G.} \bibnamefont{{Tinyakov}}}
  \bibnamefont{and} \bibinfo{author}{\bibfnamefont{I.~I.}
  \bibnamefont{{Tkachev}}}, \bibinfo{journal}{Soviet Journal of Experimental
  and Theoretical Physics Letters} \textbf{\bibinfo{volume}{74}},
  \bibinfo{pages}{445} (\bibinfo{year}{2001}), \eprint{arXiv:astro-ph/0102476}.

\bibitem[{\citenamefont{{Gorbunov} et~al.}(2004)\citenamefont{{Gorbunov},
  {Tinyakov}, {Tkachev}, and {Troitsky}}}]{GTTT04}
\bibinfo{author}{\bibfnamefont{D.~S.} \bibnamefont{{Gorbunov}}},
  \bibinfo{author}{\bibfnamefont{P.~G.} \bibnamefont{{Tinyakov}}},
  \bibinfo{author}{\bibfnamefont{I.~I.} \bibnamefont{{Tkachev}}},
  \bibnamefont{and} \bibinfo{author}{\bibfnamefont{S.~V.}
  \bibnamefont{{Troitsky}}}, \bibinfo{journal}{ArXiv Astrophysics e-prints}
  (\bibinfo{year}{2004}), \eprint{astro-ph/0406654}.

\bibitem[{\citenamefont{{Tinyakov} and {Tkachev}}(2004)}]{TT04}
\bibinfo{author}{\bibfnamefont{P.~G.} \bibnamefont{{Tinyakov}}}
  \bibnamefont{and} \bibinfo{author}{\bibfnamefont{I.~I.}
  \bibnamefont{{Tkachev}}}, \bibinfo{journal}{PRD}
  \textbf{\bibinfo{volume}{69}}, \bibinfo{pages}{128301}
  (\bibinfo{year}{2004}), \eprint{arXiv:astro-ph/0301336}.

\bibitem[{\citenamefont{{Evans} et~al.}(2004)\citenamefont{{Evans}, {Ferrer},
  and {Sarkar}}}]{EFS04}
\bibinfo{author}{\bibfnamefont{N.~W.} \bibnamefont{{Evans}}},
  \bibinfo{author}{\bibfnamefont{F.}~\bibnamefont{{Ferrer}}}, \bibnamefont{and}
  \bibinfo{author}{\bibfnamefont{S.}~\bibnamefont{{Sarkar}}},
  \bibinfo{journal}{PRD} \textbf{\bibinfo{volume}{69}}, \bibinfo{pages}{128302}
  (\bibinfo{year}{2004}), \eprint{arXiv:astro-ph/0403527}.

\bibitem[{\citenamefont{{Abbasi} et~al.}(2006)\citenamefont{{Abbasi},
  {Abu-Zayyad}, {Amann}, {Archbold}, {Belov}, {Belz}, {BenZvi}, {Bergman},
  {Blake}, {Boyer} et~al.}}]{HiRes06}
\bibinfo{author}{\bibfnamefont{R.~U.} \bibnamefont{{Abbasi}}},
  \bibinfo{author}{\bibfnamefont{T.}~\bibnamefont{{Abu-Zayyad}}},
  \bibinfo{author}{\bibfnamefont{J.~F.} \bibnamefont{{Amann}}},
  \bibinfo{author}{\bibfnamefont{G.}~\bibnamefont{{Archbold}}},
  \bibinfo{author}{\bibfnamefont{K.}~\bibnamefont{{Belov}}},
  \bibinfo{author}{\bibfnamefont{J.~W.} \bibnamefont{{Belz}}},
  \bibinfo{author}{\bibfnamefont{S.}~\bibnamefont{{BenZvi}}},
  \bibinfo{author}{\bibfnamefont{D.~R.} \bibnamefont{{Bergman}}},
  \bibinfo{author}{\bibfnamefont{S.~A.} \bibnamefont{{Blake}}},
  \bibinfo{author}{\bibfnamefont{J.~H.} \bibnamefont{{Boyer}}},
  \bibnamefont{et~al.}, \bibinfo{journal}{Astrophys.~J.~}
  \textbf{\bibinfo{volume}{636}}, \bibinfo{pages}{680} (\bibinfo{year}{2006}).

\bibitem[{\citenamefont{{Uchihori} et~al.}(2000)\citenamefont{{Uchihori},
  {Nagano}, {Takeda}, {Teshima}, {Lloyd-Evans}, and {Watson}}}]{Uea00}
\bibinfo{author}{\bibfnamefont{Y.}~\bibnamefont{{Uchihori}}},
  \bibinfo{author}{\bibfnamefont{M.}~\bibnamefont{{Nagano}}},
  \bibinfo{author}{\bibfnamefont{M.}~\bibnamefont{{Takeda}}},
  \bibinfo{author}{\bibfnamefont{M.}~\bibnamefont{{Teshima}}},
  \bibinfo{author}{\bibfnamefont{J.}~\bibnamefont{{Lloyd-Evans}}},
  \bibnamefont{and} \bibinfo{author}{\bibfnamefont{A.~A.}
  \bibnamefont{{Watson}}}, \bibinfo{journal}{Astroparticle Physics}
  \textbf{\bibinfo{volume}{13}}, \bibinfo{pages}{151} (\bibinfo{year}{2000}),
  \eprint{arXiv:astro-ph/9908193}.

\bibitem[{\citenamefont{{Farrar} et~al.}(2006)\citenamefont{{Farrar},
  {Berlind}, and {Hogg}}}]{Fea05}
\bibinfo{author}{\bibfnamefont{G.~R.} \bibnamefont{{Farrar}}},
  \bibinfo{author}{\bibfnamefont{A.~A.} \bibnamefont{{Berlind}}},
  \bibnamefont{and} \bibinfo{author}{\bibfnamefont{D.~W.}
  \bibnamefont{{Hogg}}}, \bibinfo{journal}{Astrophys. J. Lett.}
  \textbf{\bibinfo{volume}{642}}, \bibinfo{pages}{L89} (\bibinfo{year}{2006}),
  \eprint{arXiv:astro-ph/0507657}.

\bibitem[{\citenamefont{{Abbasi} et~al.}(2005)\citenamefont{{Abbasi},
  {Abu-Zayyad}, {Amann}, {Archbold}, {Atkins}, {Bellido}, {Belov}, {Belz},
  {Ben-Zvi}, {Bergman} et~al.}}]{HiRes05}
\bibinfo{author}{\bibfnamefont{R.~U.} \bibnamefont{{Abbasi}}},
  \bibinfo{author}{\bibfnamefont{T.}~\bibnamefont{{Abu-Zayyad}}},
  \bibinfo{author}{\bibfnamefont{J.~F.} \bibnamefont{{Amann}}},
  \bibinfo{author}{\bibfnamefont{G.}~\bibnamefont{{Archbold}}},
  \bibinfo{author}{\bibfnamefont{R.}~\bibnamefont{{Atkins}}},
  \bibinfo{author}{\bibfnamefont{J.~A.} \bibnamefont{{Bellido}}},
  \bibinfo{author}{\bibfnamefont{K.}~\bibnamefont{{Belov}}},
  \bibinfo{author}{\bibfnamefont{J.~W.} \bibnamefont{{Belz}}},
  \bibinfo{author}{\bibfnamefont{S.~Y.} \bibnamefont{{Ben-Zvi}}},
  \bibinfo{author}{\bibfnamefont{D.~R.} \bibnamefont{{Bergman}}},
  \bibnamefont{et~al.}, \bibinfo{journal}{Astrophys.~J.~}
  \textbf{\bibinfo{volume}{623}}, \bibinfo{pages}{164} (\bibinfo{year}{2005}).

\bibitem[{\citenamefont{Kronberg}(1994)}]{K94}
\bibinfo{author}{\bibfnamefont{P.~P.} \bibnamefont{Kronberg}},
  \bibinfo{journal}{Rep. Prog. Phys.} \textbf{\bibinfo{volume}{57}},
  \bibinfo{pages}{325} (\bibinfo{year}{1994}).

\bibitem[{\citenamefont{{Vallee}}(1997)}]{V97}
\bibinfo{author}{\bibfnamefont{J.~P.} \bibnamefont{{Vallee}}},
  \bibinfo{journal}{Fundamentals of Cosmic Physics}
  \textbf{\bibinfo{volume}{19}}, \bibinfo{pages}{1} (\bibinfo{year}{1997}).

\bibitem[{\citenamefont{Waxman and Miralda-Escud{\' e}}(1996)}]{WM96}
\bibinfo{author}{\bibfnamefont{E.}~\bibnamefont{Waxman}} \bibnamefont{and}
  \bibinfo{author}{\bibfnamefont{J.}~\bibnamefont{Miralda-Escud{\' e}}},
  \bibinfo{journal}{ApJ} \textbf{\bibinfo{volume}{472}}, \bibinfo{pages}{89}
  (\bibinfo{year}{1996}).

\bibitem[{\citenamefont{{Achterberg} et~al.}(1999)\citenamefont{{Achterberg},
  {Gallant}, {Norman}, and {Melrose}}}]{1999astro.ph..7060A}
\bibinfo{author}{\bibfnamefont{A.}~\bibnamefont{{Achterberg}}},
  \bibinfo{author}{\bibfnamefont{Y.~A.} \bibnamefont{{Gallant}}},
  \bibinfo{author}{\bibfnamefont{C.~A.} \bibnamefont{{Norman}}},
  \bibnamefont{and} \bibinfo{author}{\bibfnamefont{D.~B.}
  \bibnamefont{{Melrose}}}, \bibinfo{journal}{ArXiv Astrophysics e-prints}
  (\bibinfo{year}{1999}), \eprint{astro-ph/9907060}.

\bibitem[{\citenamefont{{Harari}
  et~al.}(2002{\natexlab{a}})\citenamefont{{Harari}, {Mollerach}, {Roulet}, and
  {S{\'a}nchez}}}]{2002JHEP...03..045H}
\bibinfo{author}{\bibfnamefont{D.}~\bibnamefont{{Harari}}},
  \bibinfo{author}{\bibfnamefont{S.}~\bibnamefont{{Mollerach}}},
  \bibinfo{author}{\bibfnamefont{E.}~\bibnamefont{{Roulet}}}, \bibnamefont{and}
  \bibinfo{author}{\bibfnamefont{F.}~\bibnamefont{{S{\'a}nchez}}},
  \bibinfo{journal}{Journal of High Energy Physics}
  \textbf{\bibinfo{volume}{3}}, \bibinfo{pages}{45}
  (\bibinfo{year}{2002}{\natexlab{a}}), \eprint{arXiv:astro-ph/0202362}.

\bibitem[{\citenamefont{{Harari}
  et~al.}(2002{\natexlab{b}})\citenamefont{{Harari}, {Mollerach}, and
  {Roulet}}}]{2002JHEP...07..006H}
\bibinfo{author}{\bibfnamefont{D.}~\bibnamefont{{Harari}}},
  \bibinfo{author}{\bibfnamefont{S.}~\bibnamefont{{Mollerach}}},
  \bibnamefont{and} \bibinfo{author}{\bibfnamefont{E.}~\bibnamefont{{Roulet}}},
  \bibinfo{journal}{Journal of High Energy Physics}
  \textbf{\bibinfo{volume}{7}}, \bibinfo{pages}{6}
  (\bibinfo{year}{2002}{\natexlab{b}}), \eprint{arXiv:astro-ph/0205484}.

\bibitem[{\citenamefont{{Wdowczyk} and
  {Wolfendale}}(1979)}]{1979Natur.281..356W}
\bibinfo{author}{\bibfnamefont{J.}~\bibnamefont{{Wdowczyk}}} \bibnamefont{and}
  \bibinfo{author}{\bibfnamefont{A.~W.} \bibnamefont{{Wolfendale}}},
  \bibinfo{journal}{Nature} \textbf{\bibinfo{volume}{281}},
  \bibinfo{pages}{356} (\bibinfo{year}{1979}).

\bibitem[{\citenamefont{{Berezinskii} et~al.}(1990)\citenamefont{{Berezinskii},
  {Grigor'eva}, and {Dogiel}}}]{1990A&A...232..582B}
\bibinfo{author}{\bibfnamefont{V.~S.} \bibnamefont{{Berezinskii}}},
  \bibinfo{author}{\bibfnamefont{S.~I.} \bibnamefont{{Grigor'eva}}},
  \bibnamefont{and} \bibinfo{author}{\bibfnamefont{V.~A.}
  \bibnamefont{{Dogiel}}}, \bibinfo{journal}{Astron.~Astrophys.~}
  \textbf{\bibinfo{volume}{232}}, \bibinfo{pages}{582} (\bibinfo{year}{1990}).

\bibitem[{\citenamefont{{Blasi} and {Olinto}}(1999)}]{1999PhRvD..59b3001B}
\bibinfo{author}{\bibfnamefont{P.}~\bibnamefont{{Blasi}}} \bibnamefont{and}
  \bibinfo{author}{\bibfnamefont{A.~V.} \bibnamefont{{Olinto}}},
  \bibinfo{journal}{PRD} \textbf{\bibinfo{volume}{59}}, \bibinfo{pages}{023001}
  (\bibinfo{year}{1999}), \eprint{arXiv:astro-ph/9806264}.

\bibitem[{\citenamefont{{Lampard} et~al.}(1997)\citenamefont{{Lampard}, {Clay},
  and {Dawson}}}]{1997APh.....7..213L}
\bibinfo{author}{\bibfnamefont{R.}~\bibnamefont{{Lampard}}},
  \bibinfo{author}{\bibfnamefont{R.~W.} \bibnamefont{{Clay}}},
  \bibnamefont{and} \bibinfo{author}{\bibfnamefont{B.~R.}
  \bibnamefont{{Dawson}}}, \bibinfo{journal}{Astroparticle Physics}
  \textbf{\bibinfo{volume}{7}}, \bibinfo{pages}{213} (\bibinfo{year}{1997}).

\bibitem[{\citenamefont{{Medina Tanco} et~al.}(1997)\citenamefont{{Medina
  Tanco}, {de Gouveia dal Pino}, and {Horvath}}}]{1997APh.....6..337M}
\bibinfo{author}{\bibfnamefont{G.~A.} \bibnamefont{{Medina Tanco}}},
  \bibinfo{author}{\bibfnamefont{E.~M.} \bibnamefont{{de Gouveia dal Pino}}},
  \bibnamefont{and} \bibinfo{author}{\bibfnamefont{J.~E.}
  \bibnamefont{{Horvath}}}, \bibinfo{journal}{Astroparticle Physics}
  \textbf{\bibinfo{volume}{6}}, \bibinfo{pages}{337} (\bibinfo{year}{1997}),
  \eprint{arXiv:astro-ph/9610172}.

\bibitem[{\citenamefont{Lemoine et~al.}(1997)\citenamefont{Lemoine, Sigl,
  Olinto, and Schramm}}]{LSOS97}
\bibinfo{author}{\bibfnamefont{M.}~\bibnamefont{Lemoine}},
  \bibinfo{author}{\bibfnamefont{G.}~\bibnamefont{Sigl}},
  \bibinfo{author}{\bibfnamefont{A.}~\bibnamefont{Olinto}}, \bibnamefont{and}
  \bibinfo{author}{\bibfnamefont{D.~N.} \bibnamefont{Schramm}},
  \bibinfo{journal}{ApJ} \textbf{\bibinfo{volume}{486}}, \bibinfo{pages}{L115}
  (\bibinfo{year}{1997}).

\bibitem[{\citenamefont{{Sigl} et~al.}(1997)\citenamefont{{Sigl}, {Lemoine},
  and {Olinto}}}]{1997PhRvD..56.4470S}
\bibinfo{author}{\bibfnamefont{G.}~\bibnamefont{{Sigl}}},
  \bibinfo{author}{\bibfnamefont{M.}~\bibnamefont{{Lemoine}}},
  \bibnamefont{and} \bibinfo{author}{\bibfnamefont{A.~V.}
  \bibnamefont{{Olinto}}}, \bibinfo{journal}{PRD}
  \textbf{\bibinfo{volume}{56}}, \bibinfo{pages}{4470} (\bibinfo{year}{1997}),
  \eprint{arXiv:astro-ph/9704204}.

\bibitem[{\citenamefont{{Clay} et~al.}(1998)\citenamefont{{Clay}, {Cook},
  {Dawson}, {Smith}, and {Lampard}}}]{1998APh.....9..221C}
\bibinfo{author}{\bibfnamefont{R.~W.} \bibnamefont{{Clay}}},
  \bibinfo{author}{\bibfnamefont{S.}~\bibnamefont{{Cook}}},
  \bibinfo{author}{\bibfnamefont{B.~R.} \bibnamefont{{Dawson}}},
  \bibinfo{author}{\bibfnamefont{A.~G.~K.} \bibnamefont{{Smith}}},
  \bibnamefont{and}
  \bibinfo{author}{\bibfnamefont{R.}~\bibnamefont{{Lampard}}},
  \bibinfo{journal}{Astroparticle Physics} \textbf{\bibinfo{volume}{9}},
  \bibinfo{pages}{221} (\bibinfo{year}{1998}).

\bibitem[{\citenamefont{{Stanev} et~al.}(2000)\citenamefont{{Stanev}, {Engel},
  {M{\"u}cke}, {Protheroe}, and {Rachen}}}]{2000PhRvD..62i3005S}
\bibinfo{author}{\bibfnamefont{T.}~\bibnamefont{{Stanev}}},
  \bibinfo{author}{\bibfnamefont{R.}~\bibnamefont{{Engel}}},
  \bibinfo{author}{\bibfnamefont{A.}~\bibnamefont{{M{\"u}cke}}},
  \bibinfo{author}{\bibfnamefont{R.~J.} \bibnamefont{{Protheroe}}},
  \bibnamefont{and} \bibinfo{author}{\bibfnamefont{J.~P.}
  \bibnamefont{{Rachen}}}, \bibinfo{journal}{PRD}
  \textbf{\bibinfo{volume}{62}}, \bibinfo{pages}{093005}
  (\bibinfo{year}{2000}), \eprint{arXiv:astro-ph/0003484}.

\bibitem[{\citenamefont{{Yoshiguchi} et~al.}(2003)\citenamefont{{Yoshiguchi},
  {Nagataki}, {Tsubaki}, and {Sato}}}]{2003ApJ...586.1211Y}
\bibinfo{author}{\bibfnamefont{H.}~\bibnamefont{{Yoshiguchi}}},
  \bibinfo{author}{\bibfnamefont{S.}~\bibnamefont{{Nagataki}}},
  \bibinfo{author}{\bibfnamefont{S.}~\bibnamefont{{Tsubaki}}},
  \bibnamefont{and} \bibinfo{author}{\bibfnamefont{K.}~\bibnamefont{{Sato}}},
  \bibinfo{journal}{Astrophys.~J.~} \textbf{\bibinfo{volume}{586}},
  \bibinfo{pages}{1211} (\bibinfo{year}{2003}),
  \eprint{arXiv:astro-ph/0210132}.

\bibitem[{\citenamefont{{Deligny} et~al.}(2004)\citenamefont{{Deligny},
  {Letessier-Selvon}, and {Parizot}}}]{2004APh....21..609D}
\bibinfo{author}{\bibfnamefont{O.}~\bibnamefont{{Deligny}}},
  \bibinfo{author}{\bibfnamefont{A.}~\bibnamefont{{Letessier-Selvon}}},
  \bibnamefont{and}
  \bibinfo{author}{\bibfnamefont{E.}~\bibnamefont{{Parizot}}},
  \bibinfo{journal}{Astroparticle Physics} \textbf{\bibinfo{volume}{21}},
  \bibinfo{pages}{609} (\bibinfo{year}{2004}), \eprint{arXiv:astro-ph/0303624}.

\bibitem[{\citenamefont{{Medina Tanco}}(1998)}]{1998ApJ...505L..79M}
\bibinfo{author}{\bibfnamefont{G.~A.} \bibnamefont{{Medina Tanco}}},
  \bibinfo{journal}{Astrophys.~J.~Lett.~} \textbf{\bibinfo{volume}{505}},
  \bibinfo{pages}{L79} (\bibinfo{year}{1998}), \eprint{arXiv:astro-ph/9808073}.

\bibitem[{\citenamefont{{Sigl} et~al.}(1999)\citenamefont{{Sigl}, {Lemoine},
  and {Biermann}}}]{1999APh....10..141S}
\bibinfo{author}{\bibfnamefont{G.}~\bibnamefont{{Sigl}}},
  \bibinfo{author}{\bibfnamefont{M.}~\bibnamefont{{Lemoine}}},
  \bibnamefont{and}
  \bibinfo{author}{\bibfnamefont{P.}~\bibnamefont{{Biermann}}},
  \bibinfo{journal}{Astroparticle Physics} \textbf{\bibinfo{volume}{10}},
  \bibinfo{pages}{141} (\bibinfo{year}{1999}), \eprint{arXiv:astro-ph/9806283}.

\bibitem[{\citenamefont{{Lemoine} et~al.}(1999)\citenamefont{{Lemoine}, {Sigl},
  and {Biermann}}}]{1999astro.ph..3124L}
\bibinfo{author}{\bibfnamefont{M.}~\bibnamefont{{Lemoine}}},
  \bibinfo{author}{\bibfnamefont{G.}~\bibnamefont{{Sigl}}}, \bibnamefont{and}
  \bibinfo{author}{\bibfnamefont{P.}~\bibnamefont{{Biermann}}},
  \bibinfo{journal}{ArXiv Astrophysics e-prints}  (\bibinfo{year}{1999}),
  \eprint{astro-ph/9903124}.

\bibitem[{\citenamefont{{Ide} et~al.}(2001)\citenamefont{{Ide}, {Nagataki},
  {Tsubaki}, {Yoshiguchi}, and {Sato}}}]{2001PASJ...53.1153I}
\bibinfo{author}{\bibfnamefont{Y.}~\bibnamefont{{Ide}}},
  \bibinfo{author}{\bibfnamefont{S.}~\bibnamefont{{Nagataki}}},
  \bibinfo{author}{\bibfnamefont{S.}~\bibnamefont{{Tsubaki}}},
  \bibinfo{author}{\bibfnamefont{H.}~\bibnamefont{{Yoshiguchi}}},
  \bibnamefont{and} \bibinfo{author}{\bibfnamefont{K.}~\bibnamefont{{Sato}}},
  \bibinfo{journal}{Publ.~Astron.~Soc.~Japan~} \textbf{\bibinfo{volume}{53}},
  \bibinfo{pages}{1153} (\bibinfo{year}{2001}),
  \eprint{arXiv:astro-ph/0106182}.

\bibitem[{\citenamefont{{Isola} et~al.}(2002)\citenamefont{{Isola}, {Lemoine},
  and {Sigl}}}]{2002PhRvD..65b3004I}
\bibinfo{author}{\bibfnamefont{C.}~\bibnamefont{{Isola}}},
  \bibinfo{author}{\bibfnamefont{M.}~\bibnamefont{{Lemoine}}},
  \bibnamefont{and} \bibinfo{author}{\bibfnamefont{G.}~\bibnamefont{{Sigl}}},
  \bibinfo{journal}{PRD} \textbf{\bibinfo{volume}{65}}, \bibinfo{pages}{023004}
  (\bibinfo{year}{2002}), \eprint{arXiv:astro-ph/0104289}.

\bibitem[{\citenamefont{{Isola} and {Sigl}}(2002)}]{2002PhRvD..66h3002I}
\bibinfo{author}{\bibfnamefont{C.}~\bibnamefont{{Isola}}} \bibnamefont{and}
  \bibinfo{author}{\bibfnamefont{G.}~\bibnamefont{{Sigl}}},
  \bibinfo{journal}{PRD} \textbf{\bibinfo{volume}{66}}, \bibinfo{pages}{083002}
  (\bibinfo{year}{2002}), \eprint{arXiv:astro-ph/0203273}.

\bibitem[{\citenamefont{{Stanev} et~al.}(2003)\citenamefont{{Stanev}, {Seckel},
  and {Engel}}}]{2003PhRvD..68j3004S}
\bibinfo{author}{\bibfnamefont{T.}~\bibnamefont{{Stanev}}},
  \bibinfo{author}{\bibfnamefont{D.}~\bibnamefont{{Seckel}}}, \bibnamefont{and}
  \bibinfo{author}{\bibfnamefont{R.}~\bibnamefont{{Engel}}},
  \bibinfo{journal}{PRD} \textbf{\bibinfo{volume}{68}}, \bibinfo{pages}{103004}
  (\bibinfo{year}{2003}), \eprint{arXiv:astro-ph/0108338}.

\bibitem[{\citenamefont{{Sigl} et~al.}(2003)\citenamefont{{Sigl}, {Miniati},
  and {En\ss{}lin}}}]{2003PhRvD..68d3002S}
\bibinfo{author}{\bibfnamefont{G.}~\bibnamefont{{Sigl}}},
  \bibinfo{author}{\bibfnamefont{F.}~\bibnamefont{{Miniati}}},
  \bibnamefont{and} \bibinfo{author}{\bibfnamefont{T.~A.}
  \bibnamefont{{En\ss{}lin}}}, \bibinfo{journal}{PRD}
  \textbf{\bibinfo{volume}{68}}, \bibinfo{pages}{043002}
  (\bibinfo{year}{2003}), \eprint{arXiv:astro-ph/0302388}.

\bibitem[{\citenamefont{{Sigl} et~al.}(2004)\citenamefont{{Sigl}, {Miniati},
  and {En{\ss}lin}}}]{2004PhRvD..70d3007S}
\bibinfo{author}{\bibfnamefont{G.}~\bibnamefont{{Sigl}}},
  \bibinfo{author}{\bibfnamefont{F.}~\bibnamefont{{Miniati}}},
  \bibnamefont{and} \bibinfo{author}{\bibfnamefont{T.~A.}
  \bibnamefont{{En{\ss}lin}}}, \bibinfo{journal}{PRD}
  \textbf{\bibinfo{volume}{70}}, \bibinfo{pages}{043007}
  (\bibinfo{year}{2004}), \eprint{arXiv:astro-ph/0401084}.

\bibitem[{\citenamefont{{Armengaud} et~al.}(2005)\citenamefont{{Armengaud},
  {Sigl}, and {Miniati}}}]{2005PhRvD..72d3009A}
\bibinfo{author}{\bibfnamefont{E.}~\bibnamefont{{Armengaud}}},
  \bibinfo{author}{\bibfnamefont{G.}~\bibnamefont{{Sigl}}}, \bibnamefont{and}
  \bibinfo{author}{\bibfnamefont{F.}~\bibnamefont{{Miniati}}},
  \bibinfo{journal}{PRD} \textbf{\bibinfo{volume}{72}}, \bibinfo{pages}{043009}
  (\bibinfo{year}{2005}), \eprint{arXiv:astro-ph/0412525}.

\bibitem[{\citenamefont{Dolag et~al.}(2004)\citenamefont{Dolag, Grasso,
  Springel, and Tkachev}}]{DGST04}
\bibinfo{author}{\bibfnamefont{K.}~\bibnamefont{Dolag}},
  \bibinfo{author}{\bibfnamefont{D.}~\bibnamefont{Grasso}},
  \bibinfo{author}{\bibfnamefont{V.}~\bibnamefont{Springel}}, \bibnamefont{and}
  \bibinfo{author}{\bibfnamefont{I.}~\bibnamefont{Tkachev}},
  \bibinfo{journal}{JKAS} \textbf{\bibinfo{volume}{37}}, \bibinfo{pages}{427}
  (\bibinfo{year}{2004}).

\bibitem[{\citenamefont{{Dolag} et~al.}(2005)\citenamefont{{Dolag}, {Grasso},
  {Springel}, and {Tkachev}}}]{DGST05}
\bibinfo{author}{\bibfnamefont{K.}~\bibnamefont{{Dolag}}},
  \bibinfo{author}{\bibfnamefont{D.}~\bibnamefont{{Grasso}}},
  \bibinfo{author}{\bibfnamefont{V.}~\bibnamefont{{Springel}}},
  \bibnamefont{and}
  \bibinfo{author}{\bibfnamefont{I.}~\bibnamefont{{Tkachev}}},
  \bibinfo{journal}{Journal of Cosmology and Astro-Particle Physics}
  \textbf{\bibinfo{volume}{1}}, \bibinfo{pages}{9} (\bibinfo{year}{2005}),
  \eprint{arXiv:astro-ph/0410419}.

\bibitem[{\citenamefont{{Kang} et~al.}(2007)\citenamefont{{Kang}, {Das}, {Ryu},
  and {Cho}}}]{KRC07}
\bibinfo{author}{\bibfnamefont{H.}~\bibnamefont{{Kang}}},
  \bibinfo{author}{\bibfnamefont{S.}~\bibnamefont{{Das}}},
  \bibinfo{author}{\bibfnamefont{D.}~\bibnamefont{{Ryu}}}, \bibnamefont{and}
  \bibinfo{author}{\bibfnamefont{J.}~\bibnamefont{{Cho}}},
  \bibinfo{journal}{ArXiv e-prints} \textbf{\bibinfo{volume}{706}}
  (\bibinfo{year}{2007}), \eprint{0706.2597}.

\bibitem[{\citenamefont{{Sigl}}(2007)}]{2007PhRvD..75j3001S}
\bibinfo{author}{\bibfnamefont{G.}~\bibnamefont{{Sigl}}},
  \bibinfo{journal}{PRD} \textbf{\bibinfo{volume}{75}}, \bibinfo{pages}{103001}
  (\bibinfo{year}{2007}), \eprint{arXiv:astro-ph/0703403}.

\bibitem[{\citenamefont{{Medina Tanco}}(1997)}]{1997astro.ph..7054M}
\bibinfo{author}{\bibfnamefont{G.~A.} \bibnamefont{{Medina Tanco}}},
  \bibinfo{journal}{ArXiv Astrophysics e-prints}  (\bibinfo{year}{1997}),
  \eprint{astro-ph/9707054}.

\bibitem[{\citenamefont{{Takami} et~al.}(2006)\citenamefont{{Takami},
  {Yoshiguchi}, and {Sato}}}]{2006ApJ...639..803T}
\bibinfo{author}{\bibfnamefont{H.}~\bibnamefont{{Takami}}},
  \bibinfo{author}{\bibfnamefont{H.}~\bibnamefont{{Yoshiguchi}}},
  \bibnamefont{and} \bibinfo{author}{\bibfnamefont{K.}~\bibnamefont{{Sato}}},
  \bibinfo{journal}{Astrophys.~J.~} \textbf{\bibinfo{volume}{639}},
  \bibinfo{pages}{803} (\bibinfo{year}{2006}), \eprint{arXiv:astro-ph/0506203}.

\bibitem[{\citenamefont{{Takami} and {Sato}}(2007{\natexlab{a}})}]{TS07}
\bibinfo{author}{\bibfnamefont{H.}~\bibnamefont{{Takami}}} \bibnamefont{and}
  \bibinfo{author}{\bibfnamefont{K.}~\bibnamefont{{Sato}}},
  \bibinfo{journal}{ArXiv e-prints} \textbf{\bibinfo{volume}{706}}
  (\bibinfo{year}{2007}{\natexlab{a}}), \eprint{0706.3666}.

\bibitem[{\citenamefont{Kotera and Lemoine}(2007)}]{KL07}
\bibinfo{author}{\bibfnamefont{K.}~\bibnamefont{Kotera}} \bibnamefont{and}
  \bibinfo{author}{\bibfnamefont{M.}~\bibnamefont{Lemoine}}
  (\bibinfo{year}{2007}), \eprint{arXiv:0706.1891 [astro-ph]}.

\bibitem[{\citenamefont{{Medina-Tanco} and {En{\ss}lin}}(2001)}]{MTE01}
\bibinfo{author}{\bibfnamefont{G.}~\bibnamefont{{Medina-Tanco}}}
  \bibnamefont{and} \bibinfo{author}{\bibfnamefont{T.~A.}
  \bibnamefont{{En{\ss}lin}}}, \bibinfo{journal}{Astroparticle Physics}
  \textbf{\bibinfo{volume}{16}}, \bibinfo{pages}{47} (\bibinfo{year}{2001}),
  \eprint{arXiv:astro-ph/0011454}.

\bibitem[{\citenamefont{{Berezinsky} et~al.}(2002)\citenamefont{{Berezinsky},
  {Gazizov}, and {Grigorieva}}}]{2002astro.ph.10095B}
\bibinfo{author}{\bibfnamefont{V.}~\bibnamefont{{Berezinsky}}},
  \bibinfo{author}{\bibfnamefont{A.~Z.} \bibnamefont{{Gazizov}}},
  \bibnamefont{and} \bibinfo{author}{\bibfnamefont{S.~I.}
  \bibnamefont{{Grigorieva}}}, \bibinfo{journal}{ArXiv Astrophysics e-prints}
  (\bibinfo{year}{2002}), \eprint{astro-ph/0210095}.

\bibitem[{\citenamefont{Widrow}(2002)}]{W02}
\bibinfo{author}{\bibfnamefont{L.~M.} \bibnamefont{Widrow}},
  \bibinfo{journal}{Rev. Mod. Phys.} \textbf{\bibinfo{volume}{74}},
  \bibinfo{pages}{775} (\bibinfo{year}{2002}).

\bibitem[{\citenamefont{{Ryu} et~al.}(1998)\citenamefont{{Ryu}, {Kang}, and
  {Biermann}}}]{1998AA...335...19R}
\bibinfo{author}{\bibfnamefont{D.}~\bibnamefont{{Ryu}}},
  \bibinfo{author}{\bibfnamefont{H.}~\bibnamefont{{Kang}}}, \bibnamefont{and}
  \bibinfo{author}{\bibfnamefont{P.~L.} \bibnamefont{{Biermann}}},
  \bibinfo{journal}{A\&A} \textbf{\bibinfo{volume}{335}}, \bibinfo{pages}{19}
  (\bibinfo{year}{1998}), \eprint{arXiv:astro-ph/9803275}.

\bibitem[{\citenamefont{Sigl et~al.}(2004)\citenamefont{Sigl, Miniati, and
  En\ss{}lin}}]{SME04}
\bibinfo{author}{\bibfnamefont{G.}~\bibnamefont{Sigl}},
  \bibinfo{author}{\bibfnamefont{F.}~\bibnamefont{Miniati}}, \bibnamefont{and}
  \bibinfo{author}{\bibfnamefont{T.~A.} \bibnamefont{En\ss{}lin}},
  \bibinfo{journal}{Phys. Rev. D} \textbf{\bibinfo{volume}{70}},
  \bibinfo{pages}{043007} (\bibinfo{year}{2004}).

\bibitem[{\citenamefont{{Dolag}}(2006)}]{D06}
\bibinfo{author}{\bibfnamefont{K.}~\bibnamefont{{Dolag}}},
  \bibinfo{journal}{Astronomische Nachrichten} \textbf{\bibinfo{volume}{327}},
  \bibinfo{pages}{575} (\bibinfo{year}{2006}), \eprint{arXiv:astro-ph/0601484}.

\bibitem[{\citenamefont{{Kronberg} et~al.}(1999)\citenamefont{{Kronberg},
  {Lesch}, and {Hopp}}}]{KLH99}
\bibinfo{author}{\bibfnamefont{P.~P.} \bibnamefont{{Kronberg}}},
  \bibinfo{author}{\bibfnamefont{H.}~\bibnamefont{{Lesch}}}, \bibnamefont{and}
  \bibinfo{author}{\bibfnamefont{U.}~\bibnamefont{{Hopp}}},
  \bibinfo{journal}{Astrophys. J.} \textbf{\bibinfo{volume}{511}},
  \bibinfo{pages}{56} (\bibinfo{year}{1999}).

\bibitem[{\citenamefont{{Birk} et~al.}(2000)\citenamefont{{Birk}, {Wiechen},
  {Lesch}, and {Kronberg}}}]{Bea00}
\bibinfo{author}{\bibfnamefont{G.~T.} \bibnamefont{{Birk}}},
  \bibinfo{author}{\bibfnamefont{H.}~\bibnamefont{{Wiechen}}},
  \bibinfo{author}{\bibfnamefont{H.}~\bibnamefont{{Lesch}}}, \bibnamefont{and}
  \bibinfo{author}{\bibfnamefont{P.~P.} \bibnamefont{{Kronberg}}},
  \bibinfo{journal}{Astron. Astrophys.} \textbf{\bibinfo{volume}{353}},
  \bibinfo{pages}{108} (\bibinfo{year}{2000}).

\bibitem[{\citenamefont{{Bertone} et~al.}(2006)\citenamefont{{Bertone}, {Vogt},
  and {En{\ss}lin}}}]{BVE06}
\bibinfo{author}{\bibfnamefont{S.}~\bibnamefont{{Bertone}}},
  \bibinfo{author}{\bibfnamefont{C.}~\bibnamefont{{Vogt}}}, \bibnamefont{and}
  \bibinfo{author}{\bibfnamefont{T.}~\bibnamefont{{En{\ss}lin}}},
  \bibinfo{journal}{Month. Not. Roy. Astron. Soc.}
  \textbf{\bibinfo{volume}{370}}, \bibinfo{pages}{319} (\bibinfo{year}{2006}),
  \eprint{arXiv:astro-ph/0604462}.

\bibitem[{\citenamefont{{Rees} and {Setti}}(1968)}]{RS68}
\bibinfo{author}{\bibfnamefont{M.~J.} \bibnamefont{{Rees}}} \bibnamefont{and}
  \bibinfo{author}{\bibfnamefont{G.}~\bibnamefont{{Setti}}},
  \bibinfo{journal}{Nature} \textbf{\bibinfo{volume}{219}},
  \bibinfo{pages}{127} (\bibinfo{year}{1968}).

\bibitem[{\citenamefont{{Furlanetto} and {Loeb}}(2001)}]{FL01}
\bibinfo{author}{\bibfnamefont{S.~R.} \bibnamefont{{Furlanetto}}}
  \bibnamefont{and} \bibinfo{author}{\bibfnamefont{A.}~\bibnamefont{{Loeb}}},
  \bibinfo{journal}{Astrophys. J.} \textbf{\bibinfo{volume}{556}},
  \bibinfo{pages}{619} (\bibinfo{year}{2001}), \eprint{arXiv:astro-ph/0102076}.

\bibitem[{\citenamefont{{Gopal-Krishna} and {Wiita}}(2001)}]{GKW01}
\bibinfo{author}{\bibnamefont{{Gopal-Krishna}}} \bibnamefont{and}
  \bibinfo{author}{\bibfnamefont{P.~J.} \bibnamefont{{Wiita}}},
  \bibinfo{journal}{Astrophys. J. Lett.} \textbf{\bibinfo{volume}{560}},
  \bibinfo{pages}{L115} (\bibinfo{year}{2001}),
  \eprint{arXiv:astro-ph/0108117}.

\bibitem[{\citenamefont{{Bertone} et~al.}(2005)\citenamefont{{Bertone},
  {Stoehr}, and {White}}}]{BSW05}
\bibinfo{author}{\bibfnamefont{S.}~\bibnamefont{{Bertone}}},
  \bibinfo{author}{\bibfnamefont{F.}~\bibnamefont{{Stoehr}}}, \bibnamefont{and}
  \bibinfo{author}{\bibfnamefont{S.~D.~M.} \bibnamefont{{White}}},
  \bibinfo{journal}{Month. Not. Roy. Astron. Soc.}
  \textbf{\bibinfo{volume}{359}}, \bibinfo{pages}{1201} (\bibinfo{year}{2005}),
  \eprint{arXiv:astro-ph/0402044}.

\bibitem[{\citenamefont{{Kronberg} et~al.}(2001)\citenamefont{{Kronberg},
  {Dufton}, {Li}, and {Colgate}}}]{KDLC01}
\bibinfo{author}{\bibfnamefont{P.~P.} \bibnamefont{{Kronberg}}},
  \bibinfo{author}{\bibfnamefont{Q.~W.} \bibnamefont{{Dufton}}},
  \bibinfo{author}{\bibfnamefont{H.}~\bibnamefont{{Li}}}, \bibnamefont{and}
  \bibinfo{author}{\bibfnamefont{S.~A.} \bibnamefont{{Colgate}}},
  \bibinfo{journal}{Astrophys.~J.~} \textbf{\bibinfo{volume}{560}},
  \bibinfo{pages}{178} (\bibinfo{year}{2001}), \eprint{arXiv:astro-ph/0106281}.

\bibitem[{\citenamefont{{Ferrarese} and {Ford}}(2005)}]{FF05}
\bibinfo{author}{\bibfnamefont{L.}~\bibnamefont{{Ferrarese}}} \bibnamefont{and}
  \bibinfo{author}{\bibfnamefont{H.}~\bibnamefont{{Ford}}},
  \bibinfo{journal}{Space Science Reviews} \textbf{\bibinfo{volume}{116}},
  \bibinfo{pages}{523} (\bibinfo{year}{2005}), \eprint{arXiv:astro-ph/0411247}.

\bibitem[{\citenamefont{{Lauer} et~al.}(2007)\citenamefont{{Lauer}, {Faber},
  {Richstone}, {Gebhardt}, {Tremaine}, {Postman}, {Dressler}, {Aller},
  {Filippenko}, {Green} et~al.}}]{2007ApJ...662..808L}
\bibinfo{author}{\bibfnamefont{T.~R.} \bibnamefont{{Lauer}}},
  \bibinfo{author}{\bibfnamefont{S.~M.} \bibnamefont{{Faber}}},
  \bibinfo{author}{\bibfnamefont{D.}~\bibnamefont{{Richstone}}},
  \bibinfo{author}{\bibfnamefont{K.}~\bibnamefont{{Gebhardt}}},
  \bibinfo{author}{\bibfnamefont{S.}~\bibnamefont{{Tremaine}}},
  \bibinfo{author}{\bibfnamefont{M.}~\bibnamefont{{Postman}}},
  \bibinfo{author}{\bibfnamefont{A.}~\bibnamefont{{Dressler}}},
  \bibinfo{author}{\bibfnamefont{M.~C.} \bibnamefont{{Aller}}},
  \bibinfo{author}{\bibfnamefont{A.~V.} \bibnamefont{{Filippenko}}},
  \bibinfo{author}{\bibfnamefont{R.}~\bibnamefont{{Green}}},
  \bibnamefont{et~al.}, \bibinfo{journal}{\apj} \textbf{\bibinfo{volume}{662}},
  \bibinfo{pages}{808} (\bibinfo{year}{2007}), \eprint{arXiv:astro-ph/0606739}.

\bibitem[{\citenamefont{{Seaquist} and {Odegard}}(1991)}]{SO91}
\bibinfo{author}{\bibfnamefont{E.~R.} \bibnamefont{{Seaquist}}}
  \bibnamefont{and}
  \bibinfo{author}{\bibfnamefont{N.}~\bibnamefont{{Odegard}}},
  \bibinfo{journal}{Astrophys. J.} \textbf{\bibinfo{volume}{369}},
  \bibinfo{pages}{320} (\bibinfo{year}{1991}).

\bibitem[{\citenamefont{{Pettini} et~al.}(2002)\citenamefont{{Pettini}, {Rix},
  {Steidel}, {Adelberger}, {Hunt}, and {Shapley}}}]{Pea02}
\bibinfo{author}{\bibfnamefont{M.}~\bibnamefont{{Pettini}}},
  \bibinfo{author}{\bibfnamefont{S.~A.} \bibnamefont{{Rix}}},
  \bibinfo{author}{\bibfnamefont{C.~C.} \bibnamefont{{Steidel}}},
  \bibinfo{author}{\bibfnamefont{K.~L.} \bibnamefont{{Adelberger}}},
  \bibinfo{author}{\bibfnamefont{M.~P.} \bibnamefont{{Hunt}}},
  \bibnamefont{and} \bibinfo{author}{\bibfnamefont{A.~E.}
  \bibnamefont{{Shapley}}}, \bibinfo{journal}{Astrophys. J.}
  \textbf{\bibinfo{volume}{569}}, \bibinfo{pages}{742} (\bibinfo{year}{2002}),
  \eprint{arXiv:astro-ph/0110637}.

\bibitem[{\citenamefont{{Adelberger} et~al.}(2003)\citenamefont{{Adelberger},
  {Steidel}, {Shapley}, and {Pettini}}}]{Aea03}
\bibinfo{author}{\bibfnamefont{K.~L.} \bibnamefont{{Adelberger}}},
  \bibinfo{author}{\bibfnamefont{C.~C.} \bibnamefont{{Steidel}}},
  \bibinfo{author}{\bibfnamefont{A.~E.} \bibnamefont{{Shapley}}},
  \bibnamefont{and}
  \bibinfo{author}{\bibfnamefont{M.}~\bibnamefont{{Pettini}}},
  \bibinfo{journal}{Astrophys. J.} \textbf{\bibinfo{volume}{584}},
  \bibinfo{pages}{45} (\bibinfo{year}{2003}), \eprint{arXiv:astro-ph/0210314}.

\bibitem[{\citenamefont{{Heckman}}(2001)}]{H01}
\bibinfo{author}{\bibfnamefont{T.~M.} \bibnamefont{{Heckman}}}, in
  \emph{\bibinfo{booktitle}{Gas and Galaxy Evolution}}, edited by
  \bibinfo{editor}{\bibfnamefont{J.~E.} \bibnamefont{{Hibbard}}},
  \bibinfo{editor}{\bibfnamefont{M.}~\bibnamefont{{Rupen}}}, \bibnamefont{and}
  \bibinfo{editor}{\bibfnamefont{J.~H.} \bibnamefont{{van Gorkom}}}
  (\bibinfo{year}{2001}), vol. \bibinfo{volume}{240} of
  \emph{\bibinfo{series}{Astronomical Society of the Pacific Conference
  Series}}, pp. \bibinfo{pages}{345--+}.

\bibitem[{\citenamefont{{Aguirre} et~al.}(2001)\citenamefont{{Aguirre},
  {Hernquist}, {Schaye}, {Weinberg}, {Katz}, and {Gardner}}}]{Aea01}
\bibinfo{author}{\bibfnamefont{A.}~\bibnamefont{{Aguirre}}},
  \bibinfo{author}{\bibfnamefont{L.}~\bibnamefont{{Hernquist}}},
  \bibinfo{author}{\bibfnamefont{J.}~\bibnamefont{{Schaye}}},
  \bibinfo{author}{\bibfnamefont{D.~H.} \bibnamefont{{Weinberg}}},
  \bibinfo{author}{\bibfnamefont{N.}~\bibnamefont{{Katz}}}, \bibnamefont{and}
  \bibinfo{author}{\bibfnamefont{J.}~\bibnamefont{{Gardner}}},
  \bibinfo{journal}{Astrophys. J.} \textbf{\bibinfo{volume}{560}},
  \bibinfo{pages}{599} (\bibinfo{year}{2001}), \eprint{arXiv:astro-ph/0006345}.

\bibitem[{\citenamefont{{Cen} et~al.}(2005)\citenamefont{{Cen}, {Nagamine}, and
  {Ostriker}}}]{Cea05}
\bibinfo{author}{\bibfnamefont{R.}~\bibnamefont{{Cen}}},
  \bibinfo{author}{\bibfnamefont{K.}~\bibnamefont{{Nagamine}}},
  \bibnamefont{and} \bibinfo{author}{\bibfnamefont{J.~P.}
  \bibnamefont{{Ostriker}}}, \bibinfo{journal}{Astrophys. J.}
  \textbf{\bibinfo{volume}{635}}, \bibinfo{pages}{86} (\bibinfo{year}{2005}),
  \eprint{arXiv:astro-ph/0407143}.

\bibitem[{\citenamefont{{Scannapieco} et~al.}(2006)\citenamefont{{Scannapieco},
  {Pichon}, {Aracil}, {Petitjean}, {Thacker}, {Pogosyan}, {Bergeron}, and
  {Couchman}}}]{Sea06}
\bibinfo{author}{\bibfnamefont{E.}~\bibnamefont{{Scannapieco}}},
  \bibinfo{author}{\bibfnamefont{C.}~\bibnamefont{{Pichon}}},
  \bibinfo{author}{\bibfnamefont{B.}~\bibnamefont{{Aracil}}},
  \bibinfo{author}{\bibfnamefont{P.}~\bibnamefont{{Petitjean}}},
  \bibinfo{author}{\bibfnamefont{R.~J.} \bibnamefont{{Thacker}}},
  \bibinfo{author}{\bibfnamefont{D.}~\bibnamefont{{Pogosyan}}},
  \bibinfo{author}{\bibfnamefont{J.}~\bibnamefont{{Bergeron}}},
  \bibnamefont{and} \bibinfo{author}{\bibfnamefont{H.~M.~P.}
  \bibnamefont{{Couchman}}}, \bibinfo{journal}{MNRAS}
  \textbf{\bibinfo{volume}{365}}, \bibinfo{pages}{615} (\bibinfo{year}{2006}),
  \eprint{arXiv:astro-ph/0503001}.

\bibitem[{\citenamefont{{Reuter} et~al.}(1992)\citenamefont{{Reuter}, {Klein},
  {Lesch}, {Wielebinski}, and {Kronberg}}}]{Rea92}
\bibinfo{author}{\bibfnamefont{H.-P.} \bibnamefont{{Reuter}}},
  \bibinfo{author}{\bibfnamefont{U.}~\bibnamefont{{Klein}}},
  \bibinfo{author}{\bibfnamefont{H.}~\bibnamefont{{Lesch}}},
  \bibinfo{author}{\bibfnamefont{R.}~\bibnamefont{{Wielebinski}}},
  \bibnamefont{and} \bibinfo{author}{\bibfnamefont{P.~P.}
  \bibnamefont{{Kronberg}}}, \bibinfo{journal}{Astron. Astrophys.}
  \textbf{\bibinfo{volume}{256}}, \bibinfo{pages}{10} (\bibinfo{year}{1992}).

\bibitem[{\citenamefont{{Clarke} et~al.}(2001)\citenamefont{{Clarke},
  {Kronberg}, and {B{\"o}hringer}}}]{2001ApJ...547L.111C}
\bibinfo{author}{\bibfnamefont{T.~E.} \bibnamefont{{Clarke}}},
  \bibinfo{author}{\bibfnamefont{P.~P.} \bibnamefont{{Kronberg}}},
  \bibnamefont{and}
  \bibinfo{author}{\bibfnamefont{H.}~\bibnamefont{{B{\"o}hringer}}},
  \bibinfo{journal}{Astrophys. J. Lett.} \textbf{\bibinfo{volume}{547}},
  \bibinfo{pages}{L111} (\bibinfo{year}{2001}),
  \eprint{arXiv:astro-ph/0011281}.

\bibitem[{\citenamefont{{Govoni} et~al.}(2006)\citenamefont{{Govoni}, {Murgia},
  {Feretti}, {Giovannini}, {Dolag}, and {Taylor}}}]{Gea06}
\bibinfo{author}{\bibfnamefont{F.}~\bibnamefont{{Govoni}}},
  \bibinfo{author}{\bibfnamefont{M.}~\bibnamefont{{Murgia}}},
  \bibinfo{author}{\bibfnamefont{L.}~\bibnamefont{{Feretti}}},
  \bibinfo{author}{\bibfnamefont{G.}~\bibnamefont{{Giovannini}}},
  \bibinfo{author}{\bibfnamefont{K.}~\bibnamefont{{Dolag}}}, \bibnamefont{and}
  \bibinfo{author}{\bibfnamefont{G.~B.} \bibnamefont{{Taylor}}},
  \bibinfo{journal}{Astron. Astrophys.} \textbf{\bibinfo{volume}{460}},
  \bibinfo{pages}{425} (\bibinfo{year}{2006}), \eprint{arXiv:astro-ph/0608433}.

\bibitem[{\citenamefont{{Br{\"u}ggen} et~al.}(2005)\citenamefont{{Br{\"u}ggen},
  {Ruszkowski}, {Simionescu}, {Hoeft}, and {Dalla Vecchia}}}]{Bea05}
\bibinfo{author}{\bibfnamefont{M.}~\bibnamefont{{Br{\"u}ggen}}},
  \bibinfo{author}{\bibfnamefont{M.}~\bibnamefont{{Ruszkowski}}},
  \bibinfo{author}{\bibfnamefont{A.}~\bibnamefont{{Simionescu}}},
  \bibinfo{author}{\bibfnamefont{M.}~\bibnamefont{{Hoeft}}}, \bibnamefont{and}
  \bibinfo{author}{\bibfnamefont{C.}~\bibnamefont{{Dalla Vecchia}}},
  \bibinfo{journal}{Astrophys. J. Lett.} \textbf{\bibinfo{volume}{631}},
  \bibinfo{pages}{L21} (\bibinfo{year}{2005}), \eprint{arXiv:astro-ph/0508231}.

\bibitem[{\citenamefont{{Doroshkevich}
  et~al.}(2001)\citenamefont{{Doroshkevich}, {Tucker}, {Fong}, {Turchaninov},
  and {Lin}}}]{Dea01}
\bibinfo{author}{\bibfnamefont{A.~G.} \bibnamefont{{Doroshkevich}}},
  \bibinfo{author}{\bibfnamefont{D.~L.} \bibnamefont{{Tucker}}},
  \bibinfo{author}{\bibfnamefont{R.}~\bibnamefont{{Fong}}},
  \bibinfo{author}{\bibfnamefont{V.}~\bibnamefont{{Turchaninov}}},
  \bibnamefont{and} \bibinfo{author}{\bibfnamefont{H.}~\bibnamefont{{Lin}}},
  \bibinfo{journal}{Month. Not. Roy. Astron. Soc.}
  \textbf{\bibinfo{volume}{322}}, \bibinfo{pages}{369} (\bibinfo{year}{2001}).

\bibitem[{\citenamefont{{Miniati} et~al.}(2000)\citenamefont{{Miniati}, {Ryu},
  {Kang}, {Jones}, {Cen}, and {Ostriker}}}]{Mea00}
\bibinfo{author}{\bibfnamefont{F.}~\bibnamefont{{Miniati}}},
  \bibinfo{author}{\bibfnamefont{D.}~\bibnamefont{{Ryu}}},
  \bibinfo{author}{\bibfnamefont{H.}~\bibnamefont{{Kang}}},
  \bibinfo{author}{\bibfnamefont{T.~W.} \bibnamefont{{Jones}}},
  \bibinfo{author}{\bibfnamefont{R.}~\bibnamefont{{Cen}}}, \bibnamefont{and}
  \bibinfo{author}{\bibfnamefont{J.~P.} \bibnamefont{{Ostriker}}},
  \bibinfo{journal}{Astrophys. J.} \textbf{\bibinfo{volume}{542}},
  \bibinfo{pages}{608} (\bibinfo{year}{2000}), \eprint{arXiv:astro-ph/0005444}.

\bibitem[{\citenamefont{{Kang} et~al.}(2005)\citenamefont{{Kang}, {Ryu}, {Cen},
  and {Song}}}]{KRCS05}
\bibinfo{author}{\bibfnamefont{H.}~\bibnamefont{{Kang}}},
  \bibinfo{author}{\bibfnamefont{D.}~\bibnamefont{{Ryu}}},
  \bibinfo{author}{\bibfnamefont{R.}~\bibnamefont{{Cen}}}, \bibnamefont{and}
  \bibinfo{author}{\bibfnamefont{D.}~\bibnamefont{{Song}}},
  \bibinfo{journal}{ApJ} \textbf{\bibinfo{volume}{620}}, \bibinfo{pages}{21}
  (\bibinfo{year}{2005}), \eprint{arXiv:astro-ph/0410477}.

\bibitem[{\citenamefont{{Kulsrud} et~al.}(1997)\citenamefont{{Kulsrud}, {Cen},
  {Ostriker}, and {Ryu}}}]{KCOR97}
\bibinfo{author}{\bibfnamefont{R.~M.} \bibnamefont{{Kulsrud}}},
  \bibinfo{author}{\bibfnamefont{R.}~\bibnamefont{{Cen}}},
  \bibinfo{author}{\bibfnamefont{J.~P.} \bibnamefont{{Ostriker}}},
  \bibnamefont{and} \bibinfo{author}{\bibfnamefont{D.}~\bibnamefont{{Ryu}}},
  \bibinfo{journal}{ApJ} \textbf{\bibinfo{volume}{480}}, \bibinfo{pages}{481}
  (\bibinfo{year}{1997}), \eprint{arXiv:astro-ph/9607141}.

\bibitem[{\citenamefont{{Loeb} and {Waxman}}(2000)}]{LW00}
\bibinfo{author}{\bibfnamefont{A.}~\bibnamefont{{Loeb}}} \bibnamefont{and}
  \bibinfo{author}{\bibfnamefont{E.}~\bibnamefont{{Waxman}}},
  \bibinfo{journal}{Nature} \textbf{\bibinfo{volume}{405}},
  \bibinfo{pages}{156} (\bibinfo{year}{2000}), \eprint{arXiv:astro-ph/0003447}.

\bibitem[{\citenamefont{{Miniati}}(2002)}]{M02}
\bibinfo{author}{\bibfnamefont{F.}~\bibnamefont{{Miniati}}},
  \bibinfo{journal}{MNRAS} \textbf{\bibinfo{volume}{337}}, \bibinfo{pages}{199}
  (\bibinfo{year}{2002}), \eprint{arXiv:astro-ph/0203014}.

\bibitem[{\citenamefont{{Keshet} et~al.}(2003)\citenamefont{{Keshet}, {Waxman},
  {Loeb}, {Springel}, and {Hernquist}}}]{Kea03}
\bibinfo{author}{\bibfnamefont{U.}~\bibnamefont{{Keshet}}},
  \bibinfo{author}{\bibfnamefont{E.}~\bibnamefont{{Waxman}}},
  \bibinfo{author}{\bibfnamefont{A.}~\bibnamefont{{Loeb}}},
  \bibinfo{author}{\bibfnamefont{V.}~\bibnamefont{{Springel}}},
  \bibnamefont{and}
  \bibinfo{author}{\bibfnamefont{L.}~\bibnamefont{{Hernquist}}},
  \bibinfo{journal}{ApJ} \textbf{\bibinfo{volume}{585}}, \bibinfo{pages}{128}
  (\bibinfo{year}{2003}), \eprint{arXiv:astro-ph/0202318}.

\bibitem[{\citenamefont{{Vink} and {Laming}}(2003)}]{VL03}
\bibinfo{author}{\bibfnamefont{J.}~\bibnamefont{{Vink}}} \bibnamefont{and}
  \bibinfo{author}{\bibfnamefont{J.~M.} \bibnamefont{{Laming}}},
  \bibinfo{journal}{ApJ} \textbf{\bibinfo{volume}{584}}, \bibinfo{pages}{758}
  (\bibinfo{year}{2003}), \eprint{arXiv:astro-ph/0210669}.

\bibitem[{\citenamefont{{Berezhko} et~al.}(2003)\citenamefont{{Berezhko},
  {Ksenofontov}, and {V{\"o}lk}}}]{BKV03}
\bibinfo{author}{\bibfnamefont{E.~G.} \bibnamefont{{Berezhko}}},
  \bibinfo{author}{\bibfnamefont{L.~T.} \bibnamefont{{Ksenofontov}}},
  \bibnamefont{and} \bibinfo{author}{\bibfnamefont{H.~J.}
  \bibnamefont{{V{\"o}lk}}}, \bibinfo{journal}{AA}
  \textbf{\bibinfo{volume}{412}}, \bibinfo{pages}{L11} (\bibinfo{year}{2003}),
  \eprint{arXiv:astro-ph/0310862}.

\bibitem[{\citenamefont{Greisen}(1966)}]{G66}
\bibinfo{author}{\bibfnamefont{K.}~\bibnamefont{Greisen}},
  \bibinfo{journal}{Phys. Rev. Lett.} \textbf{\bibinfo{volume}{16}},
  \bibinfo{pages}{748} (\bibinfo{year}{1966}).

\bibitem[{\citenamefont{Zatsepin and Kuzmin}(1966)}]{ZK66}
\bibinfo{author}{\bibfnamefont{G.~T.} \bibnamefont{Zatsepin}} \bibnamefont{and}
  \bibinfo{author}{\bibfnamefont{V.~A.} \bibnamefont{Kuzmin}},
  \bibinfo{journal}{JETP Lett.} \textbf{\bibinfo{volume}{4}},
  \bibinfo{pages}{78} (\bibinfo{year}{1966}).

\bibitem[{\citenamefont{Teyssier}(2002)}]{T02}
\bibinfo{author}{\bibfnamefont{R.}~\bibnamefont{Teyssier}},
  \bibinfo{journal}{A\&A} \textbf{\bibinfo{volume}{385}}, \bibinfo{pages}{337}
  (\bibinfo{year}{2002}).

\bibitem[{\citenamefont{{Waxman}}(1995)}]{W95}
\bibinfo{author}{\bibfnamefont{E.}~\bibnamefont{{Waxman}}},
  \bibinfo{journal}{Physical Review Letters} \textbf{\bibinfo{volume}{75}},
  \bibinfo{pages}{386} (\bibinfo{year}{1995}), \eprint{arXiv:astro-ph/9505082}.

\bibitem[{\citenamefont{{Alvarez-Mu{\~n}iz}
  et~al.}(2002)\citenamefont{{Alvarez-Mu{\~n}iz}, {Engel}, and
  {Stanev}}}]{AMES02}
\bibinfo{author}{\bibfnamefont{J.}~\bibnamefont{{Alvarez-Mu{\~n}iz}}},
  \bibinfo{author}{\bibfnamefont{R.}~\bibnamefont{{Engel}}}, \bibnamefont{and}
  \bibinfo{author}{\bibfnamefont{T.}~\bibnamefont{{Stanev}}},
  \bibinfo{journal}{Astrophys.~J.~} \textbf{\bibinfo{volume}{572}},
  \bibinfo{pages}{185} (\bibinfo{year}{2002}), \eprint{arXiv:astro-ph/0112227}.

\bibitem[{\citenamefont{{Takami} and {Sato}}(2007{\natexlab{b}})}]{TK07}
\bibinfo{author}{\bibfnamefont{H.}~\bibnamefont{{Takami}}} \bibnamefont{and}
  \bibinfo{author}{\bibfnamefont{K.}~\bibnamefont{{Sato}}},
  \bibinfo{journal}{e-prints}  (\bibinfo{year}{2007}{\natexlab{b}}),
  \eprint{arXiv:0711.2386}.

\bibitem[{\citenamefont{{Alcock} and {Hatchett}}(1978)}]{1978ApJ...222..456A}
\bibinfo{author}{\bibfnamefont{C.}~\bibnamefont{{Alcock}}} \bibnamefont{and}
  \bibinfo{author}{\bibfnamefont{S.}~\bibnamefont{{Hatchett}}},
  \bibinfo{journal}{Astrophys.~J.} \textbf{\bibinfo{volume}{222}},
  \bibinfo{pages}{456} (\bibinfo{year}{1978}).

\bibitem[{\citenamefont{Sigl et~al.}(1999)\citenamefont{Sigl, Lemoine, and
  Biermann}}]{SLB99}
\bibinfo{author}{\bibfnamefont{G.}~\bibnamefont{Sigl}},
  \bibinfo{author}{\bibfnamefont{M.}~\bibnamefont{Lemoine}}, \bibnamefont{and}
  \bibinfo{author}{\bibfnamefont{P.}~\bibnamefont{Biermann}},
  \bibinfo{journal}{Astropart. Phys.} \textbf{\bibinfo{volume}{10}},
  \bibinfo{pages}{141} (\bibinfo{year}{1999}), \eprint{astro-ph/9806283}.

\bibitem[{\citenamefont{{Waxman}}(2001)}]{W01}
\bibinfo{author}{\bibfnamefont{E.}~\bibnamefont{{Waxman}}}, in
  \emph{\bibinfo{booktitle}{Physics and Astrophysics of Ultra-High-Energy
  Cosmic Rays}}, edited by
  \bibinfo{editor}{\bibfnamefont{M.}~\bibnamefont{{Lemoine}}} \bibnamefont{and}
  \bibinfo{editor}{\bibfnamefont{G.}~\bibnamefont{{Sigl}}}
  (\bibinfo{year}{2001}), vol. \bibinfo{volume}{576} of
  \emph{\bibinfo{series}{Lecture Notes in Physics, Berlin Springer Verlag}},
  pp. \bibinfo{pages}{122--+}.

\bibitem[{\citenamefont{{Saunders} et~al.}(2000)\citenamefont{{Saunders},
  {Sutherland}, {Maddox}, {Keeble}, {Oliver}, {Rowan-Robinson}, {McMahon},
  {Efstathiou}, {Tadros}, {White} et~al.}}]{2000MNRAS.317...55S}
\bibinfo{author}{\bibfnamefont{W.}~\bibnamefont{{Saunders}}},
  \bibinfo{author}{\bibfnamefont{W.~J.} \bibnamefont{{Sutherland}}},
  \bibinfo{author}{\bibfnamefont{S.~J.} \bibnamefont{{Maddox}}},
  \bibinfo{author}{\bibfnamefont{O.}~\bibnamefont{{Keeble}}},
  \bibinfo{author}{\bibfnamefont{S.~J.} \bibnamefont{{Oliver}}},
  \bibinfo{author}{\bibfnamefont{M.}~\bibnamefont{{Rowan-Robinson}}},
  \bibinfo{author}{\bibfnamefont{R.~G.} \bibnamefont{{McMahon}}},
  \bibinfo{author}{\bibfnamefont{G.~P.} \bibnamefont{{Efstathiou}}},
  \bibinfo{author}{\bibfnamefont{H.}~\bibnamefont{{Tadros}}},
  \bibinfo{author}{\bibfnamefont{S.~D.~M.} \bibnamefont{{White}}},
  \bibnamefont{et~al.}, \bibinfo{journal}{Month. Not. Roy. Astron. Soc.}
  \textbf{\bibinfo{volume}{317}}, \bibinfo{pages}{55} (\bibinfo{year}{2000}),
  \eprint{arXiv:astro-ph/0001117}.

\bibitem[{\citenamefont{{G{\'o}rski} et~al.}(2005)\citenamefont{{G{\'o}rski},
  {Hivon}, {Banday}, {Wandelt}, {Hansen}, {Reinecke}, and
  {Bartelmann}}}]{2005ApJ...622..759G}
\bibinfo{author}{\bibfnamefont{K.~M.} \bibnamefont{{G{\'o}rski}}},
  \bibinfo{author}{\bibfnamefont{E.}~\bibnamefont{{Hivon}}},
  \bibinfo{author}{\bibfnamefont{A.~J.} \bibnamefont{{Banday}}},
  \bibinfo{author}{\bibfnamefont{B.~D.} \bibnamefont{{Wandelt}}},
  \bibinfo{author}{\bibfnamefont{F.~K.} \bibnamefont{{Hansen}}},
  \bibinfo{author}{\bibfnamefont{M.}~\bibnamefont{{Reinecke}}},
  \bibnamefont{and}
  \bibinfo{author}{\bibfnamefont{M.}~\bibnamefont{{Bartelmann}}},
  \bibinfo{journal}{Astrophys. J.} \textbf{\bibinfo{volume}{622}},
  \bibinfo{pages}{759} (\bibinfo{year}{2005}), \eprint{arXiv:astro-ph/0409513}.

\bibitem[{\citenamefont{{Harari} et~al.}(1999)\citenamefont{{Harari},
  {Mollerach}, and {Roulet}}}]{1999JHEP...08..022H}
\bibinfo{author}{\bibfnamefont{D.}~\bibnamefont{{Harari}}},
  \bibinfo{author}{\bibfnamefont{S.}~\bibnamefont{{Mollerach}}},
  \bibnamefont{and} \bibinfo{author}{\bibfnamefont{E.}~\bibnamefont{{Roulet}}},
  \bibinfo{journal}{Journal of High Energy Physics}
  \textbf{\bibinfo{volume}{8}}, \bibinfo{pages}{22} (\bibinfo{year}{1999}),
  \eprint{arXiv:astro-ph/9906309}.

\bibitem[{\citenamefont{{Aloisio} and {Berezinsky}}(2004)}]{AB04}
\bibinfo{author}{\bibfnamefont{R.}~\bibnamefont{{Aloisio}}} \bibnamefont{and}
  \bibinfo{author}{\bibfnamefont{V.}~\bibnamefont{{Berezinsky}}},
  \bibinfo{journal}{ApJ} \textbf{\bibinfo{volume}{612}}, \bibinfo{pages}{900}
  (\bibinfo{year}{2004}).

\bibitem[{\citenamefont{Berezinsky and Gazizov}(2006)}]{BG06}
\bibinfo{author}{\bibfnamefont{V.}~\bibnamefont{Berezinsky}} \bibnamefont{and}
  \bibinfo{author}{\bibfnamefont{A.}~\bibnamefont{Gazizov}},
  \bibinfo{journal}{ApJ} \textbf{\bibinfo{volume}{643}}, \bibinfo{pages}{8}
  (\bibinfo{year}{2006}).

\bibitem[{\citenamefont{Lemoine}(2005)}]{L05}
\bibinfo{author}{\bibfnamefont{M.}~\bibnamefont{Lemoine}},
  \bibinfo{journal}{Phys. Rev. D} \textbf{\bibinfo{volume}{71}},
  \bibinfo{pages}{083007} (\bibinfo{year}{2005}).

\bibitem[{\citenamefont{Aloisio and Berezinsky}(2005)}]{AB05}
\bibinfo{author}{\bibfnamefont{R.}~\bibnamefont{Aloisio}} \bibnamefont{and}
  \bibinfo{author}{\bibfnamefont{V.}~\bibnamefont{Berezinsky}},
  \bibinfo{journal}{ApJ} \textbf{\bibinfo{volume}{625}}, \bibinfo{pages}{249}
  (\bibinfo{year}{2005}).

\bibitem[{\citenamefont{{Bouchaud} and {Georges}}(1990)}]{BG90}
\bibinfo{author}{\bibfnamefont{J.-P.} \bibnamefont{{Bouchaud}}}
  \bibnamefont{and}
  \bibinfo{author}{\bibfnamefont{A.}~\bibnamefont{{Georges}}},
  \bibinfo{journal}{Phys.~Rep.} \textbf{\bibinfo{volume}{195}},
  \bibinfo{pages}{127} (\bibinfo{year}{1990}).

\bibitem[{\citenamefont{{Ball} et~al.}(1987)\citenamefont{{Ball}, {Havlin}, and
  {Weiss}}}]{BHW87}
\bibinfo{author}{\bibfnamefont{R.~C.} \bibnamefont{{Ball}}},
  \bibinfo{author}{\bibfnamefont{S.}~\bibnamefont{{Havlin}}}, \bibnamefont{and}
  \bibinfo{author}{\bibfnamefont{G.~H.} \bibnamefont{{Weiss}}},
  \bibinfo{journal}{Journal of Physics A Mathematical General}
  \textbf{\bibinfo{volume}{20}}, \bibinfo{pages}{4055} (\bibinfo{year}{1987}).

\bibitem[{\citenamefont{{Norman} et~al.}(1995)\citenamefont{{Norman},
  {Melrose}, and {Achterberg}}}]{NMA95}
\bibinfo{author}{\bibfnamefont{C.~A.} \bibnamefont{{Norman}}},
  \bibinfo{author}{\bibfnamefont{D.~B.} \bibnamefont{{Melrose}}},
  \bibnamefont{and}
  \bibinfo{author}{\bibfnamefont{A.}~\bibnamefont{{Achterberg}}},
  \bibinfo{journal}{Astrophys.~J.~} \textbf{\bibinfo{volume}{454}},
  \bibinfo{pages}{60} (\bibinfo{year}{1995}).

\bibitem[{\citenamefont{Casse et~al.}(2002)\citenamefont{Casse, Lemoine, and
  Pelletier}}]{CLP02}
\bibinfo{author}{\bibfnamefont{F.}~\bibnamefont{Casse}},
  \bibinfo{author}{\bibfnamefont{M.}~\bibnamefont{Lemoine}}, \bibnamefont{and}
  \bibinfo{author}{\bibfnamefont{G.}~\bibnamefont{Pelletier}},
  \bibinfo{journal}{Phys. Rev. D} \textbf{\bibinfo{volume}{65}},
  \bibinfo{pages}{023002} (\bibinfo{year}{2002}).

\bibitem[{\citenamefont{Gorbunov et~al.}(2007)\citenamefont{Gorbunov, Tinyakov,
  Tkachev, and Troitsky}}]{GTTT07}
\bibinfo{author}{\bibfnamefont{D.}~\bibnamefont{Gorbunov}},
  \bibinfo{author}{\bibfnamefont{P.}~\bibnamefont{Tinyakov}},
  \bibinfo{author}{\bibfnamefont{I.}~\bibnamefont{Tkachev}}, \bibnamefont{and}
  \bibinfo{author}{\bibfnamefont{S.}~\bibnamefont{Troitsky}}
  (\bibinfo{year}{2007}), \eprint{arXiv:0711.4060 [astro-ph]}.

\bibitem[{\citenamefont{{Waxman} et~al.}(1997)\citenamefont{{Waxman}, {Fisher},
  and {Piran}}}]{WFP96}
\bibinfo{author}{\bibfnamefont{E.}~\bibnamefont{{Waxman}}},
  \bibinfo{author}{\bibfnamefont{K.~B.} \bibnamefont{{Fisher}}},
  \bibnamefont{and} \bibinfo{author}{\bibfnamefont{T.}~\bibnamefont{{Piran}}},
  \bibinfo{journal}{Astrophys.~J.~} \textbf{\bibinfo{volume}{483}},
  \bibinfo{pages}{1} (\bibinfo{year}{1997}), \eprint{arXiv:astro-ph/9604005}.

\bibitem[{\citenamefont{Berezinsky et~al.}(2006)\citenamefont{Berezinsky,
  Gazizov, and Grigorieva}}]{BGG02}
\bibinfo{author}{\bibfnamefont{V.}~\bibnamefont{Berezinsky}},
  \bibinfo{author}{\bibfnamefont{A.}~\bibnamefont{Gazizov}}, \bibnamefont{and}
  \bibinfo{author}{\bibfnamefont{S.}~\bibnamefont{Grigorieva}},
  \bibinfo{journal}{Phys. Rev. D} \textbf{\bibinfo{volume}{74}},
  \bibinfo{pages}{043005} (\bibinfo{year}{2006}),
  \eprint{arXiv:hep-ph/0204357}.

\bibitem[{\citenamefont{{Harari} et~al.}(2006)\citenamefont{{Harari},
  {Mollerach}, and {Roulet}}}]{2006JCAP...11..012H}
\bibinfo{author}{\bibfnamefont{D.}~\bibnamefont{{Harari}}},
  \bibinfo{author}{\bibfnamefont{S.}~\bibnamefont{{Mollerach}}},
  \bibnamefont{and} \bibinfo{author}{\bibfnamefont{E.}~\bibnamefont{{Roulet}}},
  \bibinfo{journal}{Journal of Cosmology and Astro-Particle Physics}
  \textbf{\bibinfo{volume}{11}}, \bibinfo{pages}{12} (\bibinfo{year}{2006}),
  \eprint{arXiv:astro-ph/0609294}.

\bibitem[{\citenamefont{{Cuoco} et~al.}(2006)\citenamefont{{Cuoco},
  {D'Abrusco}, {Longo}, {Miele}, and {Serpico}}}]{2006JCAP...01..009C}
\bibinfo{author}{\bibfnamefont{A.}~\bibnamefont{{Cuoco}}},
  \bibinfo{author}{\bibfnamefont{R.}~\bibnamefont{{D'Abrusco}}},
  \bibinfo{author}{\bibfnamefont{G.}~\bibnamefont{{Longo}}},
  \bibinfo{author}{\bibfnamefont{G.}~\bibnamefont{{Miele}}}, \bibnamefont{and}
  \bibinfo{author}{\bibfnamefont{P.~D.} \bibnamefont{{Serpico}}},
  \bibinfo{journal}{Journal of Cosmology and Astro-Particle Physics}
  \textbf{\bibinfo{volume}{1}}, \bibinfo{pages}{9} (\bibinfo{year}{2006}),
  \eprint{arXiv:astro-ph/0510765}.

\bibitem[{\citenamefont{{Giacalone} and {Jokipii}}(1999)}]{GJ99}
\bibinfo{author}{\bibfnamefont{J.}~\bibnamefont{{Giacalone}}} \bibnamefont{and}
  \bibinfo{author}{\bibfnamefont{J.~R.} \bibnamefont{{Jokipii}}},
  \bibinfo{journal}{Astrophys.~J.~} \textbf{\bibinfo{volume}{520}},
  \bibinfo{pages}{204} (\bibinfo{year}{1999}).

\bibitem[{\citenamefont{Candia and Roulet}(2004)}]{CR04}
\bibinfo{author}{\bibfnamefont{J.}~\bibnamefont{Candia}} \bibnamefont{and}
  \bibinfo{author}{\bibfnamefont{E.}~\bibnamefont{Roulet}},
  \bibinfo{journal}{JCAP} \textbf{\bibinfo{volume}{0410}}, \bibinfo{pages}{007}
  (\bibinfo{year}{2004}).

\end{thebibliography}

\end{document}